\documentclass[fleqn,usenatbib,useAMS]{mnras}

\usepackage{graphicx}	
\usepackage{amsmath}	
\usepackage{amssymb}	
\usepackage{multicol}        
\usepackage{bm}		
\usepackage{pdflscape}	
\usepackage{url}
\usepackage{color}
\usepackage{comment}

\newcommand{\Lya}{Ly$\alpha$}
\newcommand{\OII}{[O\,{\sc ii}]}
\newcommand{\OIII}{[O\,{\sc iii}]}
\newcommand{\OIIIuv}{O\,{\sc iii}]}
\newcommand{\NII}{[N\,{\sc ii}]}
\newcommand{\SII}{[S\,{\sc ii}]}
\newcommand{\Ha}{H$\alpha$}
\newcommand{\Hb}{H$\beta$}

\newcommand{\CIII}{C{\sc iii}]}
\newcommand{\CIV}{C{\sc iv}}
\newcommand{\HeI}{He{\sc i}}
\newcommand{\HeII}{He{\sc ii}}
\newcommand{\NV}{N{\sc v}}

\newcommand{\HI}{H{\sc i}}
\newcommand{\HII}{H{\sc ii}}

\newcommand{\Msun}{$M_{\odot}$}
\newcommand{\Zsun}{Z$_{\odot}$}

\newcommand{\cloudy}{\textsc{Cloudy}}

\usepackage[T1]{fontenc}
\usepackage{ae,aecompl}

\usepackage{newtxtext,newtxmath}
\title[PopIII and DCBH diagnostics]{Diagnostics for PopIII galaxies and Direct Collapse Black Holes \\ in the early universe}
\author[Nakajima \& Maiolino]{K. Nakajima$^{1}$%
 and R. Maiolino$^{2,3,4}$%
\thanks{Contact e-mail: \href{mailto:kimihiko.nakajima@nao.ac.jp}{kimihiko.nakajima@nao.ac.jp}}%
\\
$^{1}$National Astronomical Observatory of Japan, 2-21-1 Osawa, Mitaka, Tokyo 181-8588, Japan\\
$^{2}$Kavli Institute for Cosmology, University of Cambridge, Madingley Road, Cambridge, CB3 0HA, UK\\
$^{3}$ Cavendish Laboratory Astrophysics Group, University of Cambridge, 19 JJ Thomson Avenue, Cambridge, CB3 0HE, UK\\
$^{4}$Department of Physics and Astronomy, University College London, Gower Street, London WC1E 6BT, UK
}

\date{Accepted for publication in MNRAS on Apr 25, 2022}

\pubyear{2022}

\begin{document}
\label{firstpage}
\pagerange{\pageref{firstpage}--\pageref{lastpage}}
\maketitle

\begin{abstract}
Forthcoming observational facilities will make the exploration of the early universe routine, likely probing large populations of galaxies at very low metallicities. It will therefore be important to have diagnostics that can solidly identify and distinguish different classes of objects in such low metallicity regimes. We use new photoionisation models to develop diagnostic diagrams involving various nebular lines. We show that combinations of these diagrams allow the identification and discrimination of the following classes of objects in the early universe: PopIII and Direct Collapse Black Holes (DCBH) in pristine environments, PopIII and DCBH embedded in slightly enriched ISM ($\rm Z\sim 10^{-5}-10^{-4}$), (metal poor) PopII and AGN in enriched ISM. Diagnostics involving rest-frame optical lines (that will be accessible by JWST) have a better discriminatory power, but also rest-frame UV diagnostics can provide very useful information. Interestingly, we find that metal lines such as \OIII$\lambda5007$ and \CIV$\lambda1549$ can remain relatively strong (about a factor of 0.1--1 relative H$\beta$ and \HeII$\lambda1640$, respectively), even in extremely metal poor environments ($\rm Z\sim 10^{-5}-10^{-4}$), which could be embedding PopIII galaxies and DCBH.
\end{abstract}




\section{Introduction}
\label{sec:introduction}

Detecting galaxies hosting the very first generation of stars, formed out of the primordial, pristine gas, and often refereed to as PopIII stars, is one of the primary, ambitious goals of modern astrophysics. These galaxies are expected to be found between the end of the ``dark ages'' (z$\sim$20--30) and the epoch of main re-ionisation (z$\sim$7) \citep[e.g.][]{abel2002,bromm2002,schneider2006,jeon2015,jaacks2018}, although some models expect that clumps of unprocessed gas may lead to formation of PopIII even at z$\sim$3 \citep{liu2020}. PopIII galaxies are expected to be very faint (low mass) and rare, since the chemical enrichment (hence the transition to PopII) proceeds very quickly \citep[e.g.][]{tornatore2007,jaacks2018}. 

Some claims of PopIII detection \citep{sobral2015} in a luminous \Lya\ emitter, CR7, at $z=6.6$ were not confirmed by later analyses \citep{bowler2017,shibuya2018_spec}. The expectation is that PopIII galaxies might be identified with the forthcoming generation of optical/near-IR observatories, especially the James Webb Space Telescope (JWST) and the Extremely Large Telescopes (ELTs), although their detectability strongly depends on the specific model and cosmological simulations and their assumptions \citep[e.g.][]{stiavelli2010,pawlik2011,inayoshi2018,visbal2020,vikaeus2021}

Another important class of primeval objects are the seeds of supermassive black holes at the center of primeval galaxies. The finding of hyper-massive black holes ($M_{\rm BH}>10^9M_{\odot}$) already in place by z$\sim$7.5, has triggered the development of numerous models and scenario about their formation and the nature of their progenitors, including massive stellar (PopIII) remnants accreting at super-Eddington rates, nuclear stellar clusters and the so-called Direct Collapse Black Holes (DCBHs), i.e. black holes with masses in the range of $\rm 10^5-10^6~M_{\odot}$ resulting from the putative direct collapse of massive gas clouds under specific conditions \citep[e.g.][]{volonteri2012,valiante2016,inayoshi2020,beckmann2019}.
 Also in this case, some models expect that DCBH or other forms of supermassive black holes seeds may be detectable with the forthcoming generation of observing facilities (\citealt{volonteri2017,valiante2018,natarajan2017,barrow2018,habouzit2019,whalen2020} and references therein).
Some potential candidates of DCBHs have actually been suggested in the past based on broad-band photometry, but which still need spectroscopic confirmation
\citep{pacucci2016}. An interpretation of the nature of CR7 as primeval DCBH \citep{pacucci2017,hartwig2016} has subsequently been questioned by subsequent analyses of this source \citep{bowler2017,shibuya2018_spec}.

Various past studies have investigated what would be the properties of these classes of primeval objects and what would be their observational signatures. As a consequence of their higher temperature, PopIII stars are expected to have slopes bluer and ionising continua harder than later generation of (PopII) stars. Along with the lack of metals, these properties are expected to result into strong hydrogen and helium line emission and, in particular, strong \HeII\ emission lines. Specifically, the presence of the \HeII$\lambda 1640$ line, along with the absence of other metal lines, has been considered a signature of ISM photoionised by PopIII stars \citep[e.g.][]{schaerer2003,raiter2010,inoue2011_metal_poor}. 
CR7 was a notable example \citep{sobral2015}, although the detection of \HeII$\lambda 1640$ and the claim of PopIII was not confirmed by subsequent studies (\citealt{bowler2017,shibuya2018_spec}; see also \citealt{sobral2019}).

\begin{figure*}
  \centering
    \begin{tabular}{c}
      \begin{minipage}{0.49\hsize}
        \begin{flushleft}
         \includegraphics[bb=23 168 550 587, width=0.95\textwidth]{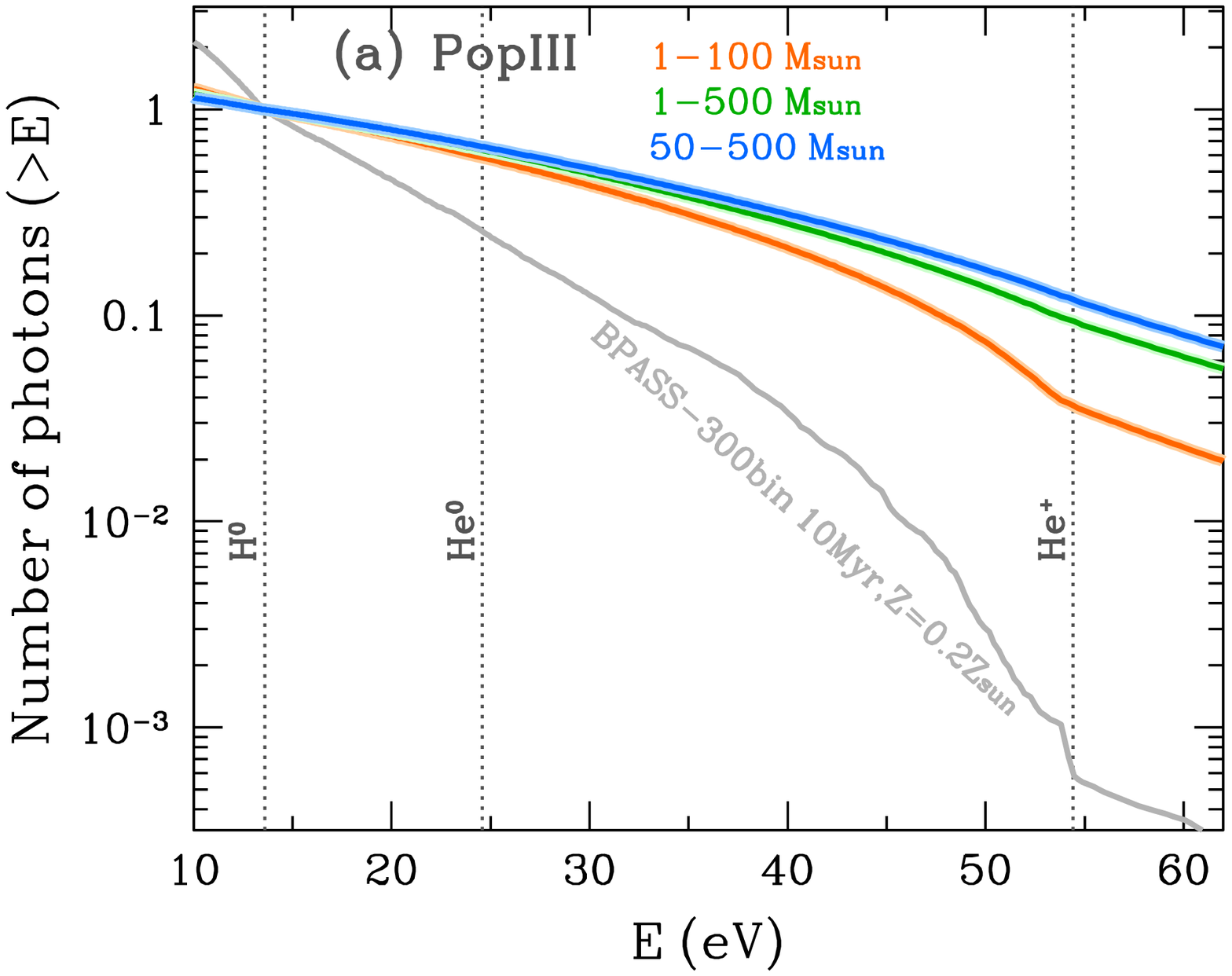}
        \end{flushleft}
      \end{minipage}
      \begin{minipage}{0.49\hsize}
        \begin{center}
         \includegraphics[bb=23 168 550 587, width=0.95\textwidth]{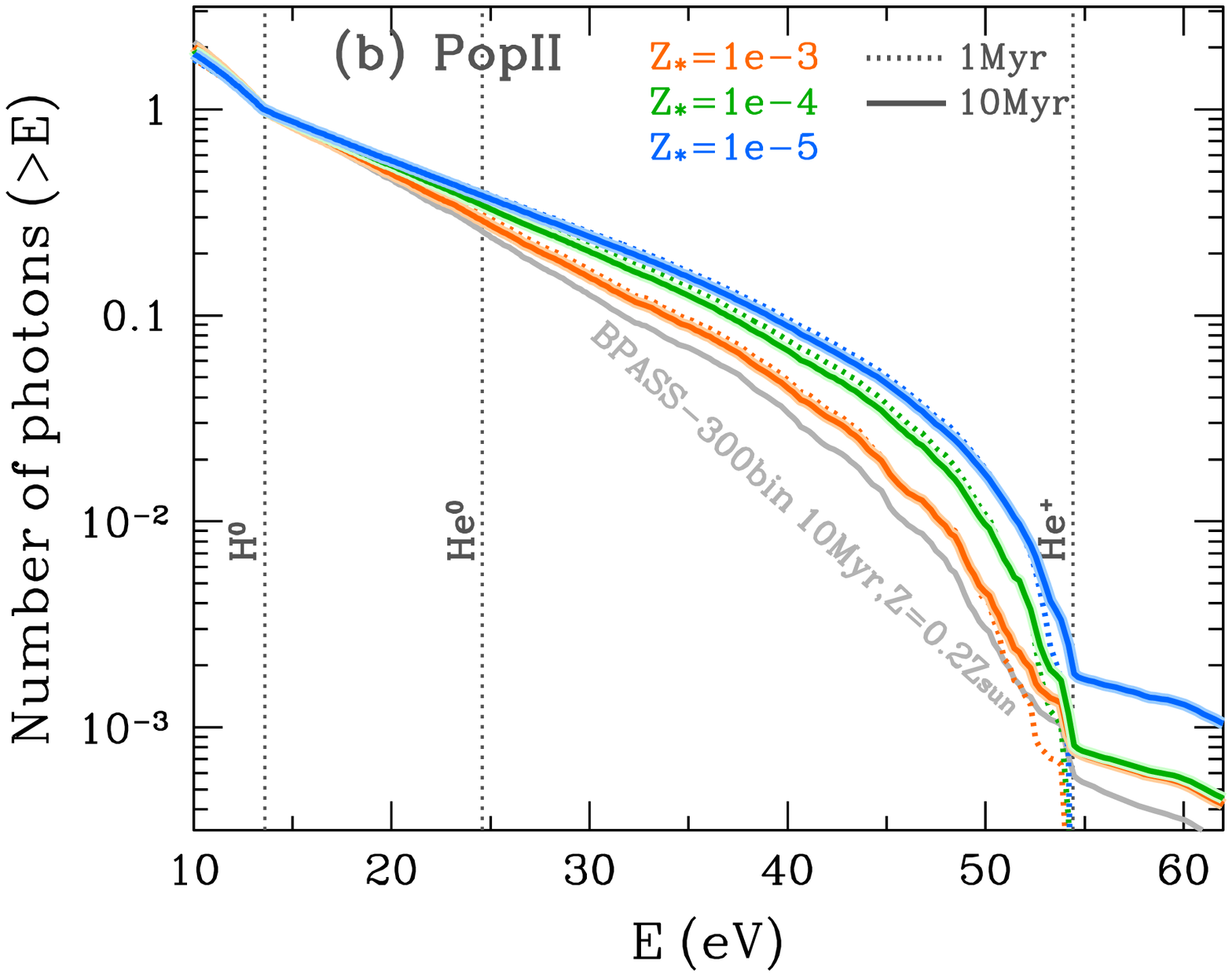}
        \end{center}
      \end{minipage}
      \\
      \begin{minipage}[b]{0.49\hsize}
        \begin{flushleft}
         \includegraphics[bb=23 168 550 587, width=0.95\textwidth]{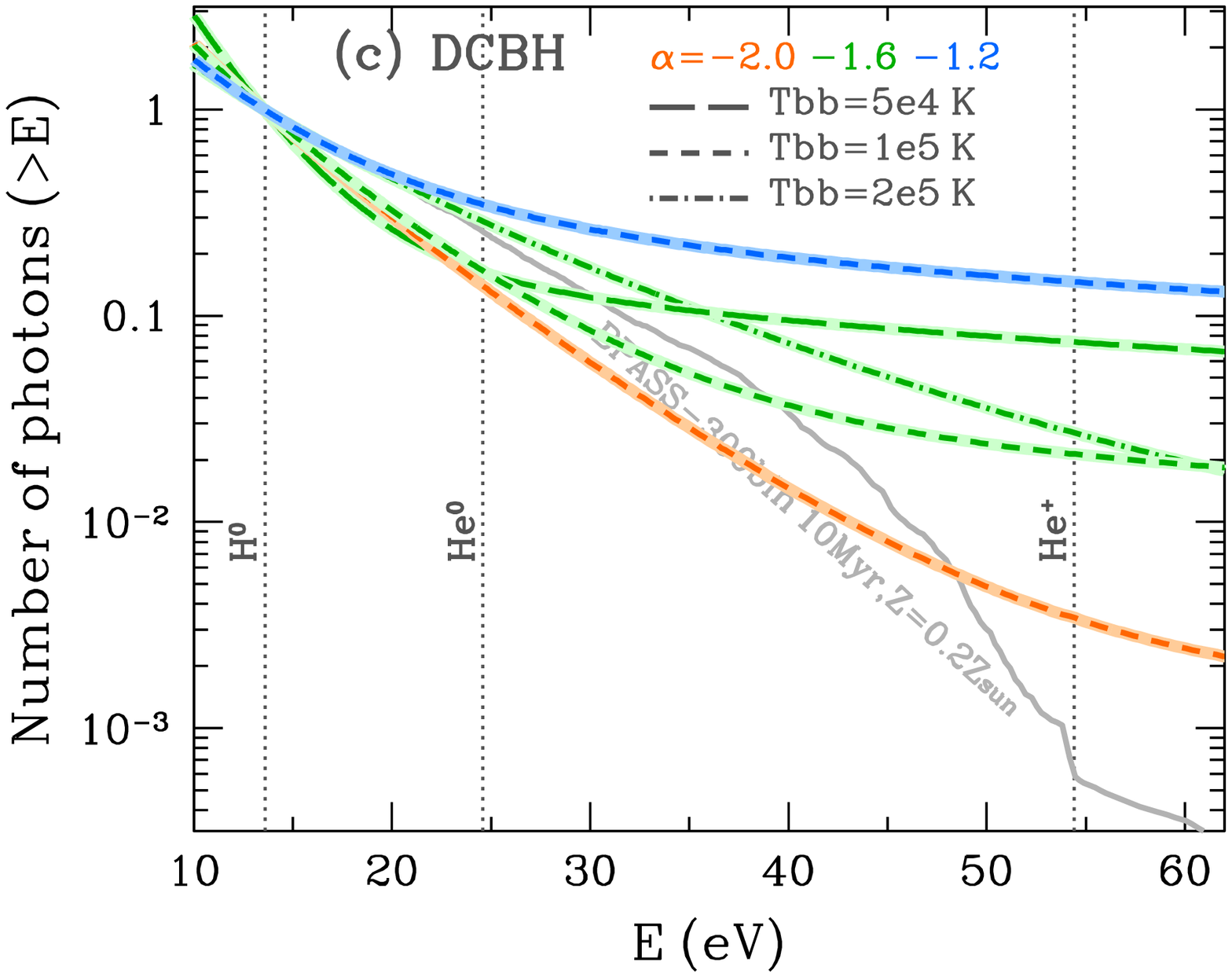}
        \end{flushleft}
      \end{minipage}
      \hspace{0.041\hsize}
      \begin{minipage}[b]{0.40\hsize}
        \begin{center}
          \caption{%
    	 Spectral shapes of the incident radiation fields used in this paper. 
    	 (a) PopIII galaxies. (b) PopII galaxies. (c) DCBHs.
    	 The cumulative number of photons emitted above the energy $E$ 
    	 is illustrated as a function of $E$.
    	 In panel (b), 
	 10\,Myr continuous star-formation as well as 1\,Myr bursty SEDs 
	 are presented.
	 Only models for IMF with $\rm M_{up}  = 300$\,\Msun\ are displayed.
    	 Differences with respect to models with IMF having $\rm M_{up}\leq100$\,\Msun\  
	 are small.
    	 In panel (c), different power-law indices are indicated with different colors
    	 (T$_{\rm bb}$ = 1e5\,K),
    	 and a different T$_{\rm bb}$ are shown with different line-styles 
    	 ($\alpha = -1.6$), as detailed in the legend.
    	 }
        \end{center}
      \end{minipage}
      \hspace{0.041\hsize}
    \end{tabular}
    \label{fig:no_photon}
\end{figure*}


The spectroscopic properties of accreting DCBH or other supermassive black hole seeds have been investigated by various studies in terms of broad photometric properties \citep{volonteri2017,valiante2018,habouzit2019}, however fewer studies have investigated their observational properties in terms of nebular emission lines \citep{pacucci2016,pacucci2017,hartwig2016}. Generally this class of objects is expected to have properties similar to PopIII, i.e. blue continua, strong nebular lines, including strong \HeII\ emission.

Despite the various efforts discussed above, past studies had several limitations. Many of them focused on the rest-frame ultra-violet (UV) diagnostics, which are those observable from ground at high-redshift; however, JWST will provide additional key information also on the rest-frame optical diagnostics. Previous studies have explored a limited range of physical parameters (e.g. in terms of ionisation parameter or shape of the ionising continuum). Moreover, past studies have generally provided the expected strength of the nebular lines in the case of PopIII stars, but have not really explored the possible range of line ratios that can be safely associated with PopIII, i.e. it is not yet clear, even in the case of detection of \HeII, what are the limits on the other lines that can unambiguously identify PopIII stars. Previous studies have only explored the case of PopIII with no metals in the ISM. However, recent theoretical works expect that pockets of pristine gas might survive amid enriched ISM \citep{liu2020}, which might result into the formation of PopIII stars photoionising gas that emits some metal lines. Finally, given the similarity between PopIII stars and accreting DCBH in terms of ionising spectra, generally past studies have not attempted to identify diagnostics that might potentially discriminate between the two classes of sources.

In this study we make an effort to tackle these previous limitations. We develop new photoionisation models using different PopIII ionising continua, corresponding to different IMFs, and different physical conditions and explore what are the nebular emission lines, both in the rest-frame UV and in the rest-frame optical, and explore how different emission line ratios and equivalent widths (EWs) can distinguish among these different cases and also how safely they can separate PopII galaxies. We also explore the new scenario of (mildly) enriched ISM photoionised by PopIII stars. Moreover, we consider the case of gas photoionised by accreting DCBH and investigate whether this can be distinguished from the case of PopIII stars. We provide diagnostic diagrams that will be useful to disentangle these various cases, especially through spectroscopic follow-up of candidate PopIII and DCBH with the forthcoming new observational facilities.

\begin{table*}
  \centering
  \caption{Main parameters of \cloudy\ photoionisation models explored in this paper.}
  \label{tbl:parameters_cloudy}
  \renewcommand{\arraystretch}{1.25}
  \begin{tabular}{@{}llll@{}}
    \hline
    Models
     & PopIII galaxies
     & PopII galaxies
     & DCBHs
    \\
    \hline\hline
    \\
    SED shape & & \\
    \hline
    Graph & Fig. \ref{fig:no_photon}(a)
     & Fig. \ref{fig:no_photon}(b)
     & Fig. \ref{fig:no_photon}(c)
    \\
    \hline
    Parameters 
     & Salpeter IMF
     & BPASS, Binary, Kroupa IMF
     & Power-law $+$ Big Bump
     \\
     & $1$--$100$\,\Msun, $1$--$500$\,\Msun, 
     & Upper mass: $100$\,\Msun\ and $300$\,\Msun\
     & $\alpha=-1.2,\, -1.6,\, -2.0$
     \\
     & $50$--$500$\,\Msun\
     & Z$_{\rm gas}$ $=$ Z$_{\star}$
     & T$_{\rm bb}$ (K) = 5e4, 1e5, 2e5
    \\
    & Zero-age
    & $1$\,Myr and $10$\,Myr (continuous) 
    &
    \\
    \hline \hline
    \\
    Gas properties
    \\
    \hline
    Z$_{\rm gas}$ 
     & 0, 1e-5,\, 1e-4
     & 1e-5,\, 1e-4,\, 1e-3
     & 0, 1e-5,\, 1e-4,\, 1e-3
    \\
     & 
     & (higher metallicities up to $0.014$ ($=1$\,\Zsun))
     & (higher metallicities up to $0.028$ ($=2$\,\Zsun))
    \\
    \hline
    $\log U$
     & \multicolumn{3}{l}{
        From $-0.5$ to $-3.0$ with a step $0.5$\,dex (down to $-3.5$ for evolved galaxies)
        }
    \\
    \hline
    $n_{\rm e}$\, (cm$^{-3}$)
     & \multicolumn{3}{l}{
        $10^2$, $10^3$, and $10^4$ 
        }
    \\
    \hline
  \end{tabular}
\end{table*}

\section{Photoionisation models} 
\label{sec:modelling}

We perform photoionisation models calculations by using \cloudy\
(version 13.05; \citealt{ferland1998,ferland2013})
to obtain spectroscopic predictions for primordial objects. 
We primarily model sources dominated by PopIII stellar populations and DCBH
to find any difference in their spectroscopic features, which would allow distinguishing between the two populations.
Moreover, we also consider evolved systems such as PopII galaxies as well as accreting supermassive black holes (AGNs) embedded in metal-enriched ISM, so to provide 
solid recipes for the identification of
extremely metal-deficient systems and their discrimination from metal enriched systems.

We adopt similar procedures as detailed in \citet{nakajima2018_vuds}
to run the \cloudy\ calculations.
Briefly as for the gas properties, 
we assume constant-density gas clouds with a plane-parallel geometry.
We do not include any dust in the ionised cloud and no elements are 
considered depleted onto dust grains except for the models of PopII galaxies
(see \S\ref{ssec:modelling_PopII})
and unless otherwise specified.
At the zero metallicity, we adopt the primordial helium abundance given by 
\citet{hsyu2020}.
In non-zero metallicities, all elements except nitrogen, carbon, and helium 
are taken to be primary nucleosynthesis elements. 
For carbon and nitrogen, we use forms given by \citet{dopita2006}
and \citet{lopez-sanchez2012}, respectively, 
to take into account their secondary nucleosynthesis production
in the high-metallicity range. 
These prescriptions are based on observations of nearby \HII\ regions,
and known to work for $z=2-3$ galaxies, on average
(e.g., \citealt{steidel2016, amorin2017, hayden-pawson2022}).
For helium, we use a form in \citet{dopita2006} but slightly updated 
to be consistent with the latest primordial value \citep{hsyu2020}.
The ionisation state is regulated by ionisation parameter ($\log U$),
whose definition in \cloudy\ is the number of hydrogen-ionising photons 
to the total hydrogen density in a dimensionless form at the radius of 
the illuminated face of the plane-parallel cloud.
The ionisation parameter $\log U$ is varied from $-3.0$ up to an extreme value 
of $-0.5$ with a step of $0.5$\,dex on the logarithmic scale.
A gas electron density of $10^3$\,cm$^{-3}$ is used as a fiducial value, 
but $10^2$ and $10^4$\,cm$^{-3}$ are also explored
and presented in Appendix \S \ref{sec_app:results_different_densities}.
We choose this range of gas density because 
(i) it is observed in ionised \HII\ regions of actively star-forming galaxies 
at $z=0-2$ \citep{shirazi2014, sanders2016}, and 
(ii) a high gas density of $\sim 10^4$\,cm$^{-3}$ is tantalisingly inferred 
in a $z=11$ galaxy \citep{oesch2016, jiang2021}. 
We adopt the same fiducial value and range of gas densities for DCBHs
and AGNs, as a comparable densities are observed 
in the narrow-line regions of typical AGNs (e.g., \citealt{nagao2006_agn, dors2014}; 
see also \citealt{kewley2013_theory}). 
Since there are not yet direct determinations of the gas electron density in the ionised 
regions surrounding DCBHs, we treat the gas density for the DCBH models in 
the same way as for the AGN models.

Three types of incident radiation field are considered here:
PopIII galaxies, PopII galaxies, and DCBHs.
Their ionising spectrum shapes are illustrated in Fig.~\ref{fig:no_photon},
and their main parameters in \cloudy\ are summarised in Table \ref{tbl:parameters_cloudy}.
We give a detailed explanation for each of the models in the followings.

We note that single-cloud models, hence adopting a single value of the ionisation parameter and gas density, are only very rough approximations of the actual distribution of clouds properties in galaxies. More realistically, even within a single \HII\ region, although chemically homogeneous, emitting gas clouds span a broad range of ionisation parameter and density, each of them emitting optimally a subset of nebular lines. This is even more true when considering the integrated emission from an entire galaxy, which includes different kind of \HII\ regions. An accurate modelling would require combining photo-ionisation models with clouds spanning a broad range of ionisation parameter and density, and combining them to reproduce the observed nebular line ratios. Failure to do so results often in inferring metallicities which can be a factor of 2--3 different from the actual values (Marconi et al., in prep.). However, while we are aware of the limitations of single-cloud, single-ionisation parameter models, in this paper we are mostly interested in the overall trend of the diagnostics ratios over a range of metallicities spanning order of magnitudes, hence the effect of considering the contribution of multiple clouds with different ionisation parameters is secondary; moreover, at a given metallicity, we will show trends for different ionisation parameters, which may give an indication of the effects resulting from the contribution of different populations of clouds.
In Appendix \S \ref{sec_app:results_different_densities} we also show photoionisation model results assuming gas densities different from the fiducial value. The results illustrate that different densities, and hence a potential variation of density in a system, do not alter the overall selection criteria and metallicity indicators that are derived based on the constant, fiducial density models as detailed below.

\subsection{PopIII galaxies} \label{ssec:modelling_PopIII}

We take three zero-metal stellar population models 
from \citet{schaerer2003} and \citet{raiter2010} as SEDs of PopIII galaxies.
They are calculated at zero-age with a \citet{salpeter1955} IMF, 
considering quite different mass ranges defined by the lower- and upper-mass
cut-offs of IMF: 
$1$--$100$\,\Msun,
$1$--$500$\,\Msun, and 
$50$--$500$\,\Msun.
Their hard ionising spectra even above the He$^{+}$-ionising potential ($E>54$\,eV)
are presented in Fig.~\ref{fig:no_photon}(a).
As illustrated by \citet{raiter2010}, the most extreme model with the mass range 
from $50$ to $500$\,\Msun\ reproduces well the most top-heavy IMFs by comparison with
 other existing models (e.g., \citealt{tumlinson2006}).
Since recent theoretical studies indicate that extremely metal-poor stellar populations should be characterised by a top-heavy IMF (e.g., \citealt{hirano2015, chon2021}), 
more top-heavy than the present-day IMF such as \citet{salpeter1955} with $1$--$100$\,\Msun,
our models would cover a reasonable range of IMF variations expected for PopIII galaxies.
A broader range of top-heavy stellar IMF for PopIII galaxies (with different analytical forms) will be explored in a forthcoming, separate paper.

We consider the zero-metal ISM as well as slightly metal-enriched cases with
Z$_{\rm gas}$ $=10^{-5}$ and $10^{-4}$.
The model calculations are stopped at the edge of the ionised nebular cloud, i.e., 
when the electron fraction, 
defined as the ratio of the number density of electrons to that of total hydrogen, 
falls below $10^{-2}$.

\subsection{PopII galaxies} \label{ssec:modelling_PopII}

To identify sources with a PopIII stellar population and understand differences
between more chemically-enriched systems, we need to also model PopII galaxies.
We adopt SEDs of BPASS (v2.2.1; \citealt{eldridge2017,stanway2018}) 
for modelling the ISM ionised by such PopII galaxies.
BPASS models might be more appropriate for moderate/low metallicity systems and 
less appropriate for high metallicity systems. 
However, here we are exploring differences between the extremely low metallicity 
regimes (akin PopIII, near-pristine systems) and low/moderate metallicity (PopII) systems,
so we believe that our choice of using the BPASS SEDs is appropriate.
We use publicly available fiducial SEDs with binary evolution
which are based on a \citet{kroupa2001} IMF 
with the two upper-mass cuts of $100$\,\Msun\ and $300$\,\Msun.
We adopt a stellar age of $10$\,Myr as fiducial models assuming continuous star formation 
to take into account contributions of hard ionising photons produced by massive Wolf-Rayet (WR) stars. 
In addition, the youngest stellar age of $1$\,Myr is also considered as commonly adopted 
to model optical emission lines in \HII\ regions (e.g., \citealt{KD2002}). 
We adopt two different stellar ages primarily to discuss the EWs of emission lines.
Usually EWs present the largest values when the stellar population is youngest
(e.g., \citealt{JR2016,nakajima2018_vuds}), but it would not always be true for the \HeII\ emission
due to the contributions of WR stars at $\sim 4$\,Myr after the onset of star formation.
The shapes of ionising spectrum for the PopII models at $10$\,Myr
as well as those at $1$\,Myr are shown in Fig.~\ref{fig:no_photon}(b)
with different stellar metallicities.
For each metallicity, the SEDs at $1$ and $10$\,Myr are almost identical
at $E\lesssim 50$\,eV, but an enhanced production of high energy ionising photons
($E\gtrsim 50$\,eV) for the $10$\,Myr model.
The model calculations are terminated at the edge of the ionised cloud.

For each of the PopII models, we assume the stellar metallicity is matched 
to the gas-phase metallicity.
This is a fair assumption as the enrichment of oxygen and other $\alpha$ elements
proceeds quickly, within less than 10\,Myr \citep{MM2019}. 
Although such a quick metal enrichment results in $\alpha$-enhanced abundance patterns
with little iron enrichment, and hence a lower opacity in young stars for their oxygen abundance, 
as observationally inferred in star-forming galaxies at $z=2-6$
\citep{steidel2016,cullen2019,harikane2020_abs}, 
this would not significantly enhance the nebular emission strengths for the PopII models
as compared to the PopIII model results. Anyway, we recall that the BPASS models adopted by us (and adequate for low metallicity systems) are already characterised by hotter atmospheres and a higher yield of ionising photons than classical, metal rich  models.
Moreover, we include dust physics in the PopII models as detailed in \citet{nakajima2018_vuds},
assuming the dust model and the depletion factors are unchanged but 
the dust abundance is scaled linearly with the gas-phase metallicity.

\begin{figure*}
  \centering
    \begin{tabular}{c}
      \begin{minipage}{0.49\hsize}
        \begin{center}
         \includegraphics[bb=18 209 555 391, width=0.85\textwidth]{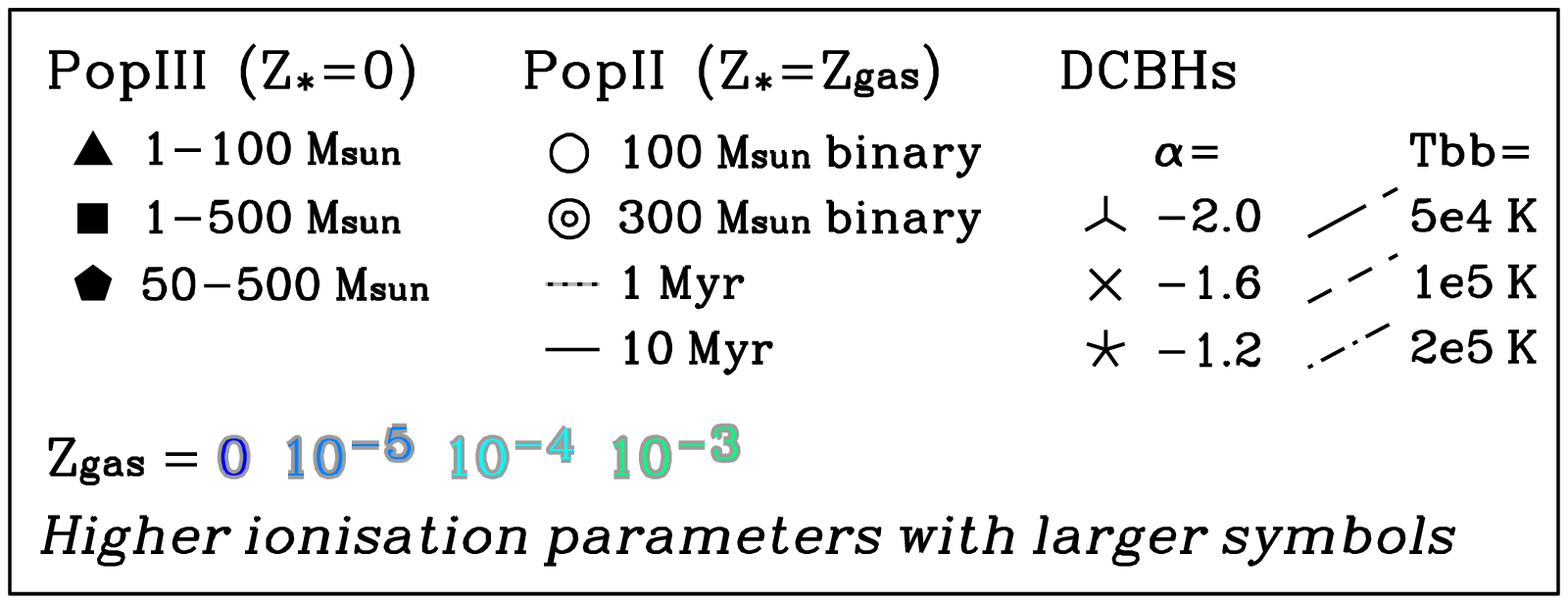}
        \end{center}
      \end{minipage}
      \begin{minipage}{0.49\hsize}
        \begin{center}
         \includegraphics[bb=18 209 555 391, width=0.85\textwidth]{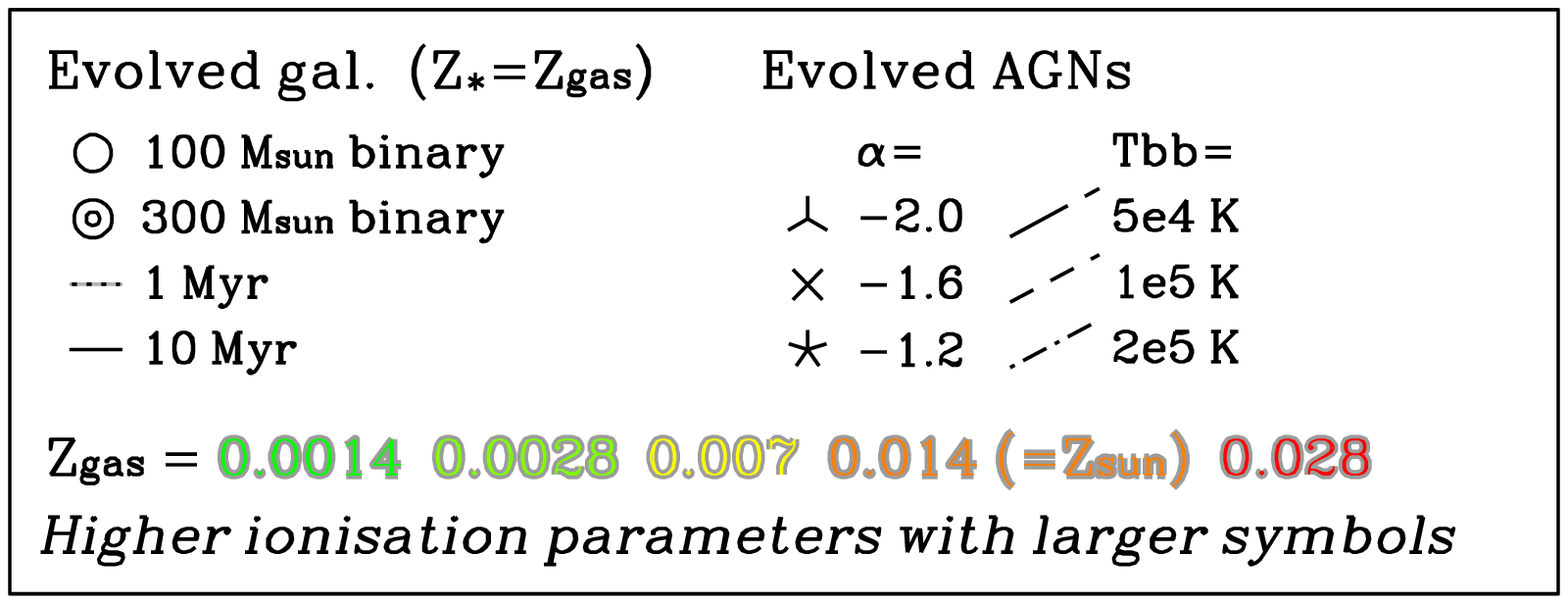}
        \end{center}
      \end{minipage}
      \\
      \begin{minipage}{0.49\hsize}
        \begin{center}
         \includegraphics[bb=18 143 555 680, width=0.85\textwidth]{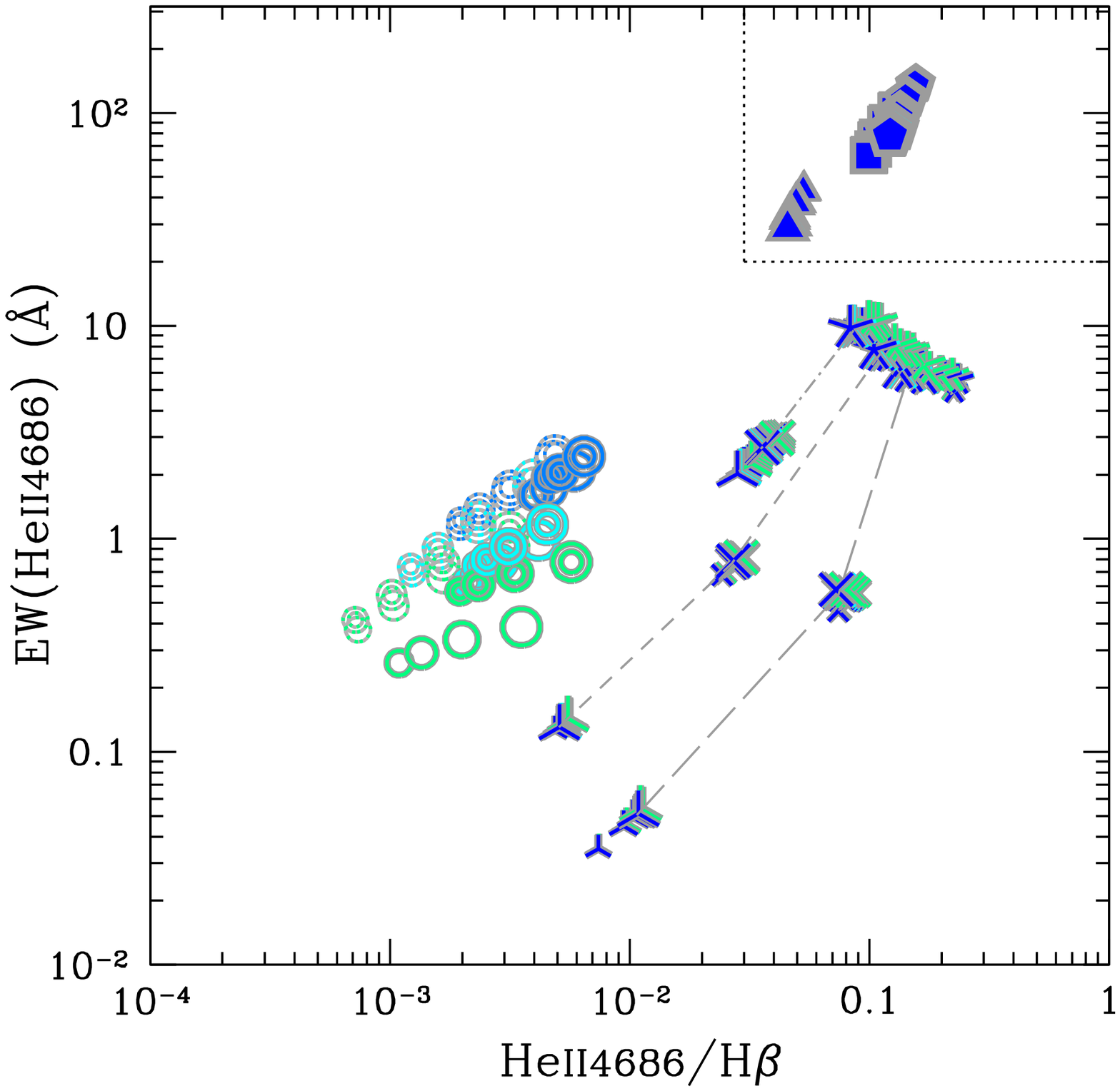}
        \end{center}
      \end{minipage}
      \begin{minipage}{0.49\hsize}
        \begin{center}
         \includegraphics[bb=18 143 555 680, width=0.85\textwidth]{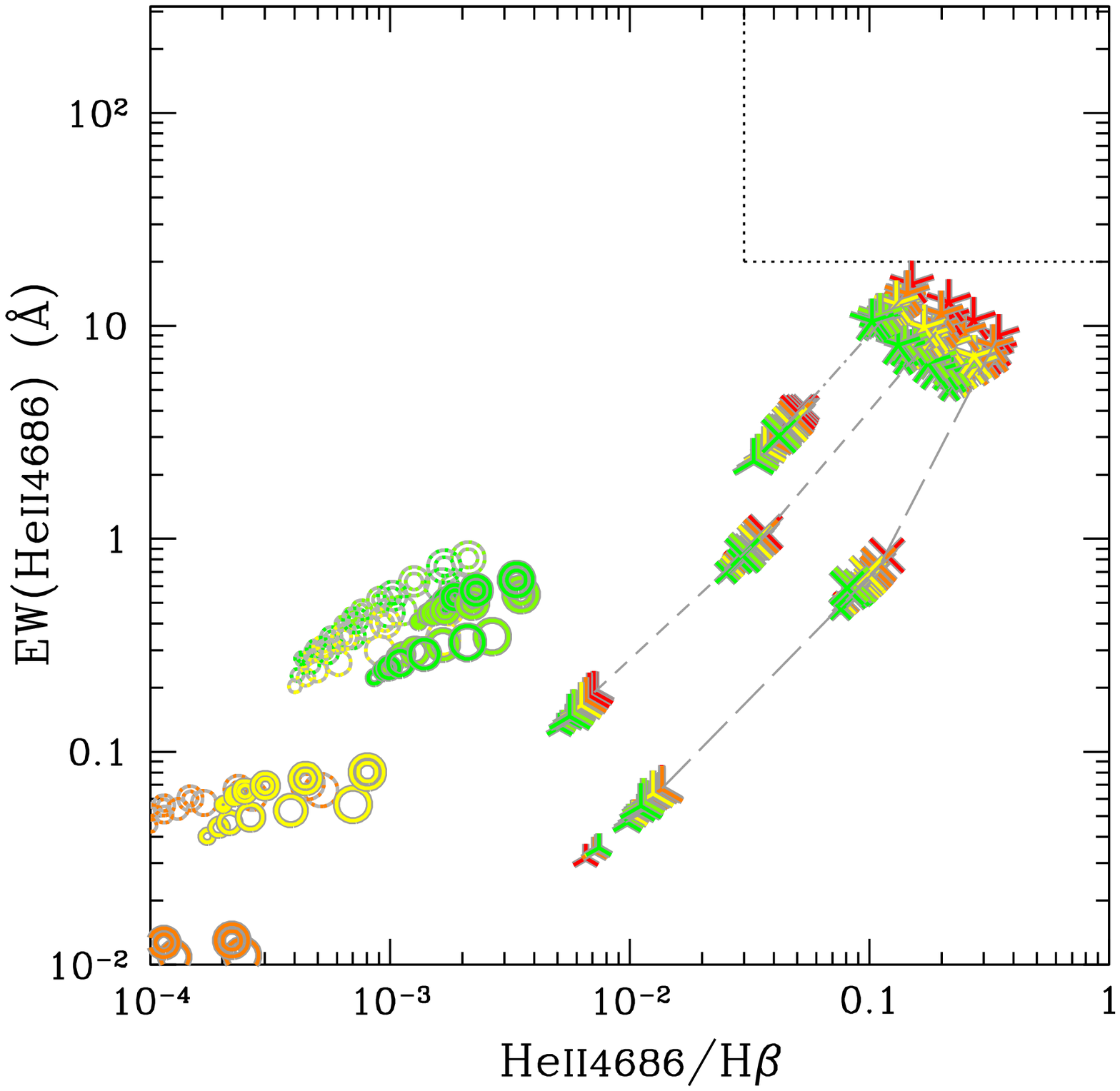}
        \end{center}
      \end{minipage}
    \end{tabular}
    \caption{%
    	Distribution of the \cloudy\ models on the EW(\HeII$\lambda 4686$) versus 
    	\HeII$\lambda4686$/\Hb\ diagram, for the primitive sources (Left) 
    	and for the chemically-evolved systems (Right). 
	As shown in the legend, 
	different symbols present different kinds of modelled objects, 
	different colors for different gas metallicities ($\rm Z=0$ (blue) to $0.028 = 2$\,\Zsun\ (red)), and 
	larger symbol sizes for higher ionisation parameters
	(Left: $\log U = -0.5$ to $-2.0$ for PopIII and PopII galaxies, $-0.5$ to $-3.0$ for DCBHs;
	Right: $\log U = -1.0$ to $-3.5$ for evolved galaxies, $-0.5$ to $-3.0$ for AGNs).
	A fiducial gas density of $10^3$\,cm$^{-3}$ is assumed.
	The black dotted lines denote the demarcation (Eq.~\ref{eq:ewhe2_he2hb})
	between PopIII and the other populations.
    } 
    \label{fig:ewhe2_he2hb}
\end{figure*}


\subsection{DCBHs} \label{ssec:modelling_DCBHs}

Our mock accreting DCBH incident radiation fields are generated by 
the \cloudy\ ``AGN'' continuum command. 
The function primarily consists of two components.
One is the so-called a Big Bump component, and another is a power-law component 
in the high energy range reaching the X-ray band. 
We have decided to adopt this continuum template as the spectrum resulting 
from the accretion on primeval black holes is expected not to be much different 
from `standard' AGN, though possibly with a broader range of the two components, as detailed below 
(\citealt{valiante2018}, and M. Volonteri, priv. comm.).
The Big Bump feature is known to appear in the UV to optical region
and would trace thermal emission from an optically thick accretion disk,  
which feeds a massive black hole.
It is parameterised by the temperature of the bump, T$_{\rm bb}$, which is 
varied between T$_{\rm bb} = 5\times10^4$\,K, $1\times10^5$\,K, and $2\times10^5$\,K.
The choice is based on the maximum temperature in the innermost region of the disc, 
T$_{\rm max} \sim 5\times 10^5$\,K for black holes with masses of $10^5-10^6$\,\Msun\, 
following the approximation of $T_{\rm max} \sim M_{\rm BH}^{-0.25}$\,keV (e.g., \citealt{yue2013}).
Because the temperature of the disc decreases from inside out such that $\sim R^{-0.75}$, 
we adopt a wide parameter range to cover an effective temperature that would be expected
in such an accreting disk around black holes with $M_{\rm BH}\sim 10^5-10^6$\,\Msun.
It is still unclear though whether the lowest temperature $\rm T_{\rm bb} = 5\times10^4~K$ can be achieved in DCBHs/AGNs (see \S\ref{sec:results}). 
We note that the black hole mass range explored in this paper covers fairly well the theoretically suggested mass range of intermediate-mass black hole seeds, $10^{4.75}-10^{6.25}$\,\Msun\ \citep{ferrara2014}. The $0.25$\,dex lower (higher) mass would increase (decrease) the maximum temperature by a factor of only $1.15$ (as $\rm T_{max}\propto M_{BH}^{-0.25}$), which does not significantly impact the resulting diagnostics for the DCBH models.

We adopt the slope of the X-ray component with the default value, but change
the power-law energy slope between the optical and X-ray bands, $\alpha_{ox}$ \citep{zamorani1981}. 
The parameter corresponds to the power-law index $\alpha$, where $f_{\nu}\propto \nu^{\alpha}$, 
determined in the range of a few to a few thousand eV, and hereafter called $\alpha$
for simplicity.
We vary the value between $\alpha=-1.2$, $-1.6$, and $-2.0$ (see \citealt{elvis2002}).
The ionising spectrum shapes for the DCBH models are shown in
Fig.~\ref{fig:no_photon}(c).

We consider cases where these DCBHs are surrounded by pristine gas (zero metallicity) as well as
mildly metal-enriched gas with $Z=10^{-5}$, $10^{-4}$, $10^{-3}$, and much 
higher metallicities up to $Z = 0.028 = 2\,Z_{\odot}$%
\footnote{A solar chemical composition of \citet{asplund2009} is assumed.
}.
The model calculations are stopped when reaching a neutral column density of 
$N$(\HI) $=10^{21}$\,cm$^{-2}$, in accordance with the AGN models 
\citep{kewley2013_theory}.

We note that the direct emission of the accretion disc may be obscured along our line of sight by the presence of a dusty torus, i.e. a `type-2' AGN configuration. While it is unlikely this happen in extremely metal poor systems, whose dust content is greatly reduced, we note this might be a caveat for the EW of the nebular lines, not their line ratios.

\section{Results} \label{sec:results}

\subsection{Optical Diagnostics} \label{ssec:results_optical}

\subsubsection{Diagnostics involving only He and H optical lines}

The rest-frame optical wavelength range will be observable for early galaxies with JWST.
This section presents the photoionisation model predictions for some diagrams 
using the optical emission line fluxes and EWs that will be useful to identify PopIII galaxies
and DCBHs.

One of the key emission lines in the optical wavelength is \HeII$\lambda 4686$
which is sensitive to the production of high energy ionising photons ($E>54$\,eV)
and thus proposed to be useful to identify PopIII galaxies. 
Particularly interesting is the \HeII$/$\Hb\ ratio, as both of them are 
recombination lines with the same dependence on density and temperature, 
and their ratio is essentially traces the ratio of ionising photons at $E>54$\,eV 
and at $E>13.6$\,eV, 
hence the shape of the ionising spectrum, nearly independently of other quantities 
such as gas metallicity, ionisation parameter and density. 
The EW(\HeII) has a similar diagnostic power, by giving an indication of 
the fraction of highly energetic photons relative to the optical, non-ionising photons, 
and hence a probe of the overall shape of the spectrum of the ionising sources.

Fig.~\ref{fig:ewhe2_he2hb} presents a diagram using the key \HeII\ 
observable quantities: EW(\HeII) vs.~\HeII$/$\Hb.
The left panel show the primitive objects cases, i.e. zero or very low metallicity. Different colors show the metallicity level (with dark blue being zero metallicity). Solid symbols are PopIII with different IMFs. Empty symbols are PopII with different IMF and ages. Starred and cross symbols are for DCBHs. Symbols of increasing size correspond to larger ionisation parameters, both for ionisation by stellar populations and in the case of DCBH. The legend gives details of the different symbols and colors coding. The most interesting feature of this figure is that PopIII models are remarkably and distinctly positioned as
having the strongest \HeII\ emission,
both in terms of EW and line ratios with respect to \Hb, 
covering a completely different and disjoint region in the diagram as compared to
the PopII models.
The PopII models generally show a larger EW(\HeII) at $1$\,Myr than $10$\,Myr
but the difference becomes small at low-metallicity, possibly as a result of the 
the increase of continuum level and the increase of \HeII\ emission at $10$\,Myr, partly cancelling out each other.
The different positions between the PopIII and PopII models on the diagram
are clearly seen irrespective of the choice of stellar age for the PopII models.
The DCBH models having the hardest ionising spectrum, with the largest power-law slope $\alpha$
and the highest T$_{\rm bb}$, present \HeII$/$\Hb\ ratios as large as those found 
in the PopIII models. However, a clear difference is seen in terms of EW(\HeII), where PopIII models show a larger EW than the DCBH models.
Although PopIII galaxies with an older stellar population present a decreased EW
than illustrated in Fig.~\ref{fig:ewhe2_he2hb}, the decrease would be within a factor of $\sim 2$
if the stellar age is younger than a few Myr (e.g., \citealt{schaerer2003}), and hence 
the diagram will remain useful to identify the primitive PopIII objects.
We propose a selection criteria for identifying PopIII galaxies:
\begin{equation}
	{\rm EW}({\rm He\,\textsc{ii}}) > 20\,\mbox{\AA}\ \&\
	{\rm He\,\textsc{ii}}/{\rm H}\beta > 0.03
	\label{eq:ewhe2_he2hb}
\end{equation}
These limits are indicated by the dotted lines in Fig.~\ref{fig:ewhe2_he2hb}.
This diagram has some potential also in discriminating between different PopIII IMFs, 
with more top heavy IMFs being characterised by higher ${\rm EW}({\rm He\,\textsc{ii}})$ 
and higher ${\rm He\,\textsc{ii}}/{\rm H}\beta$.
Appendix \S \ref{sec_app:results_different_densities} shows the same diagrams for  different gas density models and illustrating that results and selection boundaries do not change significantly.

The right panel of Fig.~\ref{fig:ewhe2_he2hb} presents the same diagram for more chemically-evolved galaxies and AGNs. Symbols are the same as in the left panel and the color-coding shows different metallicities also in this case, although the metallicity range is shifted to much higher values than in the left panel. This panel illustrates that there is
 no contamination in the region of the PopIII galaxies on the diagram
particularly for those with a top-heavy IMFs.
Clearly this diagram can separate PopIII from PopII and from DCBH and evolved AGNs. 
However, it cannot distinguish between primeval DCBHs (in a zero/low metallicity environment)
from more evolved AGNs living in a more metal enriched environment; 
disentangling these two classes of objects will require the inclusion of metal lines, 
as we will discuss later on.

It should be noted that various studies have found evidence for significant \HeII\ nebular emission in populations of metal poor, dwarf star forming galaxies \citep[e.g.][]{schaerer2019,umeda2022}. The observed equivalent width is about 1\,\AA, similar to what expected by our models in low metallicity regimes (especially if considering high ionisation parameters, which are indeed typically observed in low metallicity star forming galaxies). However, the \HeII$/$\Hb\ ratio observed in such low metallicity galaxies is often significantly higher ($\sim 0.01$) than predicted by our models. This is an issue that is faced also by other photoionisation models using stellar population templates \citep{senchyna2017,senchyna2020,chevallard2018,umeda2022}. Possible explanations that have been proposed include the contribution of X-ray binaries \citep[e.g.][]{schaerer2019} and ultraluminous X-ray sources \citep{umeda2022}, although other studies have argued against these explanations \citep{senchyna2020}. Either of these scenarios implies the presence of an additional ionising spectral component associated with accretion on compact objects and, therefore, they are essentially equivalent to adding an AGN-like component, as indeed expected by our models. It should also be noted that the presence of AGN in a sizeable number of dwarf galaxies has been found by recent observations \citep[e.g.][]{mezcua2016,mezcua2020,schutte2022} and also suggested by recent theoretical work \citep{pacucci2021}, so the high \HeII$/$\Hb\ ratio can actually be simply explained by the presence of AGN in many of them. However, regardless of the explanation, it remains true that the combination of high EW(\HeII) and high \HeII$/$\Hb\ unambiguously identifies either PopIII or DCBH with a power law index $\alpha \gtrsim -1.6$.

We finally note that these are predictions for the \HeII\ nebular lines. The stellar \HeII\ emission associated with WR stars is not considered here. However, the latter, if present, should be clearly distinguishable from its broad profile (typically of the order of $1000$\,km\,s$^{-1}$ or broader).
\\

\begin{figure}
  \centering
    \includegraphics[bb=18 143 555 680, width=0.85\columnwidth]{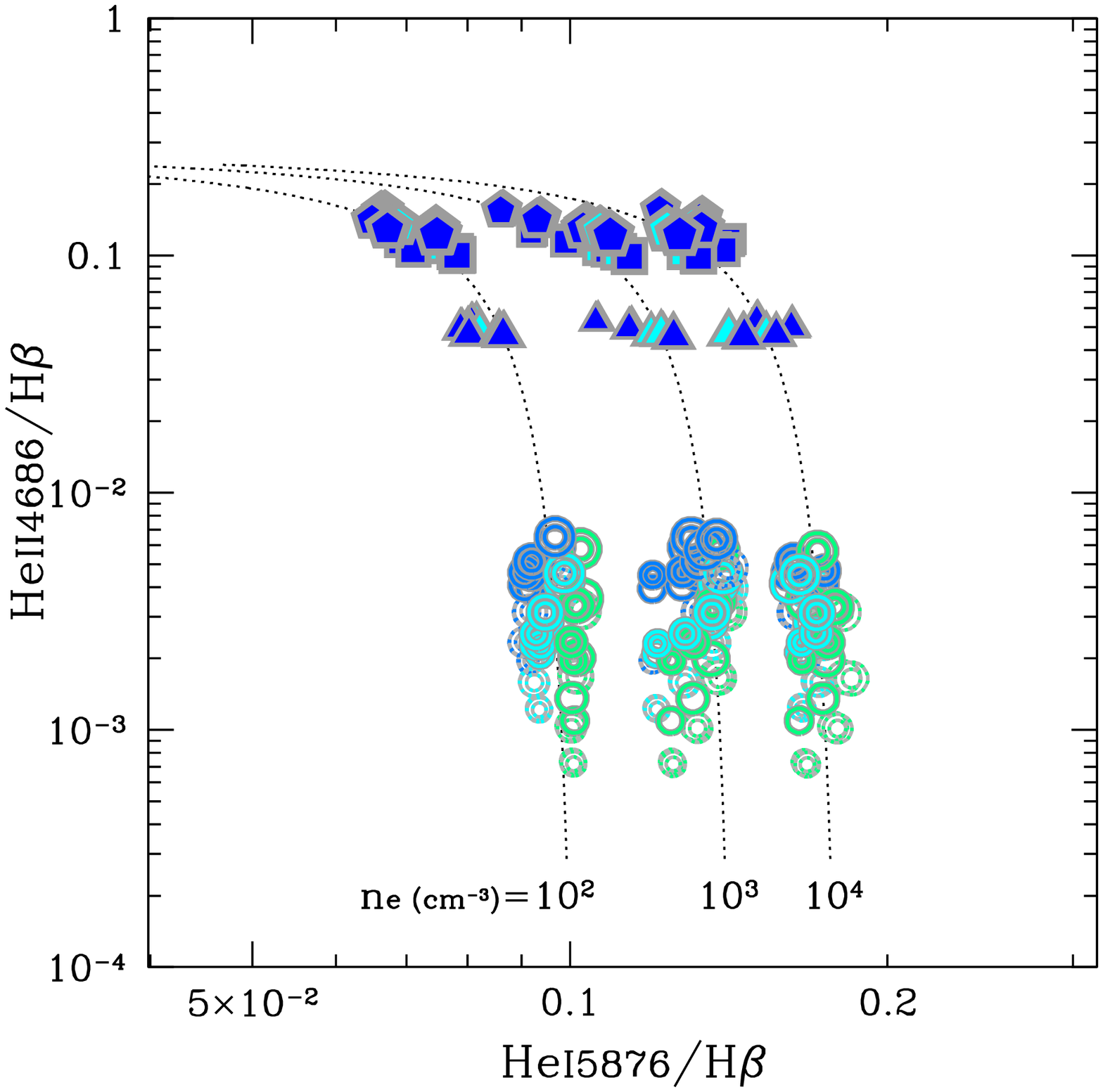}
    \caption{%
    	Distribution of the \cloudy\ models on the \HeII$\lambda4686$$/$\Hb\ versus ~\HeI$\lambda5876$$/$\Hb\ diagram,
    	only for  PopIII and PopII galaxies (Z $<10^{-3}$)
	    for three different gas density values, $n_{\rm e}=10^2$, $10^3$, 
	    and $10^4$\,cm$^{-3}$, as indicated with the dotted curves
	    (Eqs.~\ref{eq:he2hb_he1hb_5876_n2}--\ref{eq:he2hb_he1hb_5876_n4}). 
	    Symbols as in Fig.~\ref{fig:ewhe2_he2hb}.
    }
    \label{fig:he2hb_he1hb_5876_density}
\end{figure}

\begin{figure*}
  \centering
    \begin{tabular}{c}
      \begin{minipage}{0.49\hsize}
        \begin{center}
         \includegraphics[bb=18 143 555 680, width=0.85\textwidth]{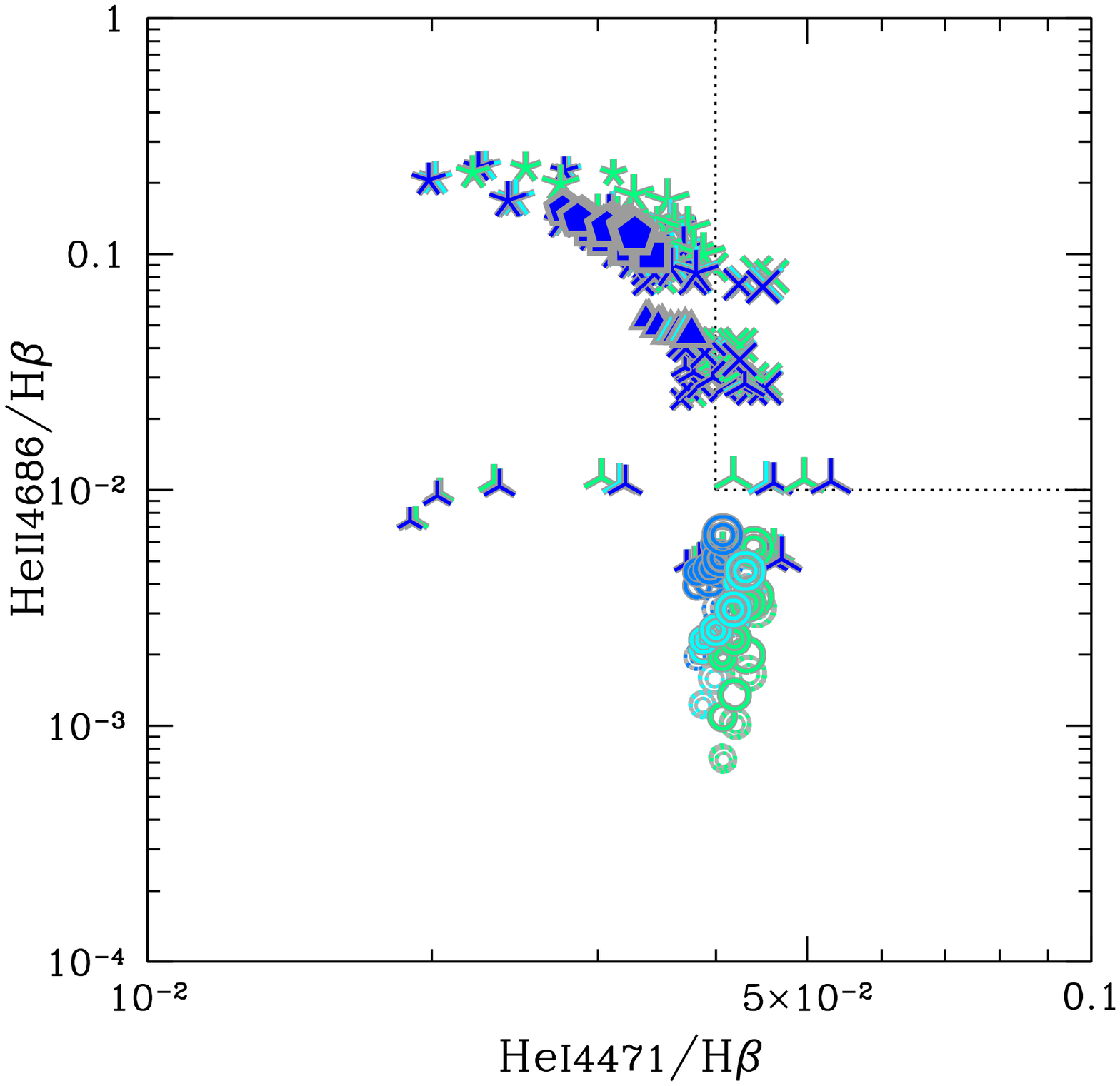}
        \end{center}
      \end{minipage}
      \begin{minipage}{0.49\hsize}
        \begin{center}
         \includegraphics[bb=18 143 555 680, width=0.85\textwidth]{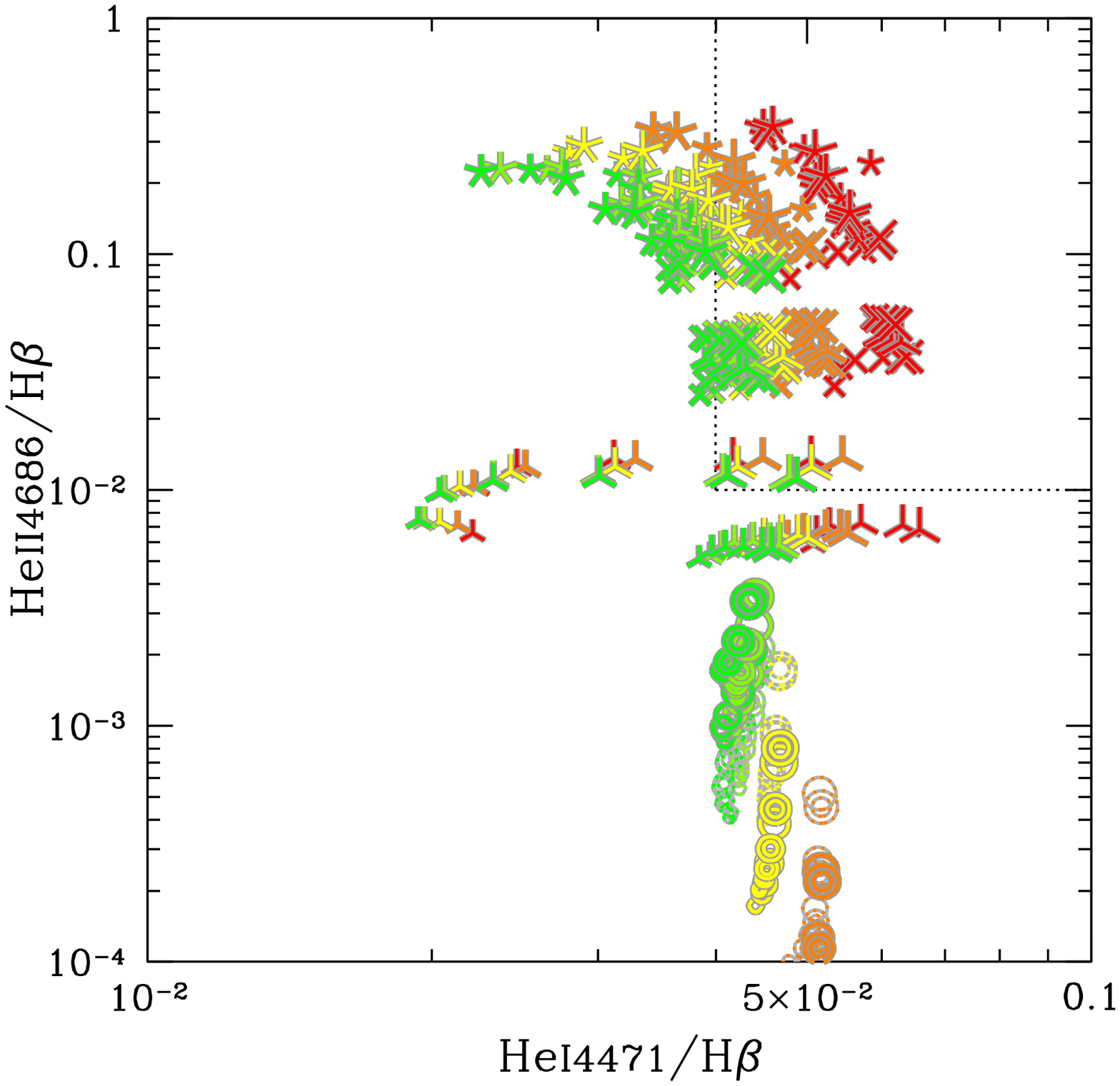}
        \end{center}
      \end{minipage}
    \end{tabular}
    \caption{%
    	Distribution of the \cloudy\ models on the \HeII$\lambda4686$$/$\Hb\ versus~\HeI$\lambda4471$$/$\Hb\ diagram,
	    for  primitive sources (Left) and 
	    for  chemically-evolved systems (Right). 
	    Symbols as in Fig.~\ref{fig:ewhe2_he2hb}.
	    The black dotted lines can be useful to disentangle some types of 
	    DCBHs (Eq.~\ref{eq:he2hb_he1hb_4471}).
	}
    \label{fig:he2hb_he1hb_4471}
\end{figure*}

Other important emission lines are those associated with \HeI, which should be 
observable even from zero-metallicity objects and can be useful to probe the 
ionising spectrum shape at the intermediate energy regime ($E\sim 24$\,eV). 
Since these are primarily recombination lines, 
together with the \HeII\ and hydrogen recombination lines, 
they would be in principle useful for determining the curvature of the ionising spectrum. 
However, the situation is not so simple in this case. 
Indeed, the relatively strong \HeI\ recombination lines result from the cascades 
of \HeI\ triplets, whose population/depopulation is affected by 
the collision effects (i.e., density and temperature) and the opacity properties 
of ISM (e.g., \citealt{izotov2014,aver2015,matsumoto2022}).
Yet, exploring the relative intensity of these lines is anyway useful.

\begin{figure*}
  \centering
    \begin{tabular}{c}
      \begin{minipage}{0.32\hsize}
        \begin{center}
         \includegraphics[bb=18 143 555 680, width=0.95\textwidth]{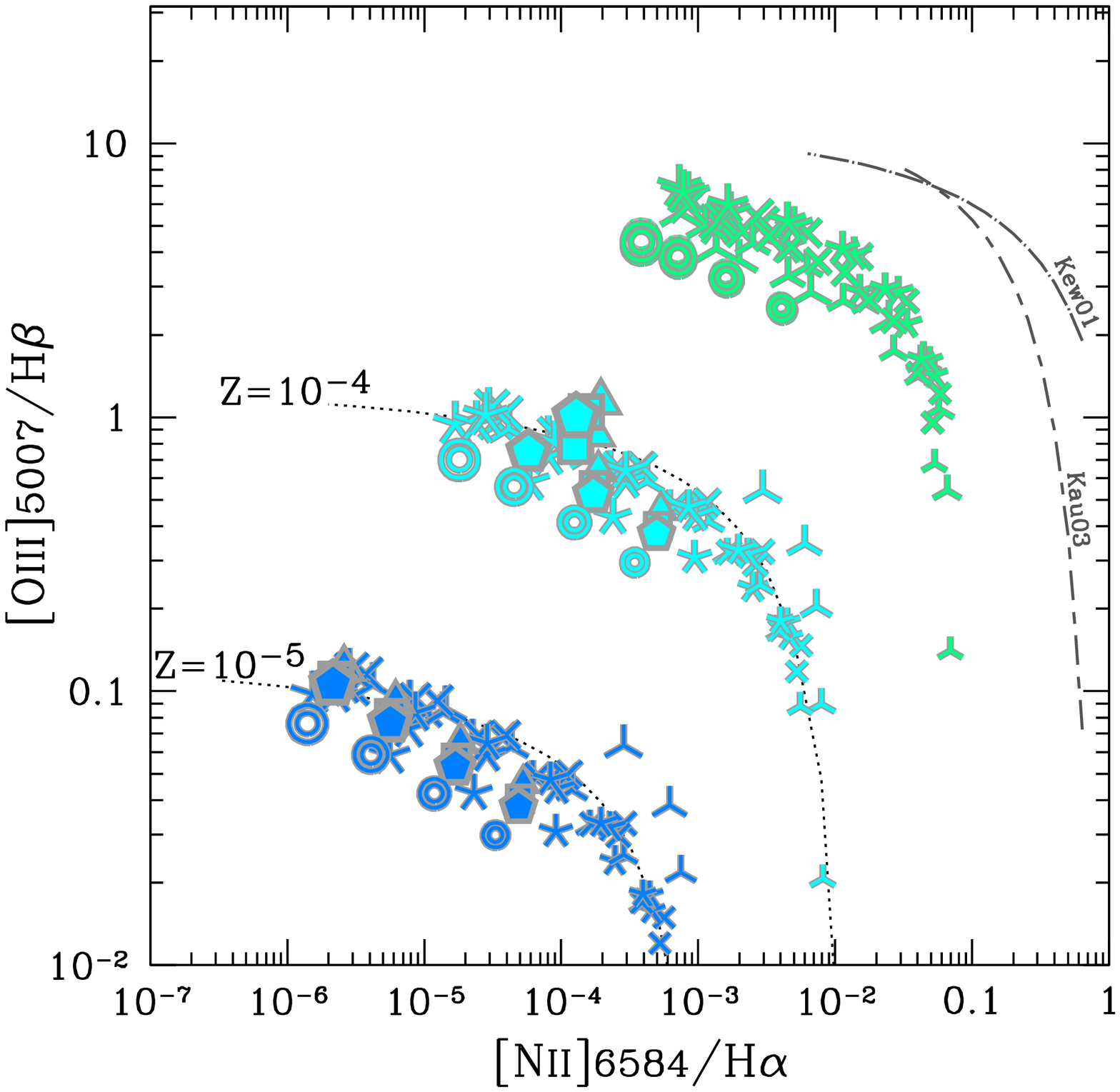}
        \end{center}
      \end{minipage}
      \begin{minipage}{0.32\hsize}
        \begin{center}
         \includegraphics[bb=18 143 555 680, width=0.95\textwidth]{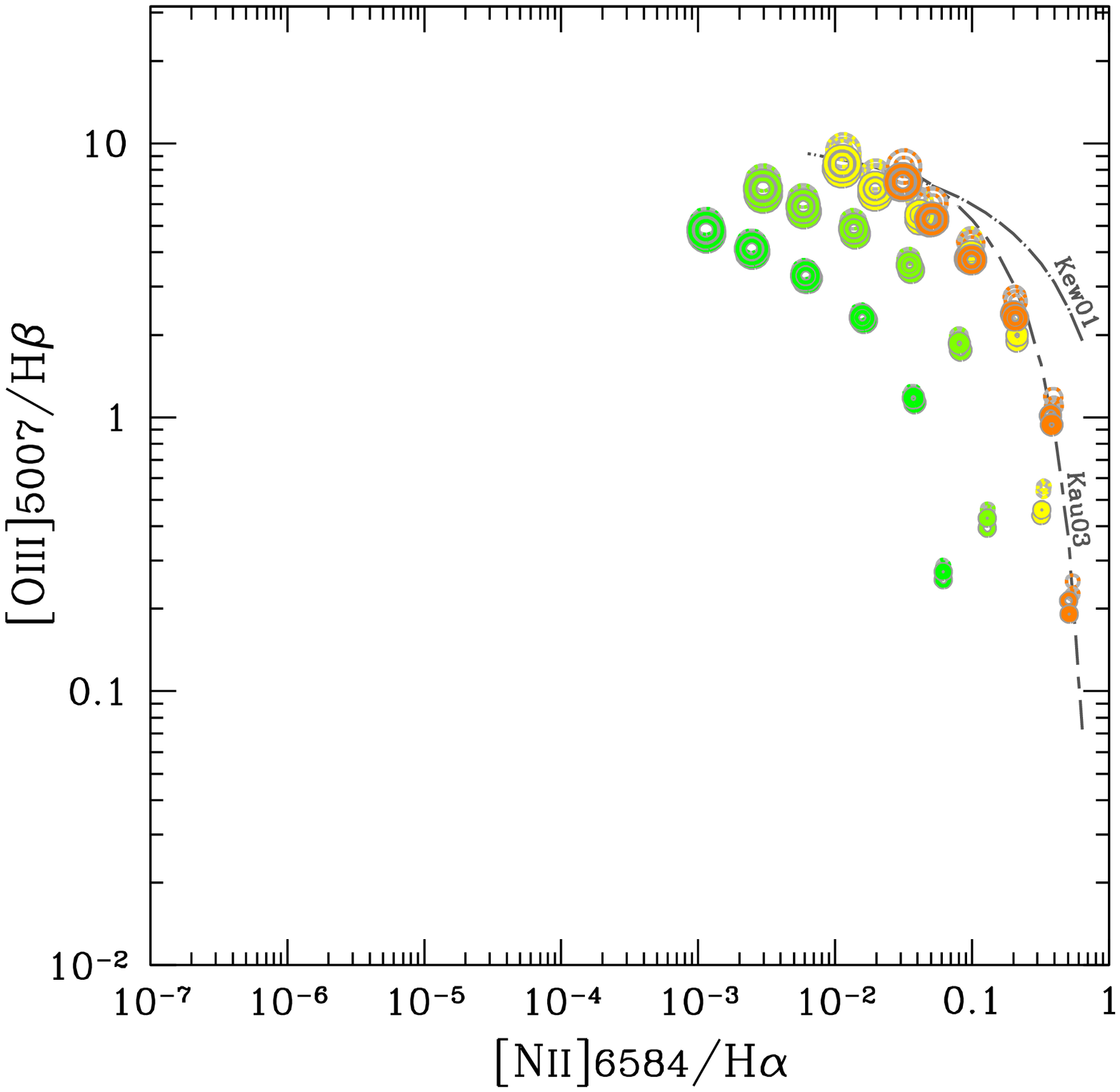}
        \end{center}
      \end{minipage}
      \begin{minipage}{0.32\hsize}
        \begin{center}
         \includegraphics[bb=18 143 555 680, width=0.95\textwidth]{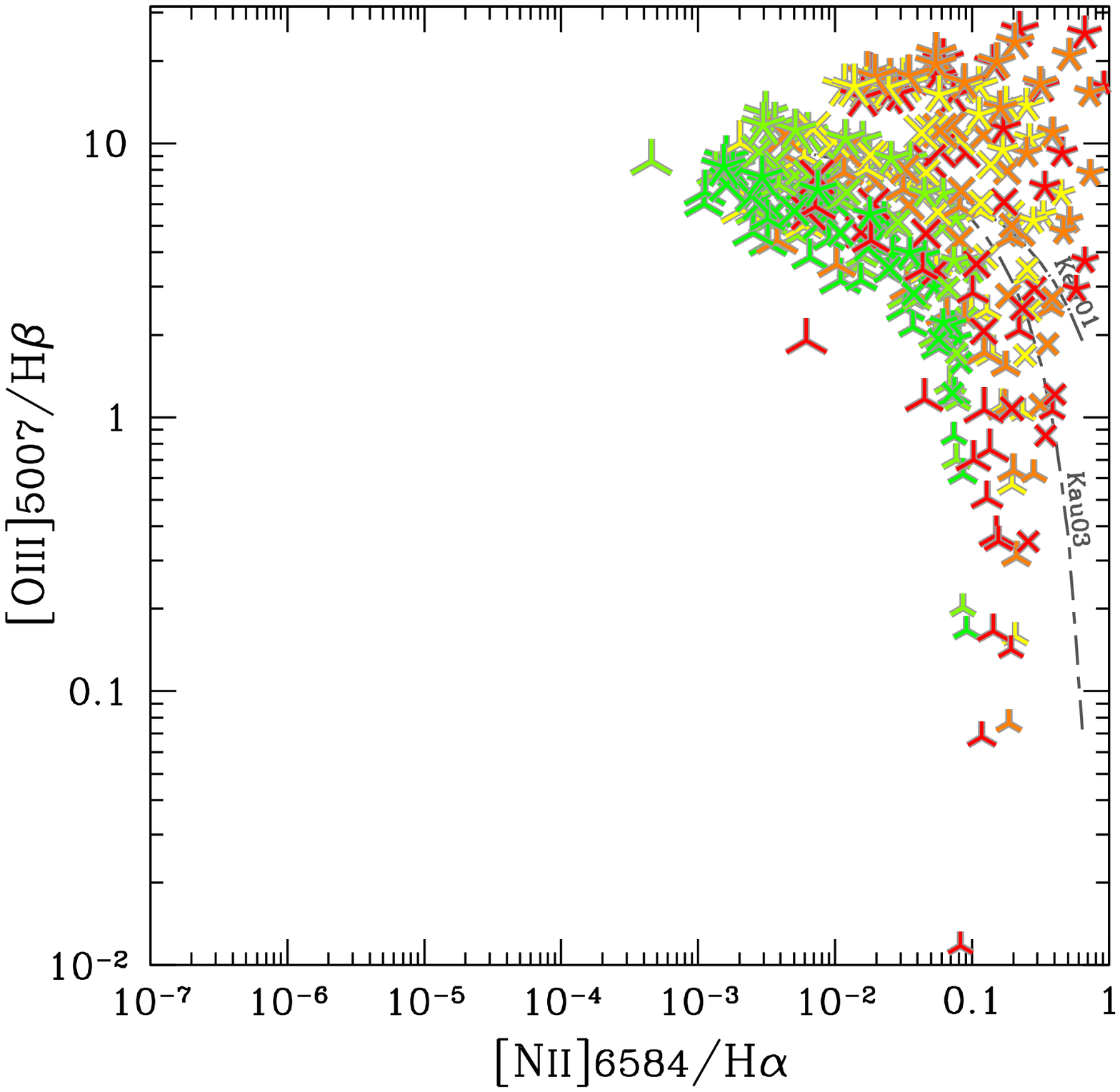}
        \end{center}
      \end{minipage}
      \\
      \begin{minipage}{0.32\hsize}
        \vspace{10pt}
        \begin{center}
         \includegraphics[bb=18 143 555 680, width=0.95\textwidth]{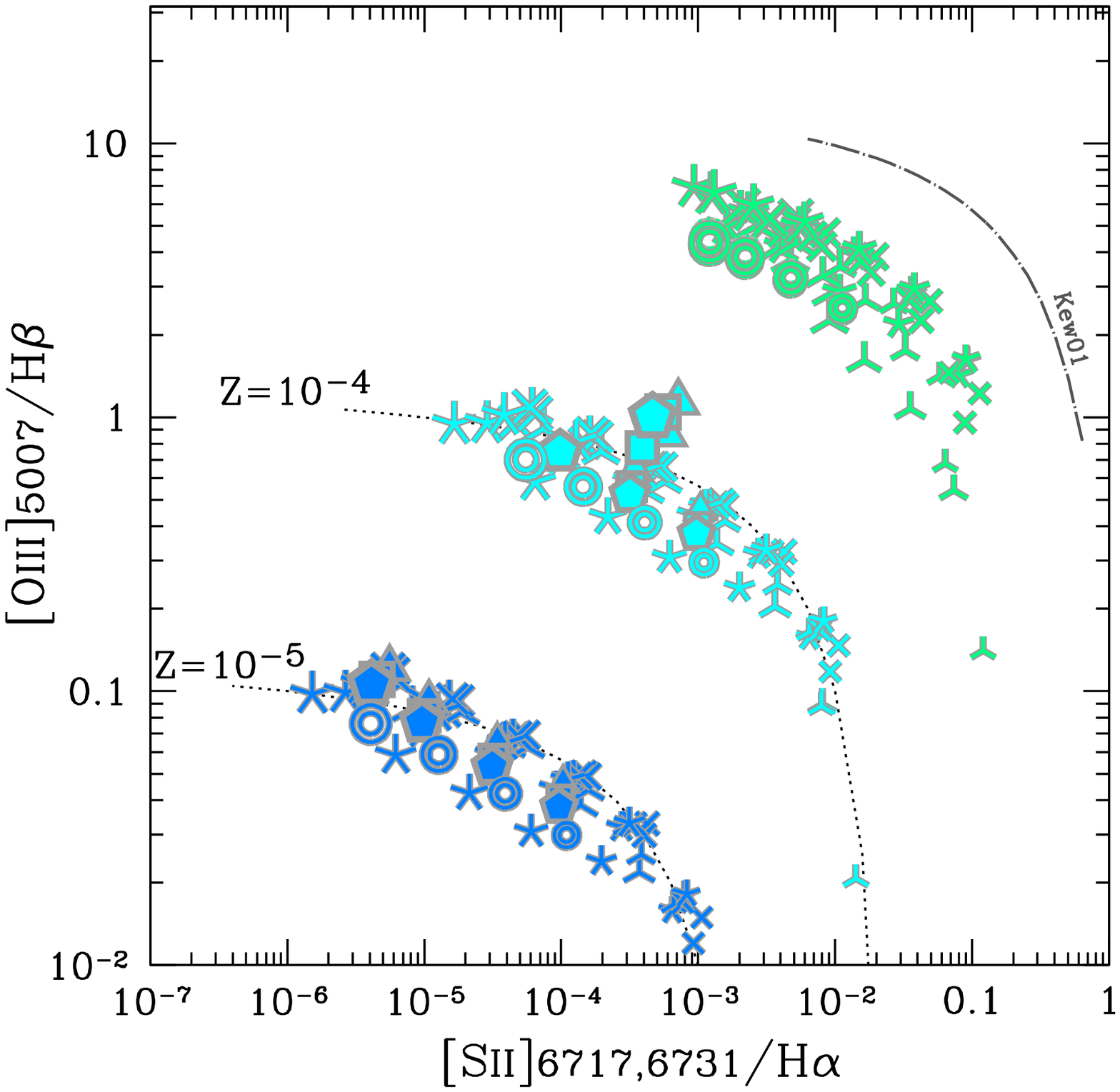}
        \end{center}
      \end{minipage}
      \begin{minipage}{0.32\hsize}
        \vspace{10pt}
        \begin{center}
         \includegraphics[bb=18 143 555 680, width=0.95\textwidth]{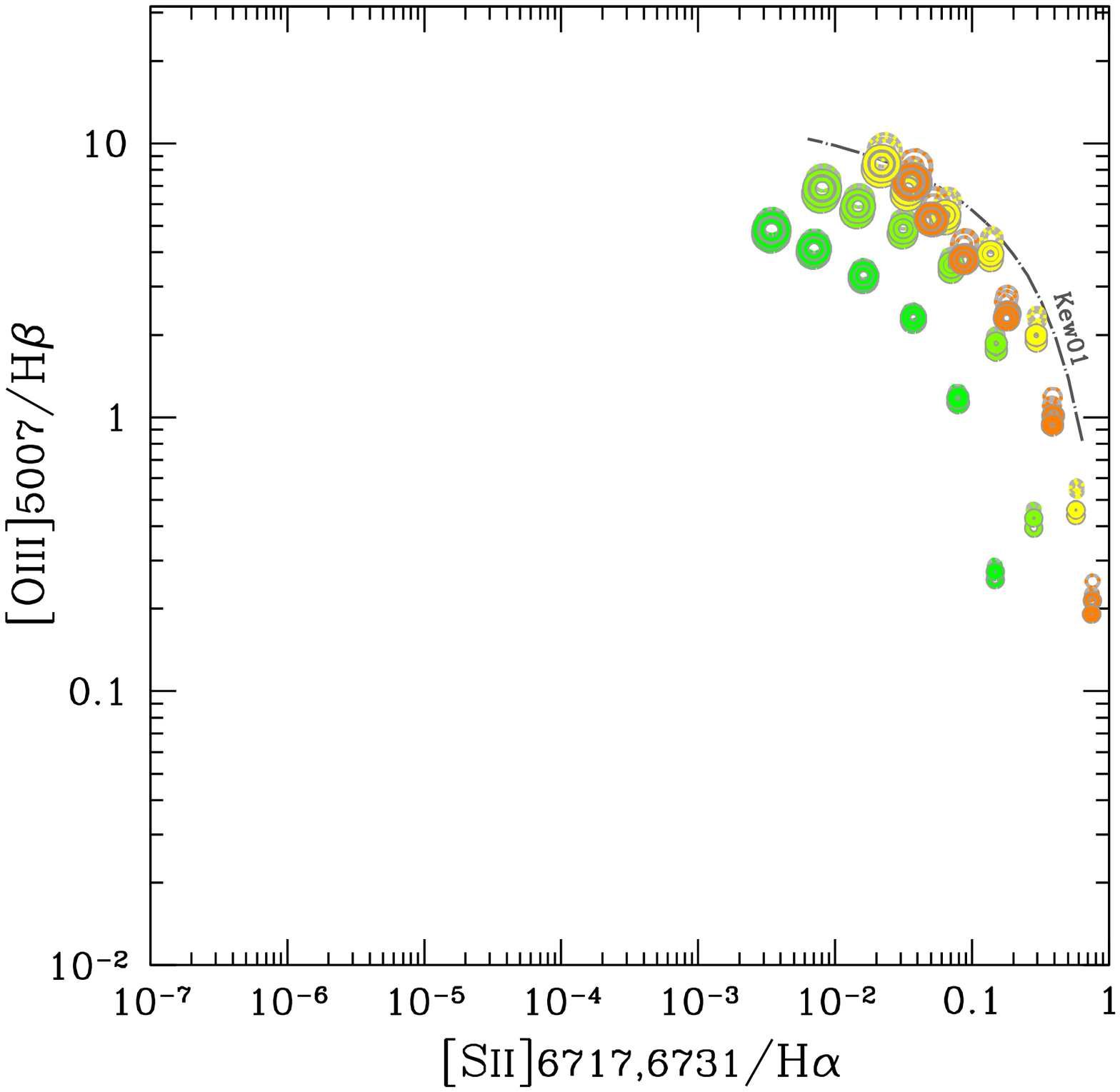}
        \end{center}
      \end{minipage}
      \begin{minipage}{0.32\hsize}
        \vspace{10pt}
        \begin{center}
         \includegraphics[bb=18 143 555 680, width=0.95\textwidth]{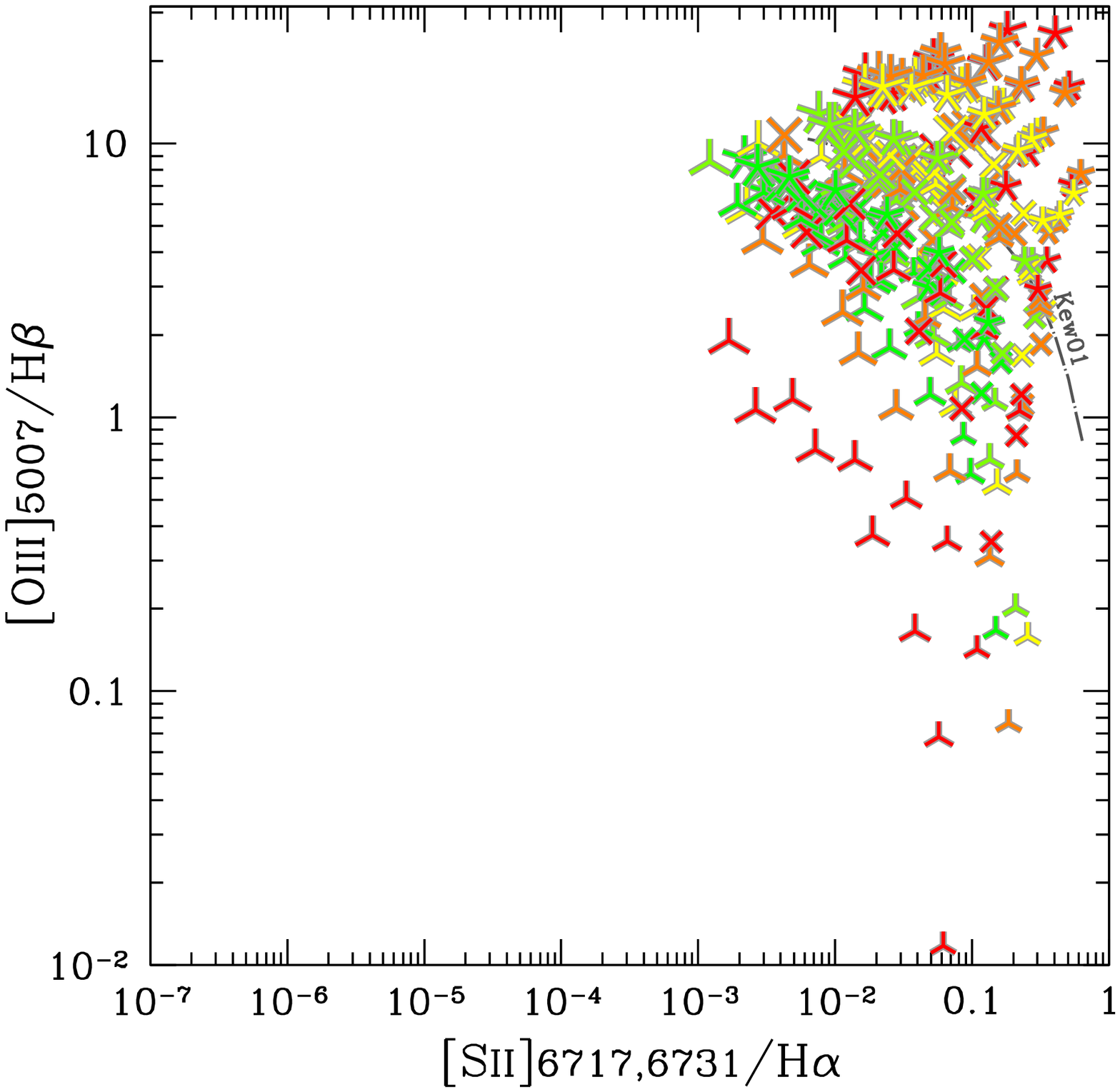}
        \end{center}
      \end{minipage}
      \\
      \begin{minipage}{0.32\hsize}
        \vspace{10pt}
        \begin{center}
         \includegraphics[bb=18 143 555 680, width=0.95\textwidth]{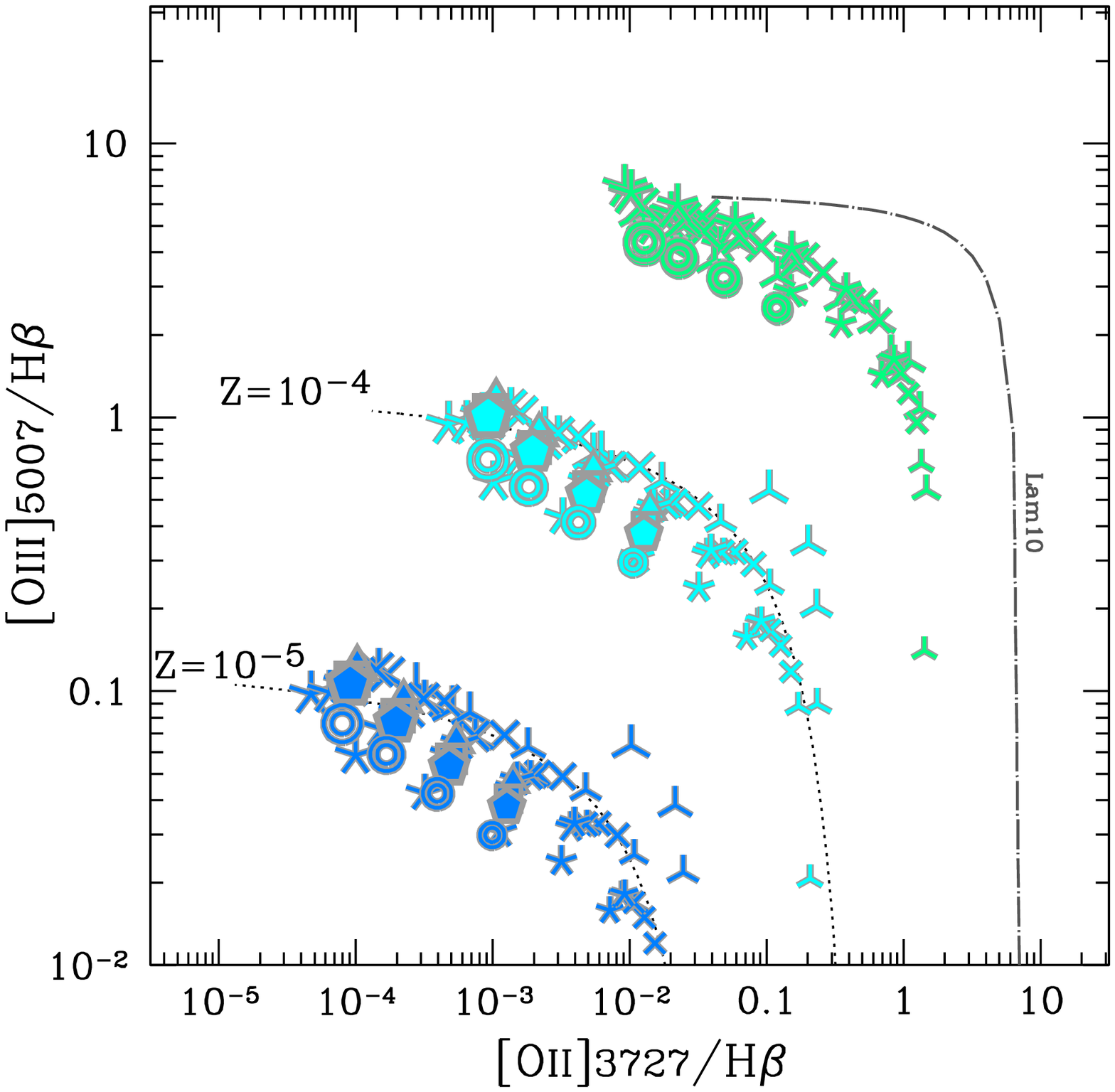}
        \end{center}
      \end{minipage}
      \begin{minipage}{0.32\hsize}
        \vspace{10pt}
        \begin{center}
         \includegraphics[bb=18 143 555 680, width=0.95\textwidth]{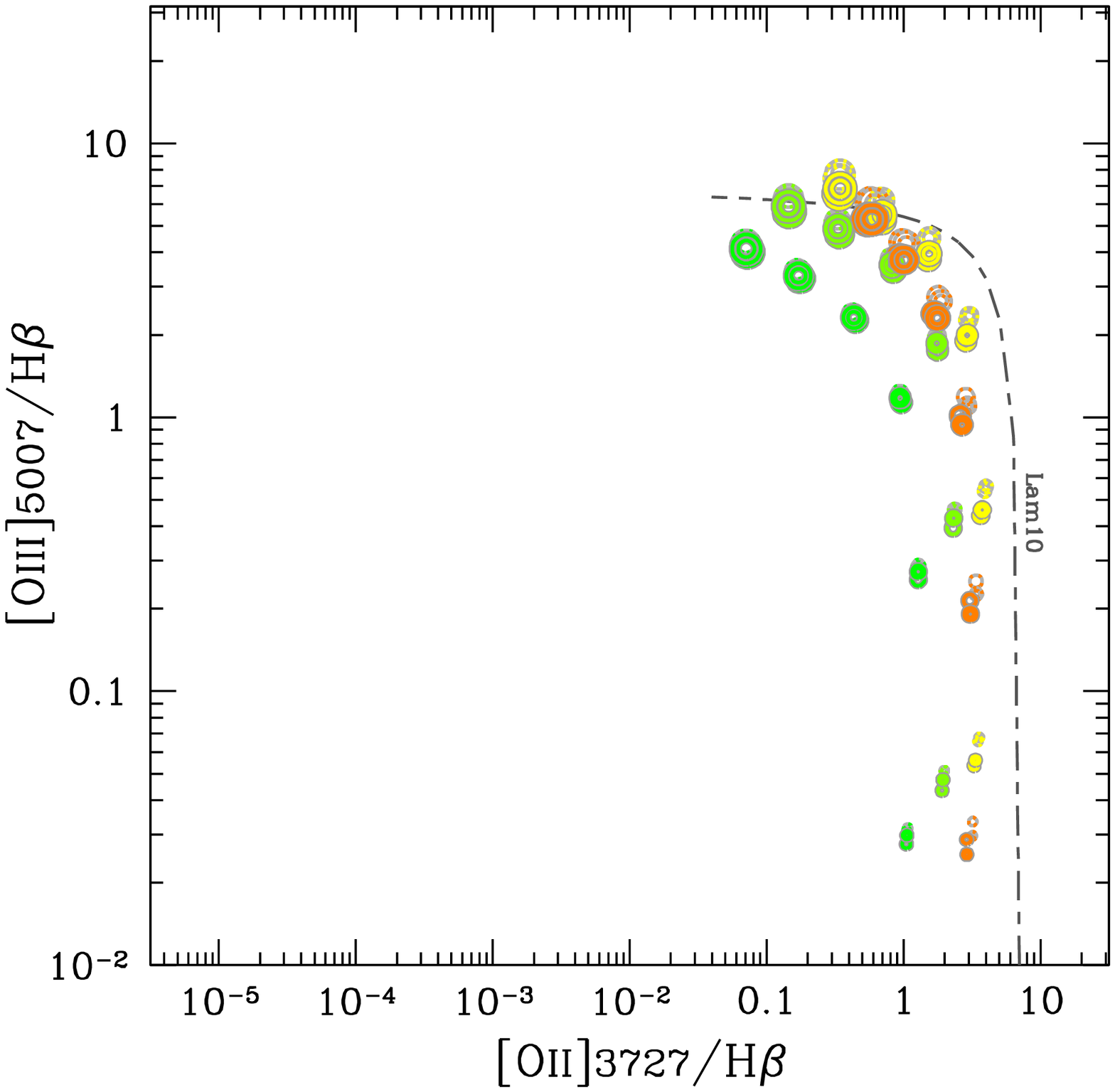}
        \end{center}
      \end{minipage}
      \begin{minipage}{0.32\hsize}
        \vspace{10pt}
        \begin{center}
         \includegraphics[bb=18 143 555 680, width=0.95\textwidth]{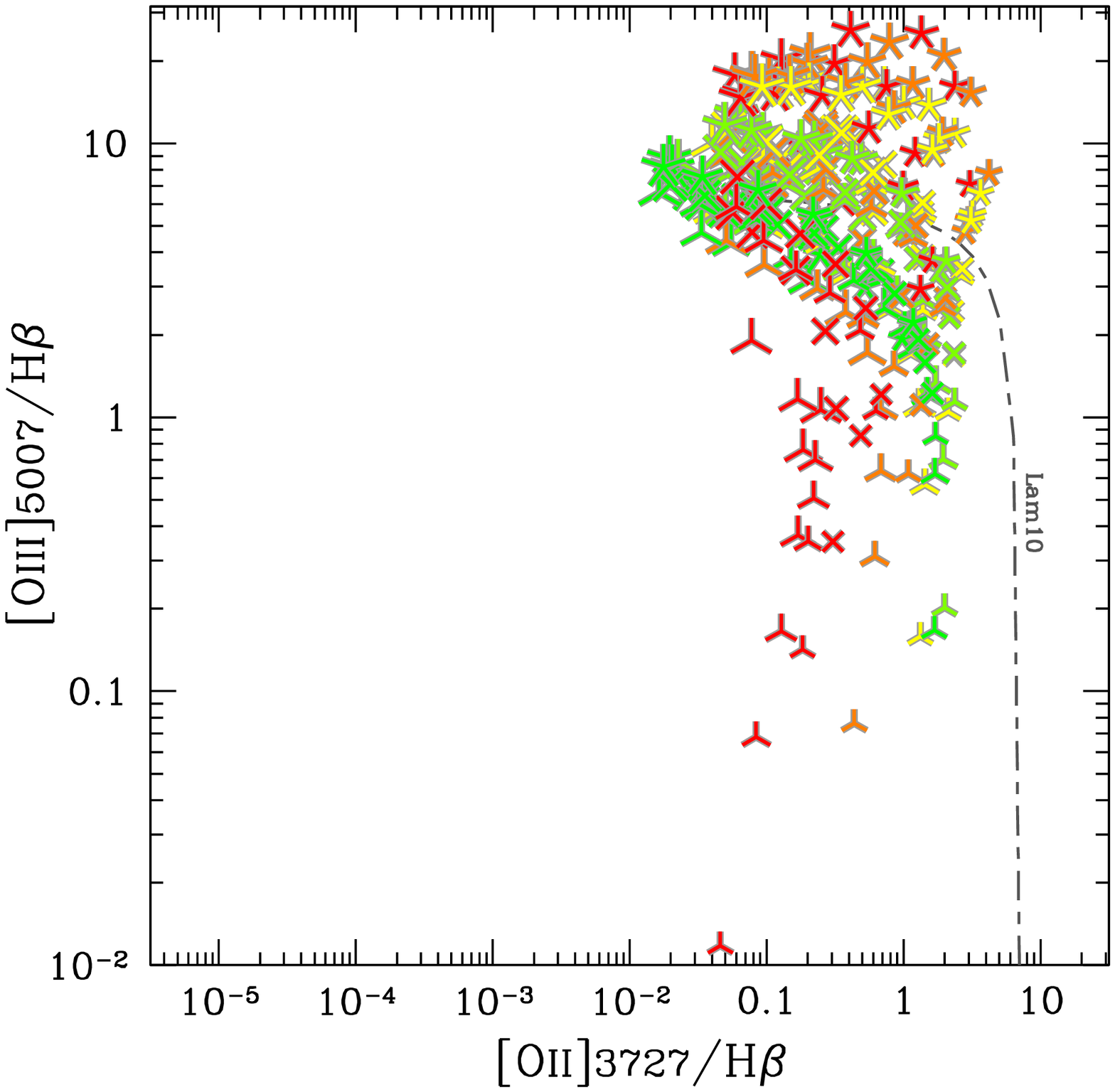}
        \end{center}
      \end{minipage}
    \end{tabular}
    \caption{%
    	Distribution of the \cloudy\ models on the optical metal-line diagnostics 
    	(\OIII$\lambda5007$$/$\Hb\ versus ~\NII$\lambda 6584$$/$\Ha\ (1st row),
    	\OIII$\lambda5007$$/$\Hb\ versus ~\SII$\lambda\lambda 6717,6731$$/$\Ha\ (second row), and 
	\OIII$\lambda5007$$/$\Hb\ versus ~\OII$\lambda 3727$$/$\Hb\ (3rd row),
	    for the primitive sources (Left)
	    and for the chemically-evolved systems of galaxies (Middle) and AGNs (Right).
	    Symbols as in Fig.~\ref{fig:ewhe2_he2hb}.
	    In the extremely low-metallicity regime (Z $<10^{-3}$; left panels), 
	    all  models follow a sequence for a given gas-phase metallicity 
	    irrespective of the different ionising sources.
	    The dotted curves show the average sequences for the models with Z $=10^{-5}$ 
	    (Eqs.~\ref{eq:o3hb_n2ha_zem5}--\ref{eq:o3hb_o2hb_zem5})
	    and $10^{-4}$ (Eqs.~\ref{eq:o3hb_n2ha_zem4}--\ref{eq:o3hb_o2hb_zem4}).
	    The conventional demarcation curves between AGNs and galaxies on the BPT-diagrams
	    are depicted
	    as introduced by \citet{kewley2001} (dot-long dashed; ``Kew01'') and 
	    \citet{kauffmann2003} (short dash-long dashed; ``Kau03'').
	    The demarcation curve of \citet{lamareille2010} is plotted on the bottom diagram
	    using \OII\ (dot-long dashed; ``Lam10''). 
	    Clearly, these curves work for isolating AGNs if the metallicity is 
	    relatively high (Z $\gtrsim 0.5$\,\Zsun)
	    and the spectrum is hard ($\alpha\gtrsim -1.6$).
    }
    \label{fig:metals_optical}
\end{figure*}


Among several \HeI\ recombination emission lines in the optical wavelength, 
we firstly test with \HeI$\lambda 5876$ as it is generally the strongest \HeI\ line. 
We find, however, that the PopIII and DCBH models show a comparable 
\HeI$\lambda 5876/$\Hb\ ratio. 
Rather, the line ratio depends more on the gas density in the ISM.
Fig.~\ref{fig:he2hb_he1hb_5876_density} shows a diagram of \HeI$\lambda 5876/$\Hb\ 
and \HeII$/$\Hb\ for the PopIII and PopII galaxies,
varying the gas density from $10^2$\,cm$^{-3}$ to $10^4$\,cm$^{-3}$ in addition
to the fiducial value of $10^3$\,cm$^{-3}$.
Clearly the \HeI$\lambda 5876/$\Hb\ ratio increases with gas density 
(see also \citealt{aver2015}).
We therefore propose that this \HeI$\lambda 5876$ diagram can be used to probe the 
gas density for metal-deficient galaxies and even for DCBHs, such that 
\begin{align}
	\log {\rm He\,\textsc{ii}}/{\rm H}\beta &= 0.07/(\log {\rm He\,\textsc{i}}/{\rm H}\beta +0.98) - 0.5
	\,(n_{\rm e}=10^2), \label{eq:he2hb_he1hb_5876_n2} \\
	&= 0.07/(\log {\rm He\,\textsc{i}}/{\rm H}\beta +0.83) - 0.5
	\,(n_{\rm e}=10^3), \label{eq:he2hb_he1hb_5876_n3} \\
	&= 0.07/(\log {\rm He\,\textsc{i}}/{\rm H}\beta +0.73) - 0.5
	\,(n_{\rm e}=10^4). \label{eq:he2hb_he1hb_5876_n4}
\end{align}
This diagram can substitute the density indicators using the metal collisional 
excitation lines such as \OII$\lambda 3726/3729$ and \SII$\lambda 6717/6731$
which will probably not be available for primitive, extremely metal-poor sources.
A caveat is that this diagram does not work accurately for chemically evolved galaxies and AGNs,
because of the secondary dependence of the \HeI\ emission on the gas temperature 
(and hence on the metallicity). For the evolved systems, the classical metal line ratios 
are preferred, as conventionally used for this purpose.

Next, we investigate a diagram using another \HeI\ line at $\lambda=4471$\,\AA.
Because \HeI$\lambda 4471$ is the recombination lines resulting from a higher level
than \HeI$\lambda 5876$, it can be less affected by ISM properties and depend 
more on the shape of ionising spectrum.
Fig.~\ref{fig:he2hb_he1hb_4471} presents the diagram \HeI$\lambda 4471/$\Hb\ 
versus \HeII$/$\Hb.
Although there is a certain amount of overlap between the PopIII galaxies and DCBHs
on the diagram, some types of DCBHs can be indeed potentially disentangled from the region of overlap; specifically, objects meeting the following criteria
\begin{align}
	{\rm He\,\textsc{ii}}/{\rm H}\beta > 0.01\ \&\ 
	{\rm He\,\textsc{i}\lambda 4471}/{\rm H}\beta > 0.04.
	\label{eq:he2hb_he1hb_4471}
\end{align}
 are most likely dominated by non-thermal hard 
ionising spectrum associated with accreting black holes. 
The separation is certainly less clear than in the diagram using the EW(\HeII)
(Fig.\ref{fig:ewhe2_he2hb}), but the diagram in Fig.~\ref{fig:he2hb_he1hb_4471}, which 
 uses only line ratios, is not affected by the potential issue of 
nuclear obscuring torus that might result into an artificially higher EW of the lines.

Once again, more evolved AGNs can also fall on the same region on this diagram 
(as shown in the right panel)
and it is hard to assess the metal-enrichment only based on the \HeI\ diagram.
One would need other diagrams using the metal lines as discussed below.

\begin{figure*}
  \centering
    \begin{tabular}{c}
      \begin{minipage}{0.49\hsize}
        \begin{center}
         \includegraphics[bb=18 143 555 680, width=0.85\textwidth]{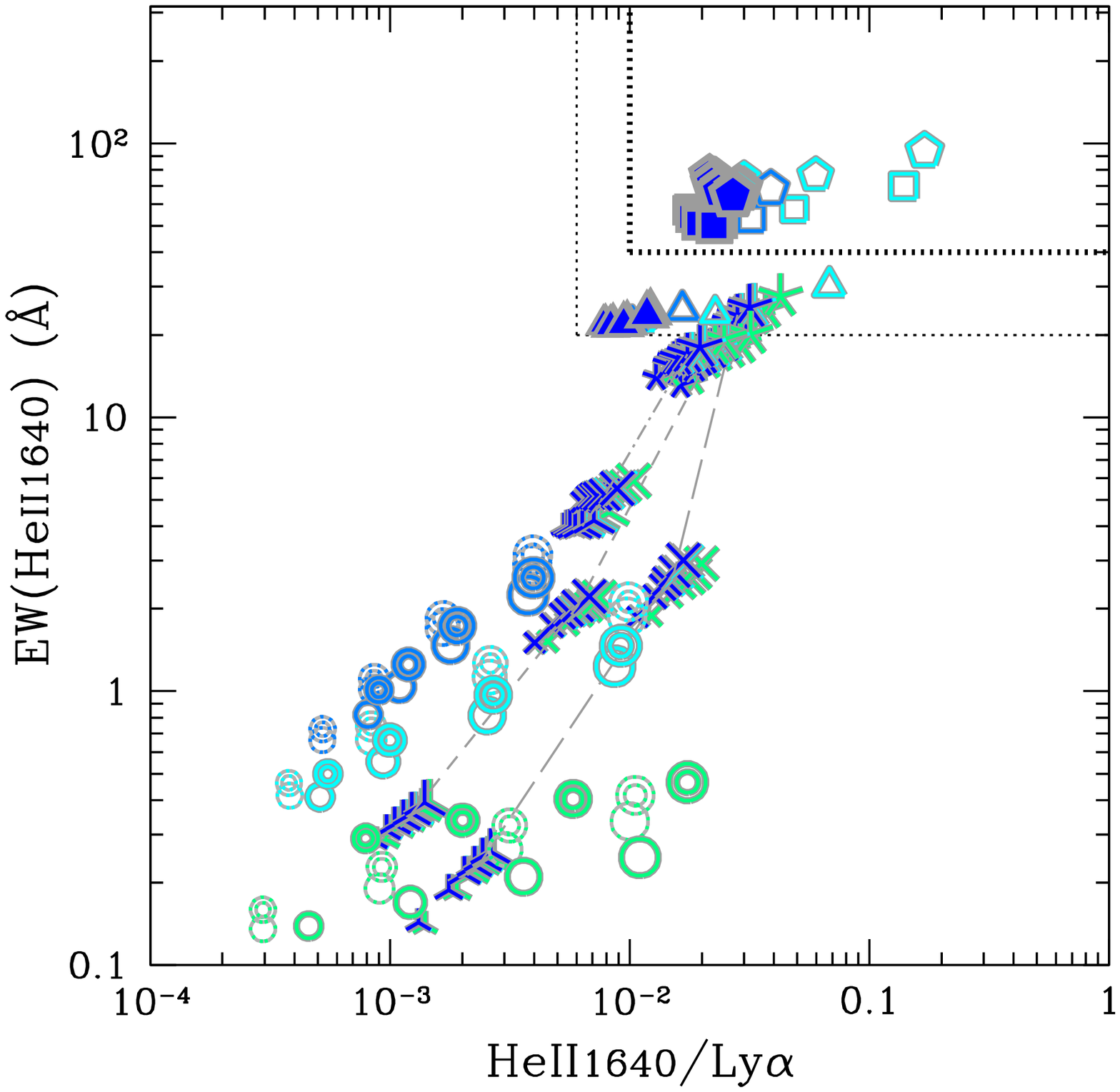}
        \end{center}
      \end{minipage}
      \begin{minipage}{0.49\hsize}
        \begin{center}
         \includegraphics[bb=18 143 555 680, width=0.85\textwidth]{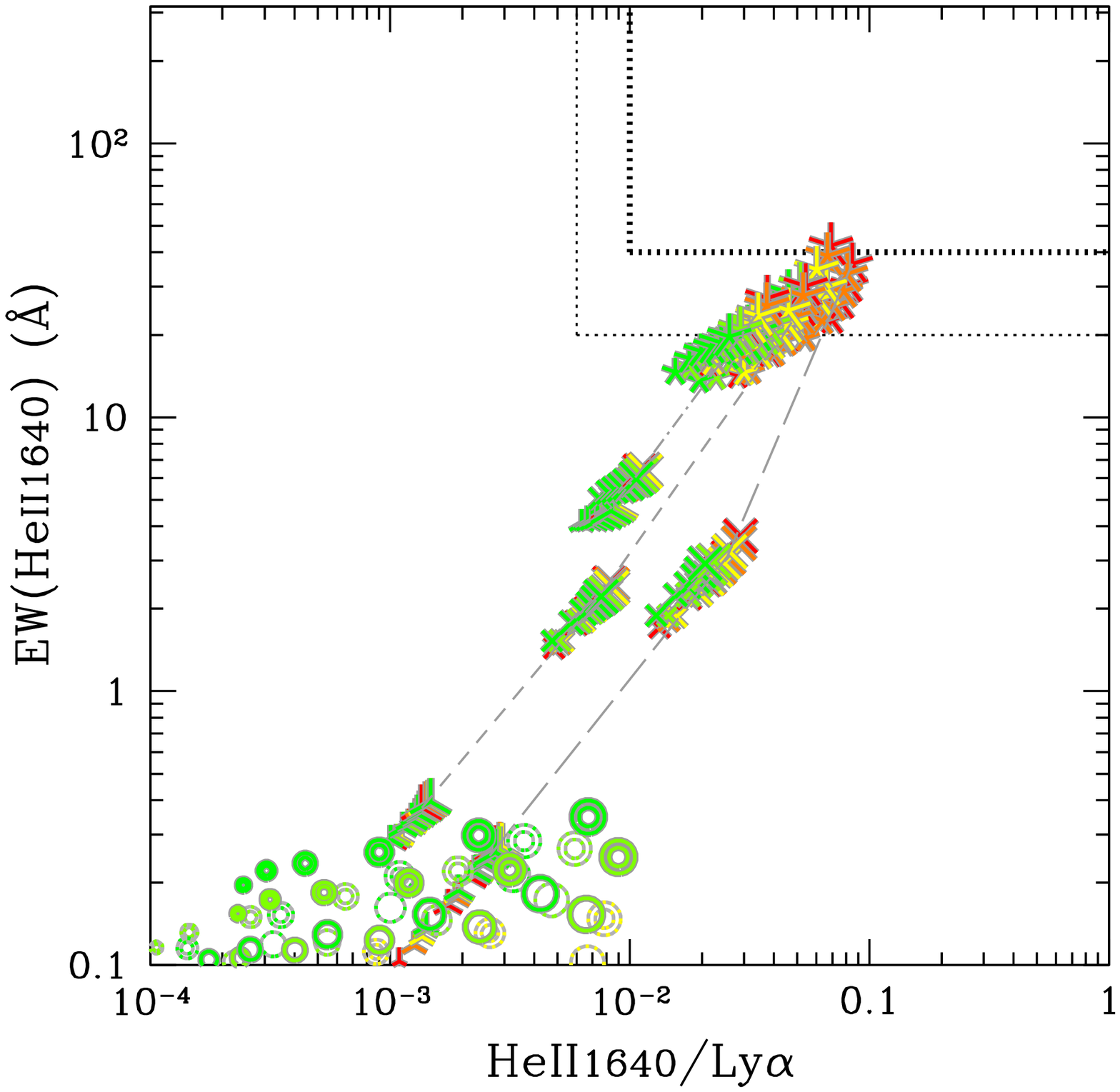}
        \end{center}
      \end{minipage}
    \end{tabular}
    \caption{%
    	\cloudy\ models of EW(\HeII$\lambda 1640$) as a function of \HeII$\lambda1640$$/$\Lya\
	    for  primitive sources (Left) and 
	    for the chemically-evolved systems (Right). 
	    Symbols as in Fig.~\ref{fig:ewhe2_he2hb}.
	    For PopIII galaxies, dust-included models are supplementarily shown 
	    with the open symbols, whose shapes and colors follow the original 
	    dust-free models. 
	    \Lya\ would be further attenuated by IGM. 
	    The \Lya\ strengths predicted by our \cloudy\ models are therefore to be considered
	     upper-limits (i.e. lower limits on the \HeII$\lambda1640$$/$\Lya\ ratio).
	    The dotted lines present the proposed criteria to identify PopIII galaxies, 
	    optimised for the purity (thick; Eq.~\ref{eq:ewhe2_he2lya_strict}) and 
	    completeness (thin; Eq.~\ref{eq:ewhe2_he2lya_relaxed}).
	    The hardest DCBH models can contaminate into the less strict region 
	    on this diagram. 
    }
    \label{fig:ewhe2_he2lya}
\end{figure*}

\subsubsection{Diagnostics involving metal optical lines}

The diagrams discussed above, which use only the Hydrogen and Helium recombination lines,
are useful to identify PopIII galaxies and DCBHs, 
but not sufficient to assess how much the surrounding ISM gas is chemically-enriched. 
Also, as already mentioned, DCBHs and more evolved AGNs, living in enriched ISM, 
would not be separable. 
To discriminate  PopIII galaxies and DCBHs in a primitive environment from those embedded in a weakly, or even strongly, enriched ISM,
 optical metal lines are also needed. 
Fig.~\ref{fig:metals_optical} presents three diagrams showing the line ratios of 
\OIII$\lambda 5007$$/$\Hb\ vs.~\NII$\lambda 6584$$/$\Ha\ (1st row), 
\OIII$\lambda 5007$$/$\Hb\ vs.~\SII$\lambda\lambda 6717,6731$$/$\Ha\ (2nd row), and 
\OIII$\lambda 5007$$/$\Hb\ vs.~\OII$\lambda 3727$$/$\Hb\ (3rd row)
\footnote{%
\OII$\lambda3727$ denotes the sum of the doublet: \OII$\lambda\lambda 3726,3729$.
}. The former two are commonly called BPT-diagrams \citep{BPT1981}
both widely used to separate AGNs from star-forming galaxies
(e.g., \citealt{kewley2001,kauffmann2003,kewley2013_theory}).
The left panel show the case for primeval, extremely metal-poor cases, the central panels show the case of evolved (significantly metal enriched) PopII galaxies, and the right panels show the case of chemically evolved AGNs.

In the central panels of the BPT-diagrams, we confirm that evolved star-forming galaxies 
follow or fall below the demarcation curves.
We note that the BPT diagnostic line ratios are not significantly influenced by the choice
of stellar age, and mostly governed by the youngest stellar population
(see also Fig.~\ref{fig:no_photon}b for the almost identical SEDs at $1$ and $10$\,Myr 
for a given metallicity at $E\lesssim 50$\,eV).
On the other hand as evidenced in the right panels, 
not all of the evolved AGN models are separable on the BPT-diagrams. 
AGNs with a gas-phase metallicity below Z $\lesssim 0.5$\,\Zsun\ can spill 
(and hence `contaminate') into the star-forming galaxy region of the diagrams, 
as already indicated by \citet{kewley2013_theory}.
Moreover, the models with the least hard spectrum ($\alpha=-2.0$), if they really exist,  
could not be classified as AGNs with the classical BPT diagnostics even with a 
super-solar metallicity gas.
The diagram in the 3rd row makes use of the \OII\ and \Hb\ emission in place of 
\NII, \SII, and \Ha. This will become particularly useful when the wavelengths around \Ha\
is not observationally available, e.g., for sources at $z\gtrsim7$ with JWST/NIRSpec.

We note the evolved star-forming galaxy models in the middle panels of the \NII\ BPT-diagram
shape a model grid consistent with \citet{kewley2013_theory} and other models results \citep[e.g.][]{gutkin2016}, 
and can reproduce the observed metallicity sequence of star-forming galaxies 
on the diagrams in the local universe (e.g., \citealt{maiolino2008,curti2017,curti2021}),
following the trend that the ionisation parameter typically gets larger 
in a more metal-deficient galaxy 
(e.g., \citealt{AM2013, sanders2020}; Nakajima et al. 2021 submitted).
A similar but offsetted sequence toward a larger \OIII$/$\Hb\ ratio is reported 
for $z=2-3$ galaxies in the \NII\ BPT-diagram (e.g., \citealt{steidel2014,shapley2015,strom2017}). 
The $z=2-3$ sequence is also compatible with our model grid, 
although the physical origins of the evolution remain under debate 
(e.g., \citealt{curti2021}).

We also note that the sequence for evolved AGN found by us is also broadly consistent with other AGN photoionisation models \citep[e.g.][]{groves2006,feltre2016}.

In the extremely low-metallicity regime in the left panels, 
all the models of PopIII galaxies, PopII galaxies, and DCBHs follow a similar sequence
for a given gas-phase metallicity with the ionisation condition varied. 
Although these optical metal-line diagrams have, therefore, no power to distinguish between the objects 
powered by different ionising sources in such a low metallicity regime below 
Z $\lesssim 10^{-3}$, 
they are very useful to infer the gas-phase metallicity, in a way similar to the strong line 
metallicity diagnostics (e.g., \citealt{MM2019}; Nakajima et al. 2021 submitted), 
but independently of the source of ionisation, and also allow the identification of sources 
that are not chemically-evolved at all.

Using the low-metallicity models found in the left panels, we define the average 
relationships as follows:
\begin{align}
	\log {\rm [O\,\textsc{iii}]}/{\rm H}\beta &= 0.61/(\log {\rm [N\,\textsc{ii}]}/{\rm H}\alpha + 2.7) - 0.8 \label{eq:o3hb_n2ha_zem5} \\
	&= 0.72/(\log {\rm [S\,\textsc{ii}]}/{\rm H}\alpha +2.4) - 0.8 \label{eq:o3hb_s2ha_zem5} \\
	&= 0.65/(\log {\rm [O\,\textsc{ii}]}/{\rm H}\beta +1.2) - 0.8 \label{eq:o3hb_o2hb_zem5}
\end{align}
for a metallicity of $Z=10^{-5}$, and 
\begin{align}
	\log {\rm [O\,\textsc{iii}]}/{\rm H}\beta &= 0.61/(\log {\rm [N\,\textsc{ii}]}/{\rm H}\alpha + 1.7) + 0.2 \label{eq:o3hb_n2ha_zem4} \\
	&= 0.72/(\log {\rm [S\,\textsc{ii}]}/{\rm H}\alpha +1.4) + 0.2 \label{eq:o3hb_s2ha_zem4} \\
	&= 0.65/(\log {\rm [O\,\textsc{ii}]}/{\rm H}\beta +0.2) + 0.2 \label{eq:o3hb_o2hb_zem4}
\end{align}
for $Z=10^{-4}$.

Very interestingly,
we note that the \OIII$\lambda 5007$ line remains relatively strong 
(at the level of one tenth of H$\beta$; see also \citealt{inoue2011_metal_poor}) 
even at extremely low metallicities, down to $\rm Z=10^{-5}$,
probably as a consequence of the fact that even a small amount of oxygen 
rapidly makes its collisionally excited transitions good coolant of the ISM.
Objects that fall below the $Z=10^{-5}$ curves are good candidates of PopIII galaxies
or DCBHs just formed in a totally primitive environment. 
As mentioned, even galaxies at $\rm Z=10^{-5}$ or $\rm Z=10^{-4}$ can actually be 
hosting PopIII galaxies or DCBHs embedded in slightly enriched ISM.
The EW(\HeII) diagram (Fig.\ref{fig:ewhe2_he2hb}) and the \HeI\ diagram (Fig.\ref{fig:he2hb_he1hb_4471})
will then be powerful to distinguish between the two ionising spectral shapes, 
i.e., PopIII stars vs.~DCBHs.

\subsection{UV Diagnostics} \label{ssec:results_uv}

\subsubsection{Diagnostics using only He and H UV lines}
\label{sssec:results_uv_h_he}

In this section we present the photoionisation model results for the UV emission lines.
A major advantage of using the UV emission lines is that they are available for
high-redshift galaxies even with the ground-based NIR instruments.

Fig.~\ref{fig:ewhe2_he2lya} focuses on the UV \HeII\ emission at $1640$\,\AA, showing
EW(\HeII) as a function of \HeII$/$\Lya.
One caveat is that the \Lya\ strengths used here are simply predicted by 
our \cloudy\ models truncated at the edge of the ionised nebula (\S\ref{sec:modelling}), 
and thus supposed to be the maximum value emitted from the cloud.
Neutral hydrogen gas and dust in the systems would weaken the \Lya\ 
emission.
Indeed, the dust-included PopIII models supplementarily plotted in Fig.~\ref{fig:ewhe2_he2lya} (hollow triangles, squares and pentagons)
show a weakened \Lya\ and thus a higher \HeII$/$\Lya\ ratio, 
even higher by a factor of $\sim 10$ for the Z $=10^{-4}$ models.
EW(\HeII) would also become enhanced if dust is included due to a weaker continuum level, 
but the effect is expected to be minor, by a factor of $<2$, as this is a secondary, differential effect.
\Lya\ would also be absorbed by the inter-galactic medium (IGM) at high redshift, in particular for sources 
at, or beyond, the reionisation epoch ($z>6$). 
The \HeII$/$\Lya\ ratios presented in Fig.~\ref{fig:ewhe2_he2lya} are therefore 
regarded as the lower-limits.

\begin{figure*}
  \centering
    \begin{tabular}{c}
      \begin{minipage}{0.49\hsize}
        \begin{center}
         \includegraphics[bb=18 143 555 680, width=0.85\columnwidth]{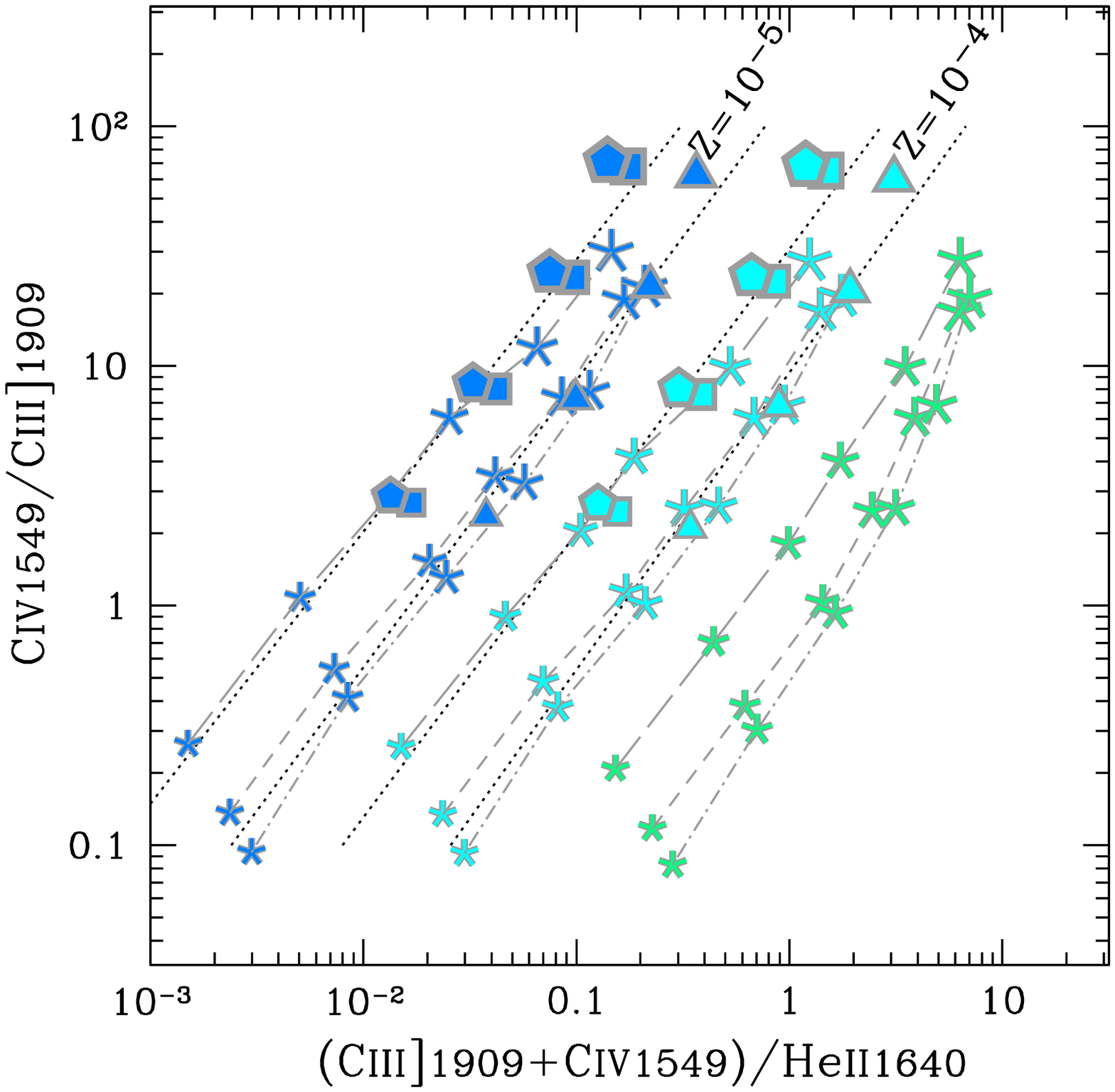}
        \end{center}
      \end{minipage}
      \begin{minipage}{0.49\hsize}
        \begin{center}
         \includegraphics[bb=18 143 555 680, width=0.85\columnwidth]{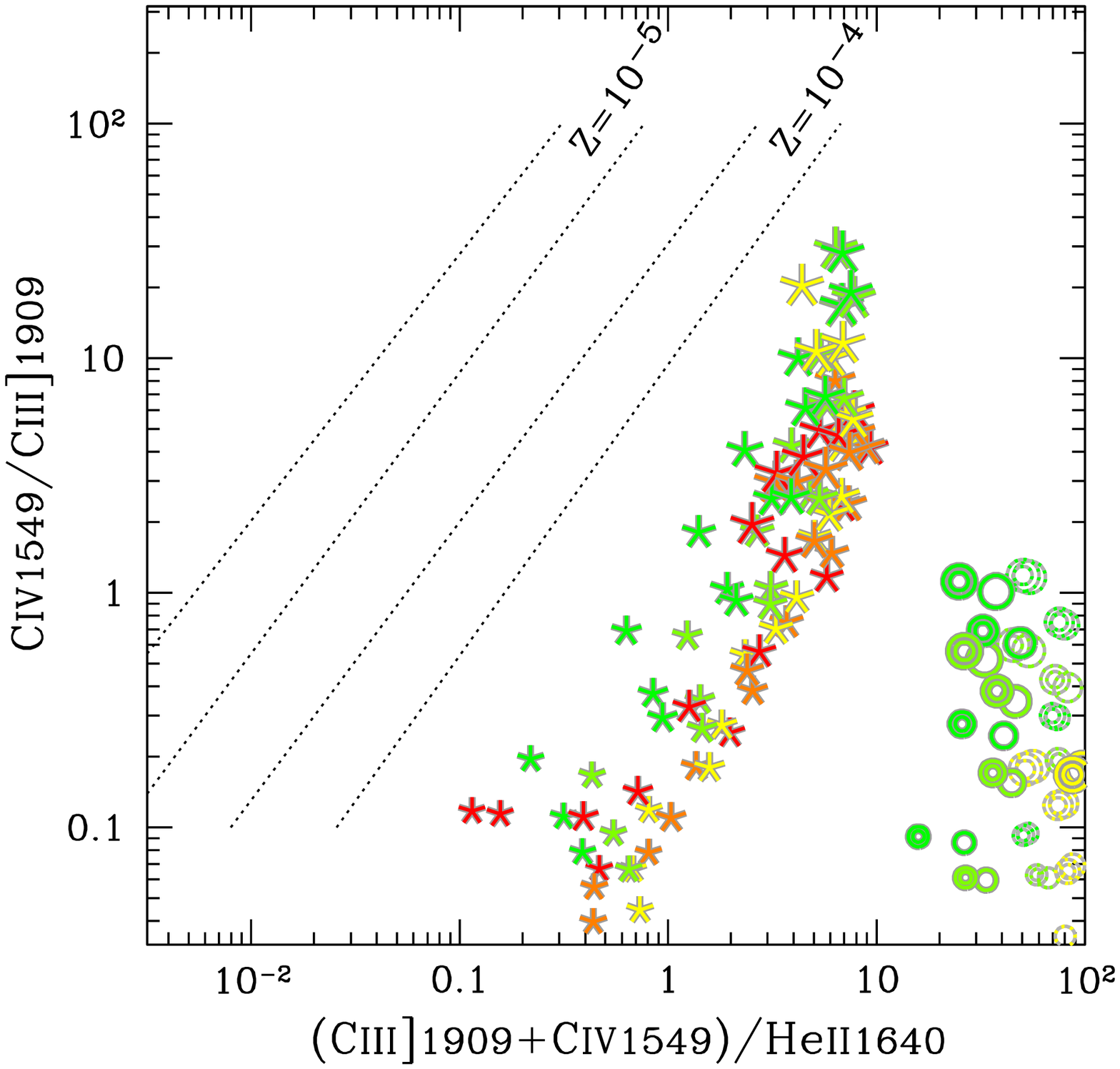}
        \end{center}
      \end{minipage}
    \end{tabular}
    \caption{%
    		\cloudy\ models of (\CIII$+$\CIV)$/$\HeII\ vs.~\CIV$/$\CIII\ 	
		for the PopIII galaxies and DCBHs (Z $\leq 10^{-3}$) with 
	    	the hardest ionising spectrum ($\alpha=-1.2$)
	    	that are supposed to be selected with the EW(\HeII$\lambda 1640$) diagnostic (Left).
	    	Symbols as in Fig.~\ref{fig:ewhe2_he2hb}.
	    	 PopIII models with top-heavy IMF ($\rm M_{up}=$ $500$\,\Msun) and  DCBH models 
	    	with $\rm T_{\rm bb}= 5~10^4\,K$ (Group A) follow a well defined sequence for a given metallicity. 
	    	The other models (Group B) fall on another metallicity sequence 
	    	which is offset toward a weaker \HeII\ and/or a lower \CIV$/$\CIII\ ratio.
	    	The sequences of $\rm Z =10^{-5}$ and $10^{-4}$ for both groups are 
        	illustrated with black dotted lines 
		(Eqs.~\ref{eq:c4c3_c34_A_zem5}--\ref{eq:c4c3_c34_B_zem5} and 
        	Eqs.~\ref{eq:c4c3_c34_A_zem4}--\ref{eq:c4c3_c34_B_zem4}, respectively).
		In the right panel, metal-rich models for AGNs ($\alpha=-1.2$) and PopII galaxies 
		are shown on the same diagram, and are confirmed not to contaminate the 
		low-metallicity sequences.
	}
    \label{fig:c4c3_c34}
\end{figure*}

Using the predicted emission line strengths for the populations on this diagram,
we propose the following selection criteria for selecting PopIII galaxies:
\begin{equation}
	{\rm EW}({\rm He\,\textsc{ii}}\lambda 1640) > 40\,\mbox{\AA}\ \&\
	{\rm He\,\textsc{ii}\lambda 1640}/{\rm Ly}\alpha > 0.01
	\label{eq:ewhe2_he2lya_strict}
\end{equation}
which are traced by thick dotted lines in Fig.~\ref{fig:ewhe2_he2lya}.
These are based on the top-heavy IMF models reaching the maximum mass-cut of 
$500$\,\Msun\ which are located in a distinct region on this diagram.
PopIII galaxies following such a top-heavy IMF, including the extreme one 
containing only massive stars of $50-500$\,\Msun, are separable from the other population
on this UV \HeII\ diagram.
The PopIII models with the moderate ($1-100$\,\Msun\ Salpeter) IMF, 
on the other hand, present a maximum EW(\HeII) of $\sim 20-30$\,\AA, 
which is comparable to the predicted EWs for the hardest DCBH/AGN models, 
and hence indistinguishable from each other on this diagram.
The criteria of Eq.~(\ref{eq:ewhe2_he2lya_strict}) is thus optimised 
to identify PopIII galaxies and minimise contaminations by DCBHs.
In order to build a more complete sample of PopIII galaxies with a certain 
contamination by DCBHs allowed, the following, less strict criteria 
can also be useful:
\begin{equation}
	{\rm EW}({\rm He\,\textsc{ii}}\lambda 1640) > 20\,\mbox{\AA}\  \&\
	{\rm He\,\textsc{ii}\lambda 1640}/{\rm Ly}\alpha > 0.006.
	\label{eq:ewhe2_he2lya_relaxed}
\end{equation}
The choice of stellar age for the PopII models does not affect the PopIII selection
criterion on the diagram, as it was the case for the optical EW(\HeII) diagnostic (Fig.~\ref{fig:ewhe2_he2hb}).
In short, the UV \HeII\ emission provides a good diagnostic to distinguish between
PopIII galaxies and DCBHs, although the optical \HeII\ has a better and more sharp 
discriminatory power.
\\

\begin{figure*}
  \centering
    \begin{tabular}{c}
      \begin{minipage}{0.24\hsize}
        \begin{center}
         \includegraphics[bb=18 143 555 680, width=0.95\textwidth]{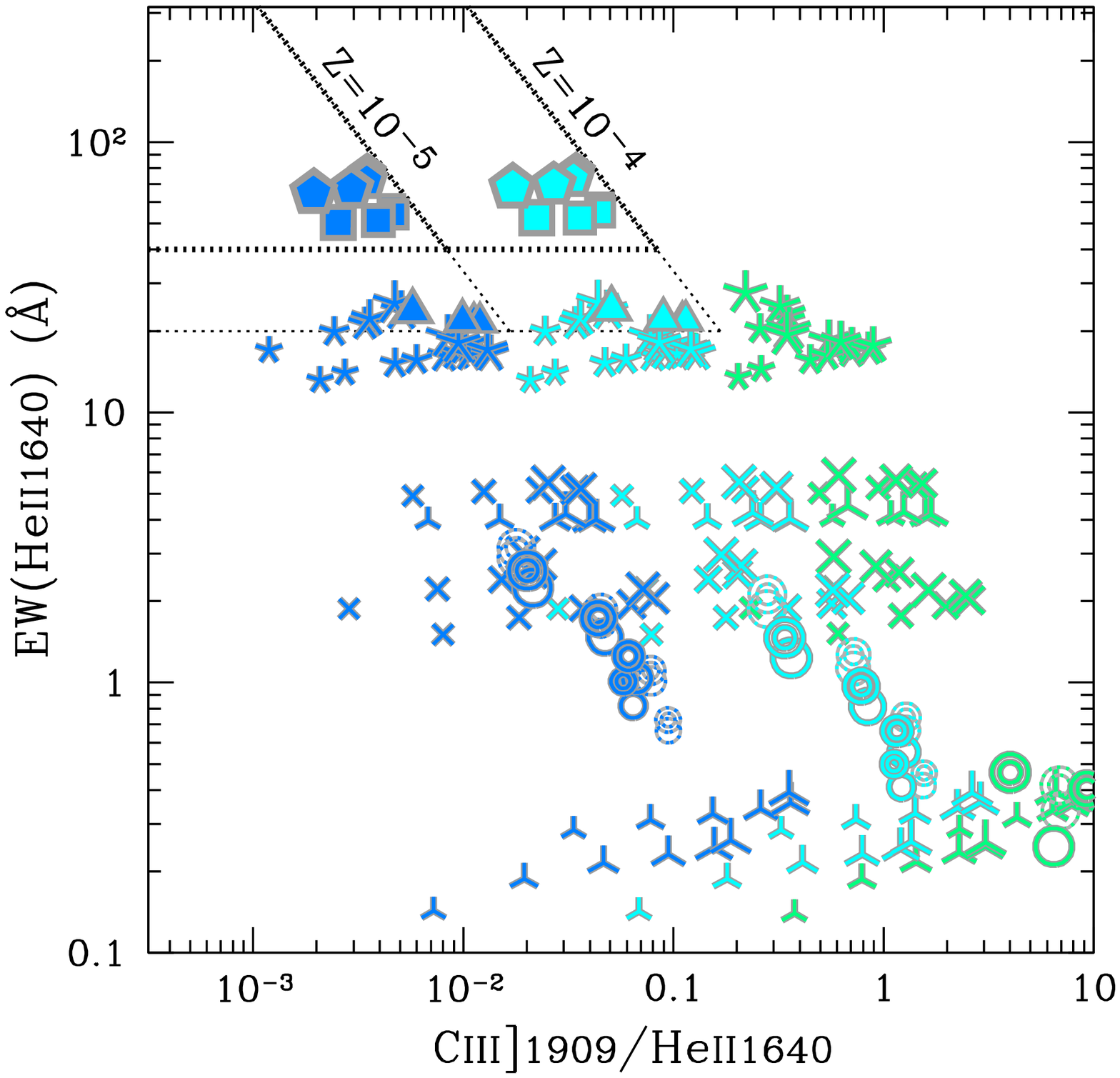}
        \end{center}
      \end{minipage}
      \begin{minipage}{0.24\hsize}
        \begin{center}
         \includegraphics[bb=18 143 555 680, width=0.95\textwidth]{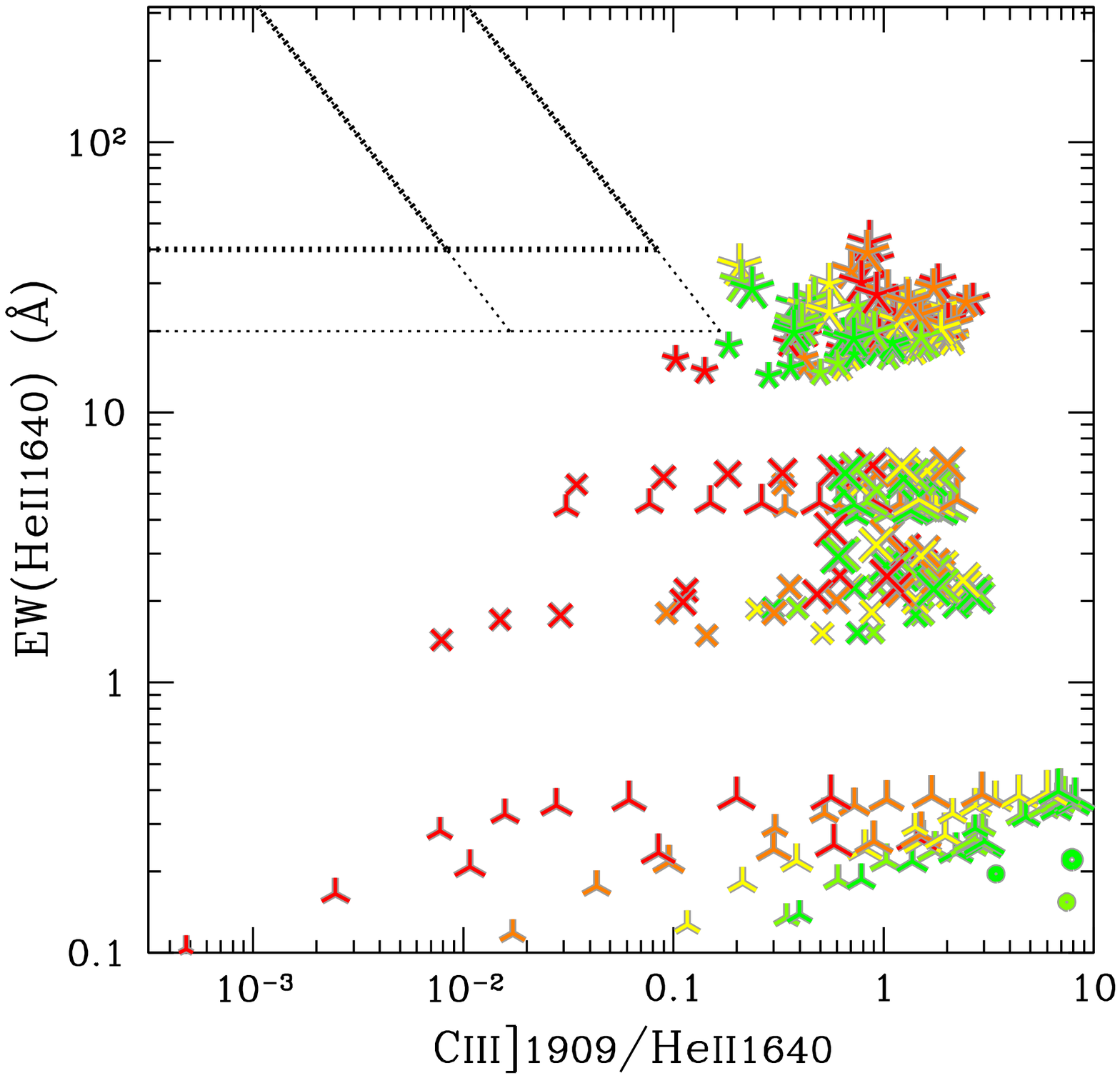}
        \end{center}
      \end{minipage}
      \hspace{0.02\hsize}
      \begin{minipage}{0.24\hsize}
        \begin{center}
         \includegraphics[bb=18 143 555 680, width=0.95\textwidth]{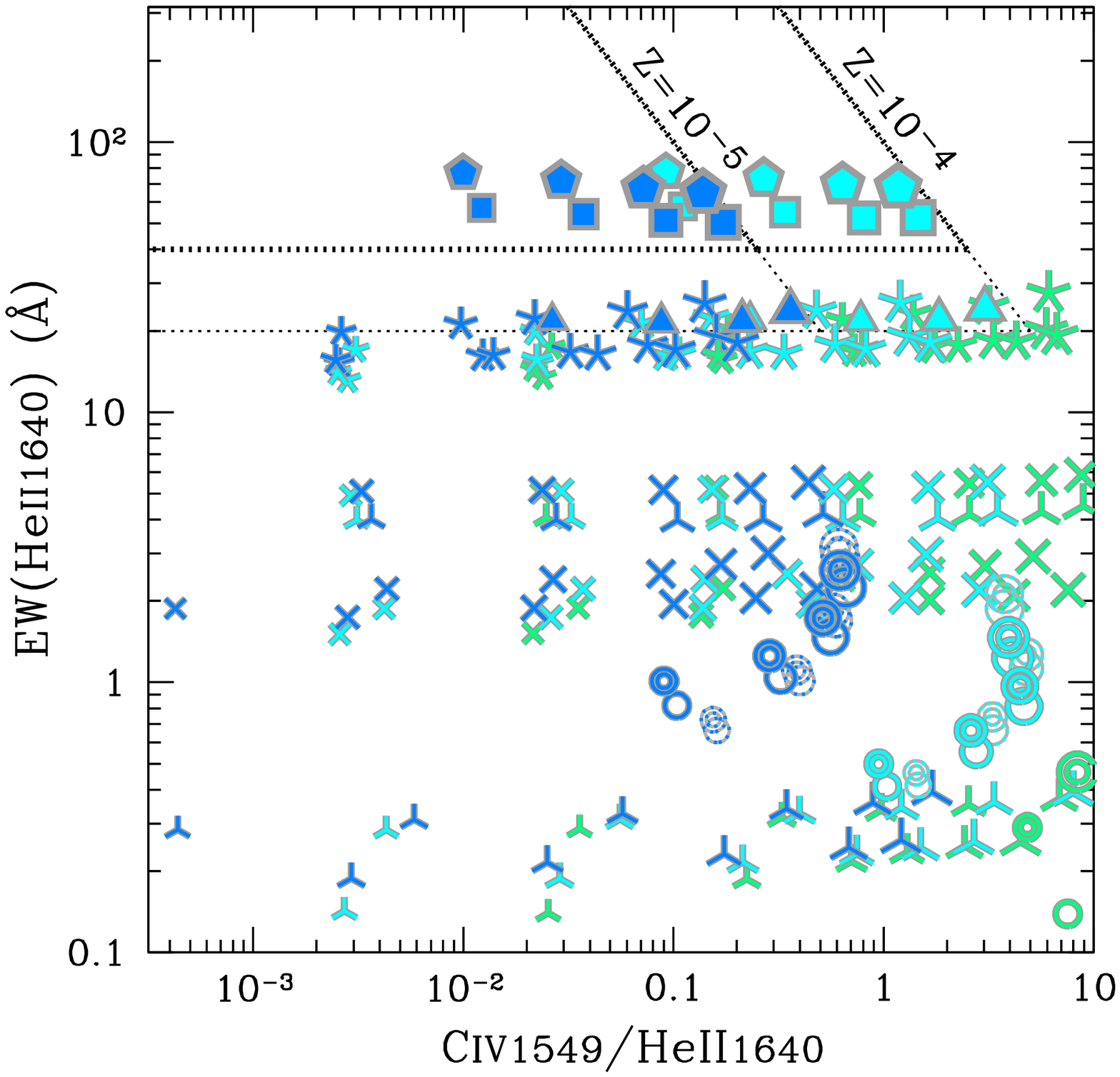}
        \end{center}
      \end{minipage}
      \begin{minipage}{0.24\hsize}
        \begin{center}
         \includegraphics[bb=18 143 555 680, width=0.95\textwidth]{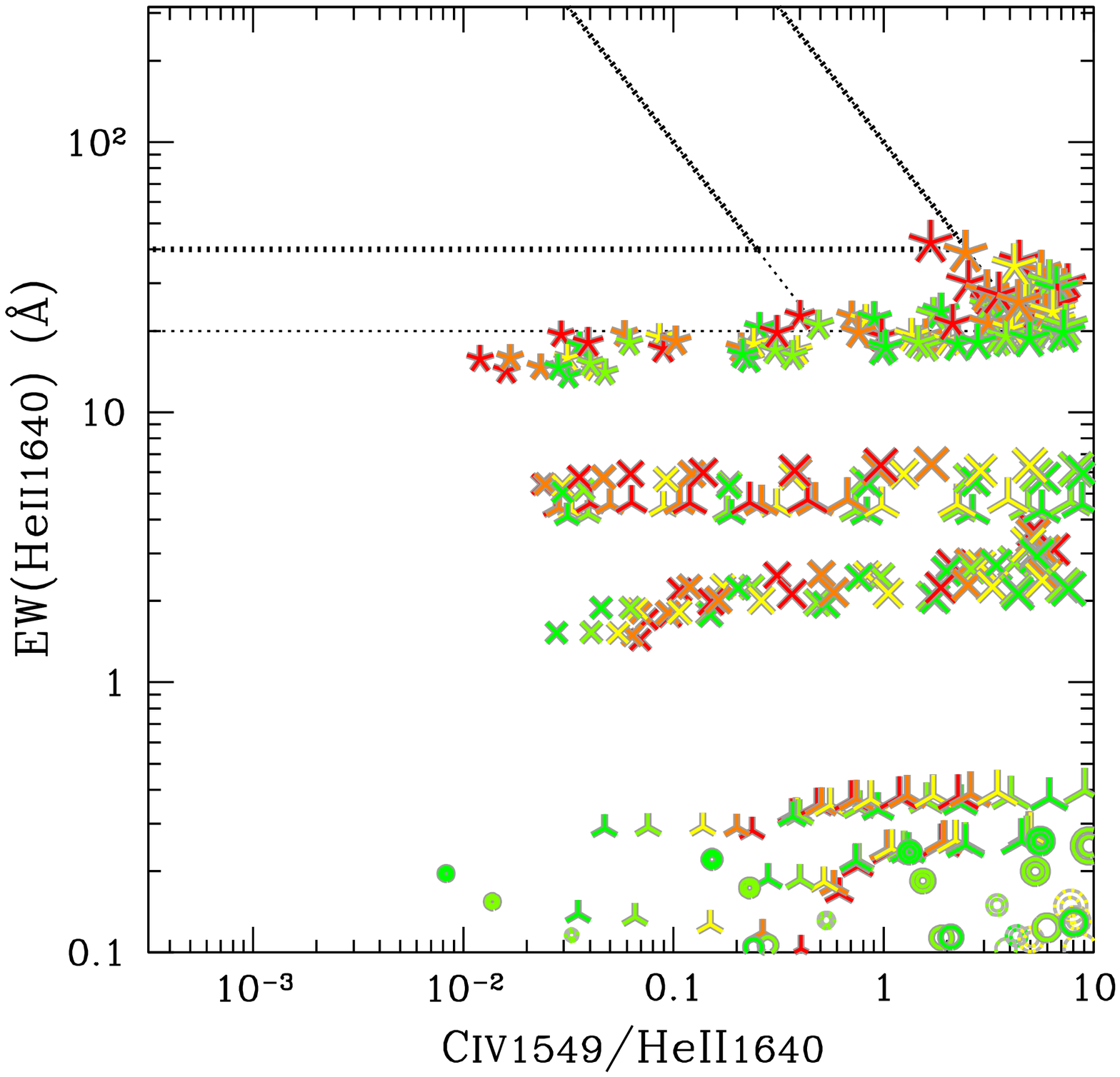}
        \end{center}
      \end{minipage}
      \\
      \begin{minipage}{0.24\hsize}
        \vspace{10pt}
        \begin{center}
         \includegraphics[bb=18 143 555 680, width=0.95\textwidth]{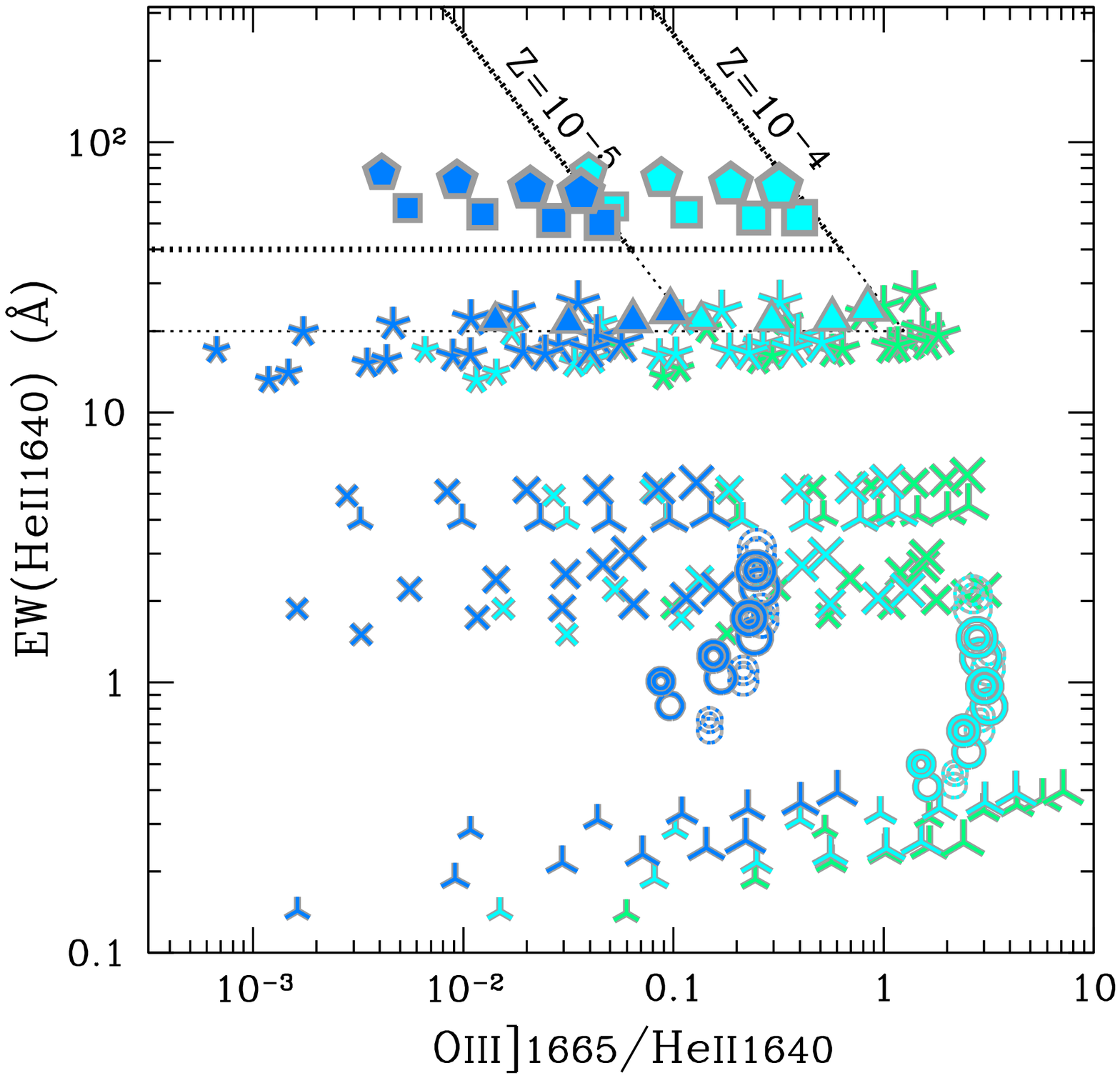}
        \end{center}
      \end{minipage}
      \begin{minipage}{0.24\hsize}
        \vspace{10pt}
        \begin{center}
         \includegraphics[bb=18 143 555 680, width=0.95\textwidth]{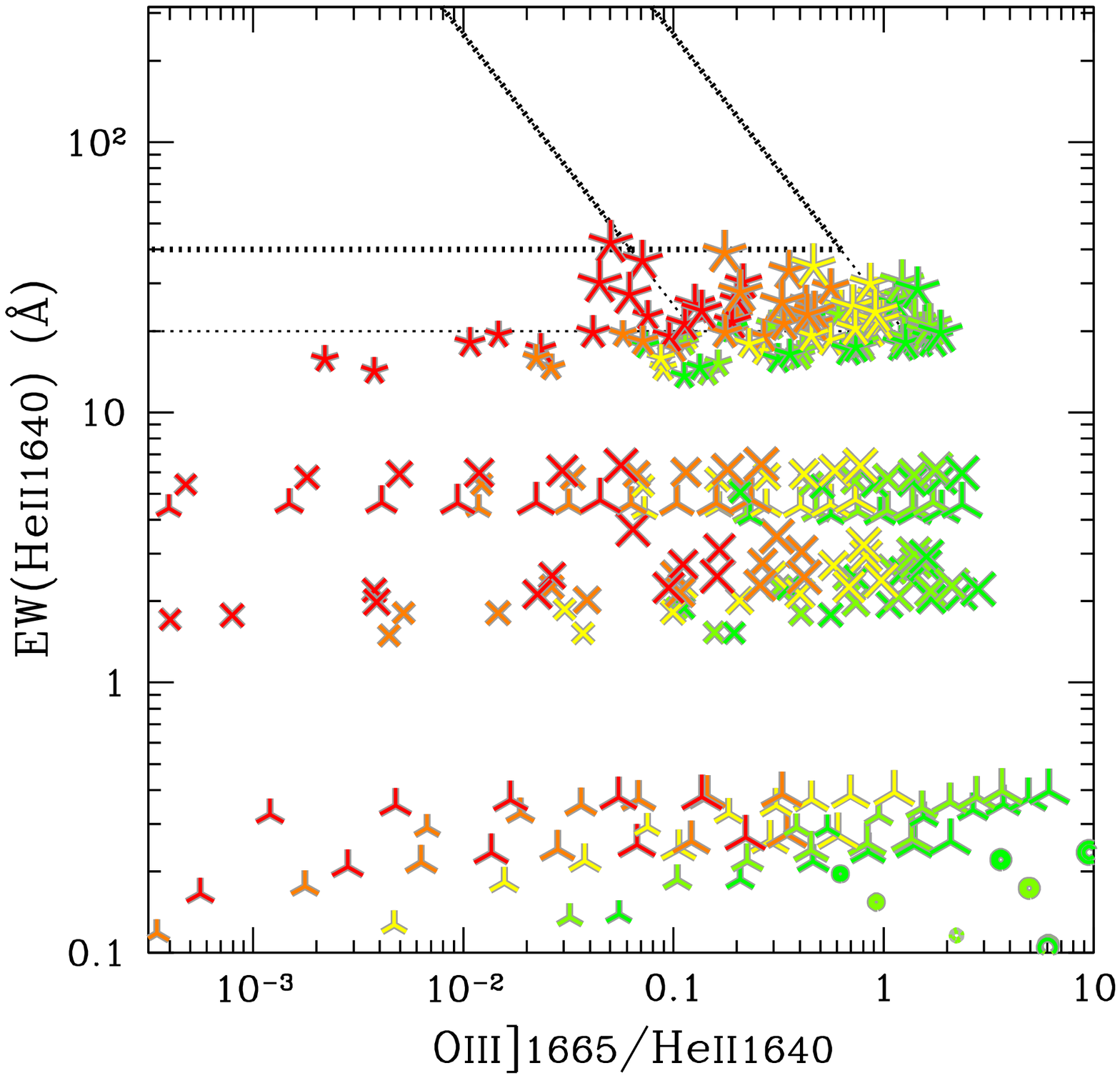}
        \end{center}
      \end{minipage}
      \hspace{0.02\hsize}
      \begin{minipage}{0.24\hsize}
        \vspace{10pt}
        \begin{center}
         \includegraphics[bb=18 143 555 680, width=0.95\textwidth]{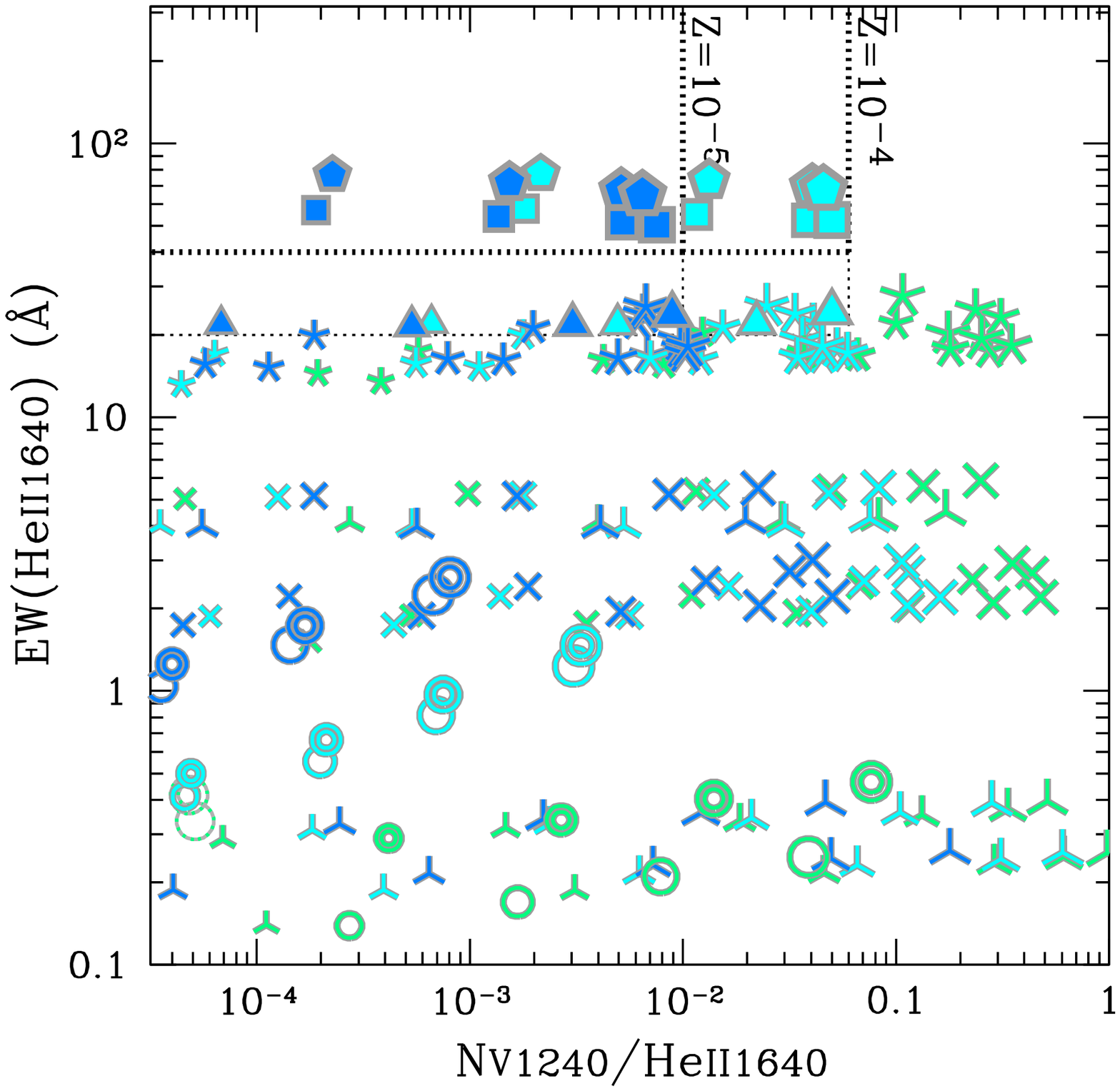}
        \end{center}
      \end{minipage}
      \begin{minipage}{0.24\hsize}
        \vspace{10pt}
        \begin{center}
         \includegraphics[bb=18 143 555 680, width=0.95\textwidth]{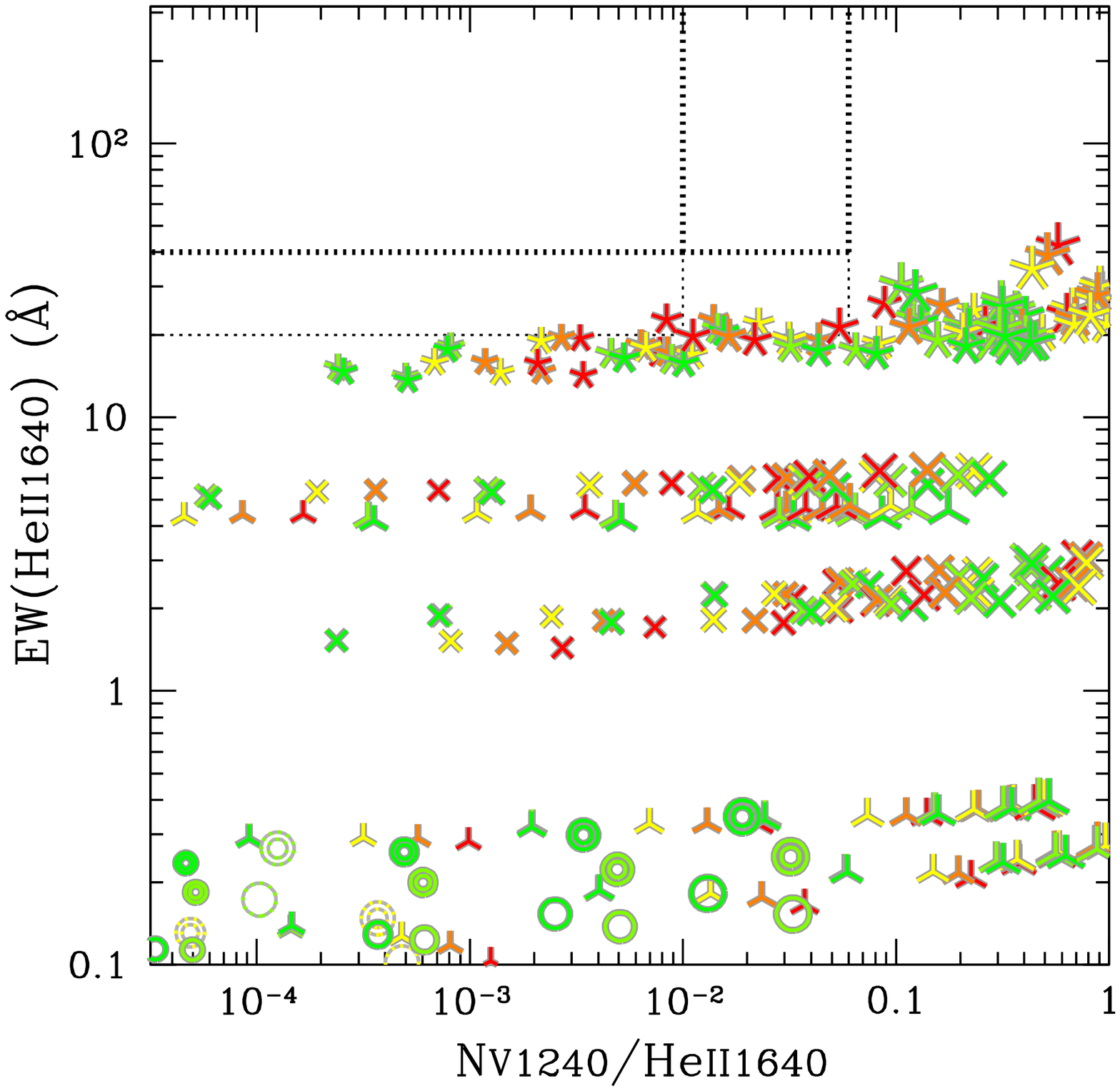}
        \end{center}
      \end{minipage}
    \end{tabular}
    \caption{%
    	Distribution of the \cloudy\ models in diagrams showing EW(\HeII$\lambda 1640$) as functions of 
	    \CIII$\lambda 1909$/\HeII$\lambda1640$ (top left 2 panels), 
	    \CIV$\lambda 1549$/\HeII$\lambda1640$ (top right), 
	    \OIIIuv$\lambda 1665$/\HeII$\lambda1640$ (bottom left), and 
	    \NV$\lambda 1240$/\HeII$\lambda1640$ (bottom right).
	    For each metal line, the primitive sources are displayed on the left-hand side,
	    and the chemically-evolved systems on the right.
	    Symbols as in Fig.~\ref{fig:ewhe2_he2hb}.
	    The dotted lines show the metallicity demarcations 
	    for PopIII galaxies 
	    (with the two EW(\HeII$\lambda 1640$) criteria; 
	    Eqs.~\ref{eq:ewhe2_he2lya_strict}--\ref{eq:ewhe2_he2lya_relaxed}) for  
	    Z $=10^{-5}$ (Eqs.~\ref{eq:ewhe2_he2c3_zem5}--\ref{eq:ewhe2_he2n5_zem5})
	    and 
	    $=10^{-4}$ (Eqs.~\ref{eq:ewhe2_he2c3_zem4}--\ref{eq:ewhe2_he2n5_zem4}).
    }
    \label{fig:metals_uv}
\end{figure*}

\subsubsection{Diagnostics using UV metal lines}
\label{sssec:results_uv_metals}

To diagnose the chemical enrichment of the ISM 
(even in the case of mildly enriched gas photoionised by PopIII stars) 
using the UV emission lines alone,
we propose to combine \HeII$\lambda 1640$ with the UV metal lines of 
[C\,{\sc iii}]$\lambda 1907$, C\,{\sc iii}]$\lambda 1909$
(hereafter their sum is referred to as \CIII$\lambda1909$, or simply \CIII)
and \CIV$\lambda \lambda 1548,1550$ (hereafter \CIV$1549$, or \CIV).
Fig.~\ref{fig:c4c3_c34} displays the diagram correlating 
the (\CIII$+$\CIV)$/$\HeII\ ratio and the \CIV$/$\CIII\ ratio, 
each of which is named C34 and C4C3, respectively.
In the left panel, we only plot  PopIII galaxies and the DCBH models 
with the hardest ionising spectrum ($\alpha=-1.2$) that are to be selected 
with the above EW(\HeII$\lambda 1640$) diagnostic 
(Eqs.~\ref{eq:ewhe2_he2lya_strict} or \ref{eq:ewhe2_he2lya_relaxed}).
The C34-index is informative to probe the gas metallicity, although
the secondary dependence on the ionisation condition is not negligible. 
The C4C3-index on the ordinate axis is then helpful to resolve the degeneracy and 
allows this diagram to infer the gas metallicity together with the ionisation parameter 
(see also, e.g., \citealt{nakajima2018_vuds}).

The diagram indeed shows the power to diagnose the gas metallicity.
Following the distributions of the models on the left diagram,
we define two metallicity sequences for these primitive sources:
Group A is for the PopIII galaxies with top-heavy IMF ($\rm M_{up}=500$\,\Msun, 
(pentagons and squares) and DCBHs with T$_{\rm bb}=$ 5e4\,K 
(star symbols connected with a long-dashed line),
and another group (Group B) for the other models.
Each group shapes a relatively tight sequence on this diagram for a given metallicity, 
with Group A showing a systematically stronger \HeII\ and/or a higher \CIV$/$\CIII\ ratio
than Group B.
The relationships are formulated as follows:
\begin{align}
	\log {\rm C4C3} &= 1.2 \log {\rm C34} + 2.6 \,\,{\rm (A)} \label{eq:c4c3_c34_A_zem5} \\
		&= 1.2 \log {\rm C34} + 2.1 \,\,{\rm (B)} \label{eq:c4c3_c34_B_zem5} 
\end{align}
to trace sources with Z $=10^{-5}$, and 
\begin{align}
	\log {\rm C4C3} &= 1.2 \log {\rm C34} + 1.5 \,\,{\rm (A)} \label{eq:c4c3_c34_A_zem4} \\
		&= 1.2 \log {\rm C34} + 1.0 \,\,{\rm (B)} \label{eq:c4c3_c34_B_zem4}
\end{align}
for Z $=10^{-4}$.

In the right panel of the same figure, metal-rich models of AGNs ($\alpha=-1.2$) and PopII galaxies 
are shown on the diagram. They present a larger C34 value,
as expected, than the primitive sources for a given C4C3, and well apart from the 
$\rm Z=10^{-5}$ and $10^{-4}$ sequences above.
The relationships thus provide a good measure of gas metallicity in such a low metallicity regime 
if all the necessary emission lines are available.
Particularly, PopIII galaxies with top-heavy IMF would be 
robustly selected following the strict selection criterion (Eq.~\ref{eq:ewhe2_he2lya_strict}) and
whose metallicity be inferred following the above relationships of Group A.

It is also expected that not all of the necessary lines are always available
from sources in the early universe.
To provide a substitute to probe the chemical enrichment of the ISM
in an efficient way with limited spectroscopic measurements,  
we summarise in Fig.~\ref{fig:metals_uv} the model distributions on 
 diagrams which employ a single metal line each, in addition to \HeII.
The considered metal lines include
\OIIIuv$\lambda\lambda 1661,1666$ (hereafter \OIIIuv$1665$, or \OIIIuv) and 
\NV$\lambda\lambda 1239, 1243$(hereafter \NV$1240$, or \NV) 
in addition to \CIII\ and \CIV.
For each of the diagrams, we can define the diagnostic regions defined by
\begin{align}
       \log {\rm C\,\textsc{iii}]} / {\rm He\,\textsc{ii}}\lambda 1640
		&< - \log [{\rm EW}({\rm He\,\textsc{ii}}\lambda 1640) \times 3.0] 
		\label{eq:ewhe2_he2c3_zem5} \\
	\log {\rm C\,\textsc{iv}} / {\rm He\,\textsc{ii}}\lambda 1640
		&< - \log [{\rm EW}({\rm He\,\textsc{ii}}\lambda 1640) \times 0.1] 
		\label{eq:ewhe2_he2c4_zem5} \\
	\log {\rm O\,\textsc{iii}]} / {\rm He\,\textsc{ii}}\lambda 1640
		&< - \log [{\rm EW}({\rm He\,\textsc{ii}}\lambda 1640) \times 0.4] 
		\label{eq:ewhe2_he2o3_zem5} \\
	{\rm N\,\textsc{v}} / {\rm He\,\textsc{ii}}\lambda 1640
		&< 0.01 \label{eq:ewhe2_he2n5_zem5} 
\end{align}
to select galaxies with a gas phase metallicity below $Z=10^{-5}$, and 
\begin{align}
       \log {\rm C\,\textsc{iii}]} / {\rm He\,\textsc{ii}}\lambda 1640
		&< - \log [{\rm EW}({\rm He\,\textsc{ii}}\lambda 1640) \times 0.3] 
		\label{eq:ewhe2_he2c3_zem4} \\
	\log {\rm C\,\textsc{iv}} / {\rm He\,\textsc{ii}}\lambda 1640
		&< - \log [{\rm EW}({\rm He\,\textsc{ii}}\lambda 1640) \times 0.01] 
		\label{eq:ewhe2_he2c4_zem4} \\
	\log {\rm O\,\textsc{iii}]} / {\rm He\,\textsc{ii}}\lambda 1640
		&< - \log [{\rm EW}({\rm He\,\textsc{ii}}\lambda 1640) \times 0.04] 
		\label{eq:ewhe2_he2o3_zem4} \\
	{\rm N\,\textsc{v}} / {\rm He\,\textsc{ii}}\lambda 1640
		&< 0.06 \label{eq:ewhe2_he2n5_zem4} 
\end{align}
for galaxies below $Z=10^{-4}$. Note that EW(\HeII) in the above equations 
should be in units of \AA\ in the rest frame.

\begin{figure*}
  \centering
    \begin{tabular}{c}
      \begin{minipage}{0.24\hsize}
        \begin{center}
         \includegraphics[bb=18 143 555 680, width=0.95\textwidth]{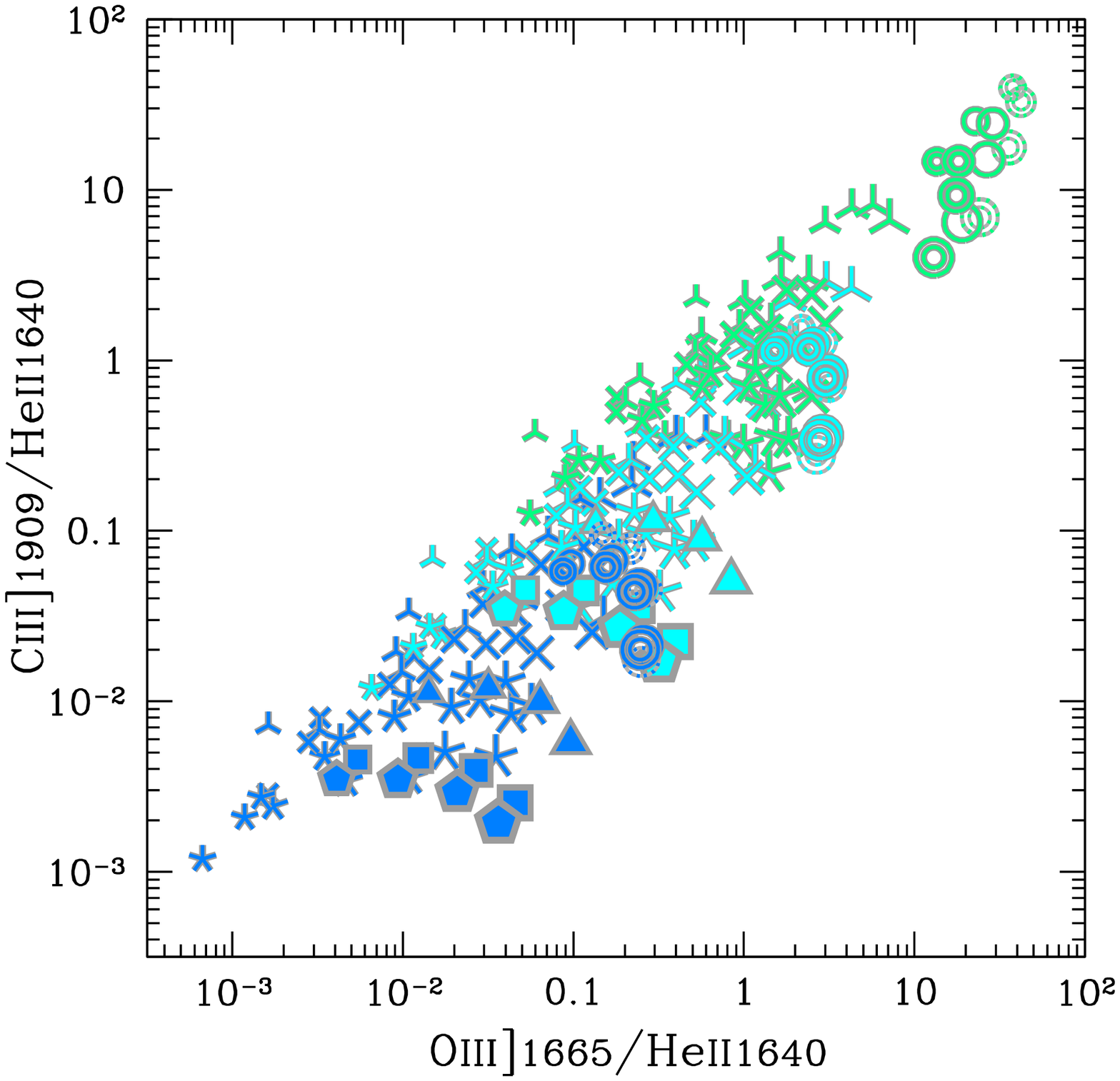}
        \end{center}
      \end{minipage}
      \begin{minipage}{0.24\hsize}
        \begin{center}
         \includegraphics[bb=18 143 555 680, width=0.95\textwidth]{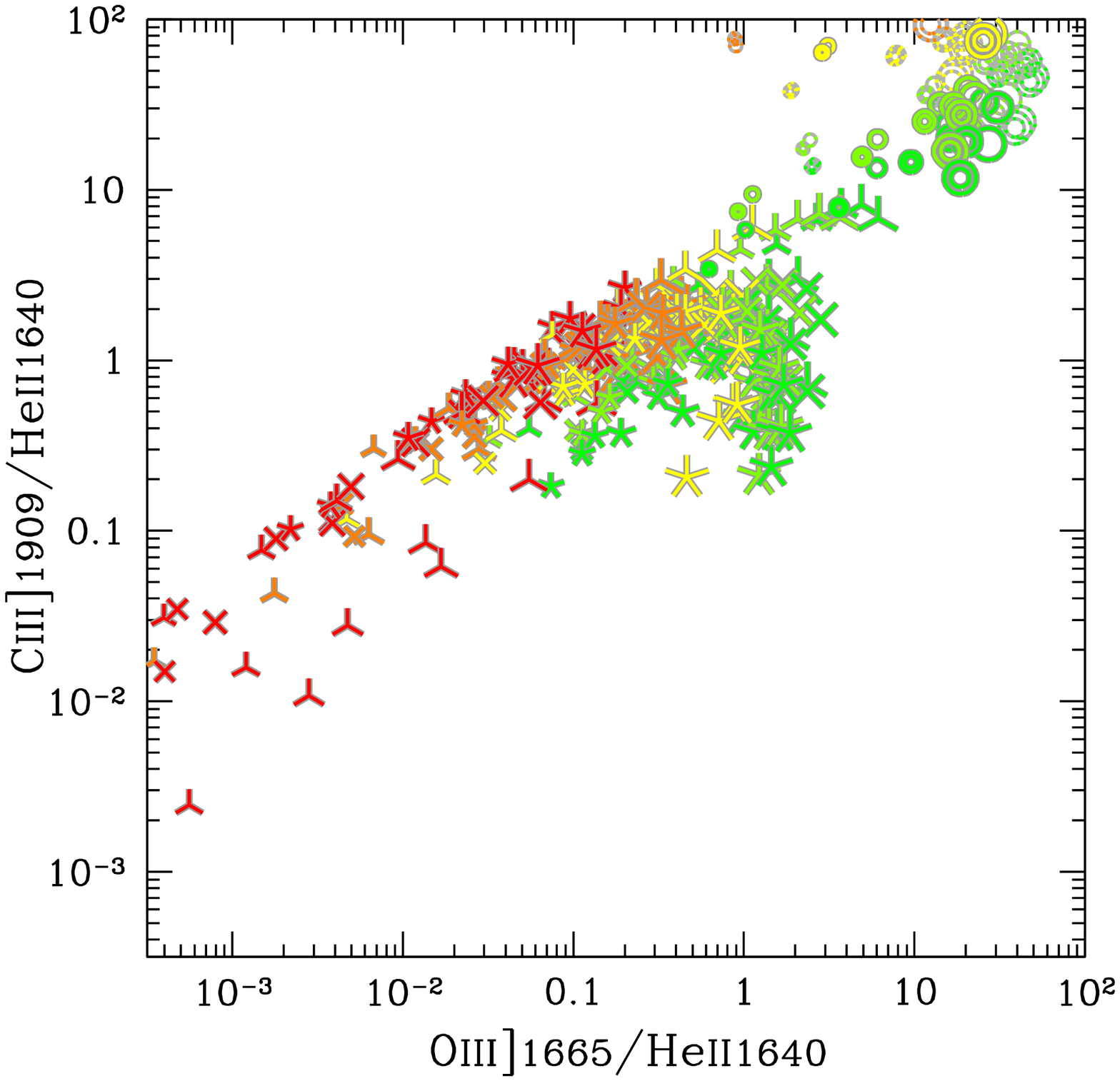}
        \end{center}
      \end{minipage}
      \hspace{0.02\hsize}
      \begin{minipage}{0.24\hsize}
        \begin{center}
         \includegraphics[bb=18 143 555 680, width=0.95\textwidth]{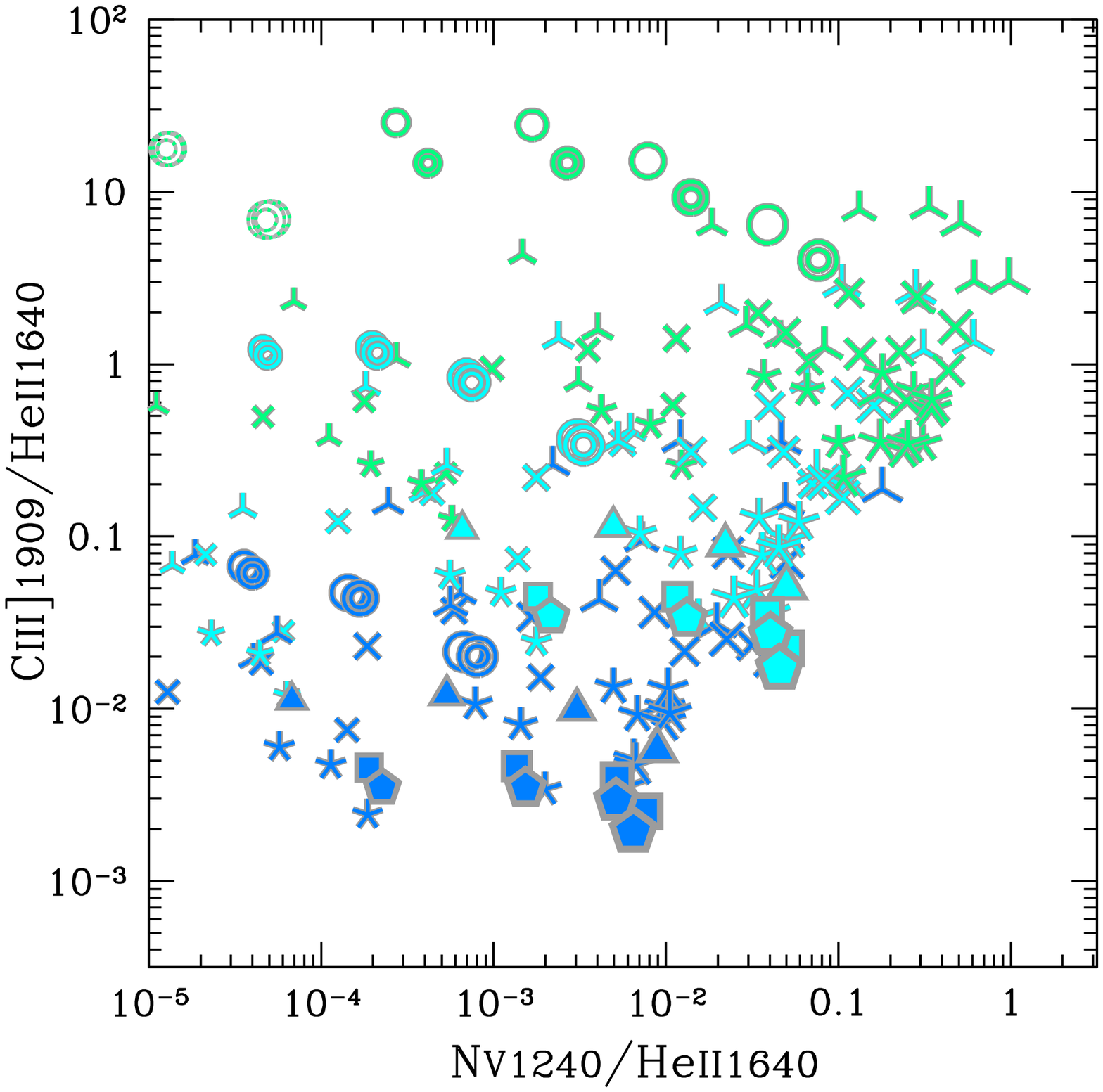}
        \end{center}
      \end{minipage}
      \begin{minipage}{0.24\hsize}
        \begin{center}
         \includegraphics[bb=18 143 555 680, width=0.95\textwidth]{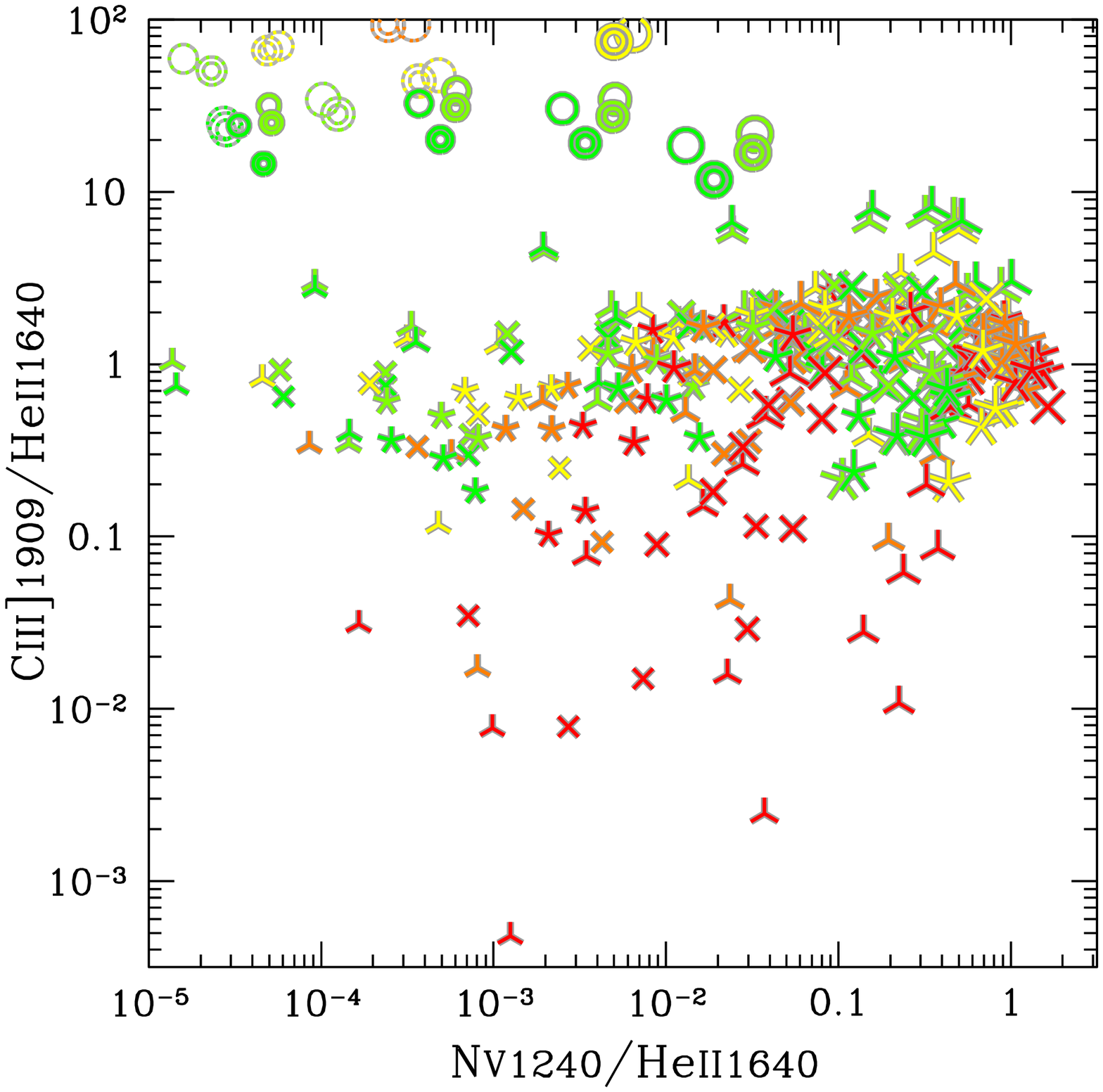}
        \end{center}
      \end{minipage}
    \end{tabular}
    \caption{%
    	Distribution of \cloudy\ models on the
	    \CIII$\lambda 1909$/\HeII$\lambda1640$ versus ~\OIIIuv$\lambda 1665$/\HeII$\lambda1640$ diagram
	    (left 2 panels) and on the
	    \CIII$\lambda 1909$/\HeII$\lambda1640$ versus ~\NV$\lambda 1240$/\HeII$\lambda1640$ diagram
	    (right 2 panels).
	    For each diagram, the primitive sources are displayed on the left-hand side,
	    and the chemically-evolved systems on the right.
	    Symbols as in Fig.~\ref{fig:ewhe2_he2hb}.
    }
    \label{fig:c3he2_o3he2_n5he2}
\end{figure*}

We highlight the very interesting result that the \CIV\ emission remains as strong as \HeII\ even if the gas metallicity 
falls below $Z\leq10^{-4}$, potentially at the level of about one tenth of HeII even at $Z=10^{-5}$, although the strength largely depends on the 
ionisation state of ISM as seen with different symbols sizes 
as the ionisation parameter varies.
Moreover, the line ratio would not always work for DCBHs, 
especially for those whose EW(\HeII) falls below $20$\,\AA, 
where metal-enriched AGNs even with a super-solar metallicity 
can be found to have a comparable \CIV$/$\HeII\ ratio.
A similar situation applies to the \OIIIuv\ and \NV\ diagrams.
The \CIII\ diagram appears to provide a more clean metallicity measurement 
for PopIII galaxies and DCBHs with smaller uncertainties associated with ionisation condition 
variations and contamination by metal-rich sources, 
although the \CIII\ line is generally predicted to be weak and its observation 
becomes challenging from the primitive sources.
\\

Finally, we note that
the C4C3-C34 diagram (Fig.~\ref{fig:c4c3_c34}) in \citet{nakajima2018_vuds} is partly 
used to separate AGNs from star-forming galaxies. 
We reaffirm that the separation method works for galaxies with a gas metallicity 
higher than $\rm Z \geq 0.01$\,\Zsun, and that sources dominated by PopIII stars and AGNs
are not distinguishable, as cautioned by \citet{nakajima2018_vuds} 
and illustrated in Fig.~\ref{fig:c4c3_c34} (left).

In Fig.~\ref{fig:c3he2_o3he2_n5he2}, we also examine the other popular UV diagnostics 
combining \CIII, \OIIIuv, \NV, and \HeII\ 
(\citealt{feltre2016,gutkin2016,laporte2017_agn,hirschmann2019,saxena2020}).
As seen on the the C4C3-C34 diagram, the PopIII models are not well discriminated 
from other classes of metal-poor DCBHs and/or AGNs, 
and, more generally, discriminating different classes of targets and metallicities is difficult  on these UV diagrams. Yet, it is confirmed that the \NV/\HeII\ ratio (right panels in Fig.~\ref{fig:c3he2_o3he2_n5he2}) is powerful in discriminating AGN and PopII galaxies in the metal rich regime.

As an additional caveat of these diagrams, one should take into account that they depend on the assumptions of metal abundance ratios 
such as C/O and N/O.
Although our adopted prescriptions of the metal abundances are known to be generally valid 
for local \HII\ regions and $z=2-3$ star-forming galaxies on average \citep[][]{hayden-pawson2022}, they are subject to significant 
 dispersion, which could be increased
by some systematic uncertainties at high-redshift.
It would be recommended to directly constrain the metal abundance ratios  
using appropriate sets of emission lines if all the necessary lines are available 
\citep[e.g.,][]{perez-montero2017,hayden-pawson2022} to correctly interpret observations.
Yet, our assumptions provide a reasonable first step to explore the behaviours
of these nebular emission lines, 
offering the diagnostics which will be useful even when limited sets of emission lines 
are observationally available.

\section{Discussion and Summary} \label{sec:discussion_summary}
We have presented \cloudy\ photoionisation models predictions of 
the rest-frame optical and UV emission line strengths for PopIII galaxies and DCBHs, 
and we have proposed several diagnostic diagrams to select and distinguish between 
these early sources and more-evolved systems.
The modelling reveals that it is possible to separate PopIII galaxies from 
other kind of primeval object, and in particular primeval black hole seeds 
(Direct Collapse Black Holes, DCBHs)
by utilising the \HeII\ emission lines, both in the optical and UV, 
in conjunction with the other lines probing a wider range of ionising spectral shapes.
Moreover, metal line diagrams allow the characterisation of the ISM properties, including 
gas metallicity and density in extreme conditions, including the exploration of 
scenarios in which PopIII or DCBH are embedded in slightly enriched ISM, 
and paving the way for an exploration of primitive sources 
(Z $\lesssim 10^{-5}$) in the early universe.

Among the most important diagnostics that we have identified, and interesting findings, 
we summarise the following:

\begin{itemize}

    \item The EW of the \HeII\ optical line at 4686\AA\ is the best discriminator of PopIII galaxies. A threshold of EW(\HeII$\lambda 4686$) $>$20\AA\ discriminates well PopIII from PopII and from DCBH. Along with the \HeII$/$\Hb\ ratio, the EW(\HeII$\lambda 4686$) can also discriminate among PopIII with different IMFs. The EW of the \HeII\ UV transition at 1640\AA\ has a similar good diagnostic power, although not as sharp as the optical line.
    
    \item DCBHs can be distinguished well from PopIII and PopII on the EW(\HeII$\lambda 4686$) vs.~\HeII$\lambda 4686$$/$\Hb\ diagram. The equivalent diagram in the UV, i.e. EW(\HeII$\lambda 1640$) vs.~\HeII$\lambda 1640$ $/$\Lya\ diagram, unfortunately does not have the same discriminatory power, although still useful to discriminate PopIII with top-heavy IMF.
    
    \item Very low metallicity systems, including PopIII and DCBH embedded in low metallicity ISM ($\rm Z\sim 10^{-5}-10^{-4}$), can be distinguished well from more evolved, moderate/high metallicity systems (PopII and AGN in metal-rich environments) through the optical diagrams \OIII$/$\Hb\ vs.~\NII$/$\Ha, or vs. \SII$/$\Ha, or vs. \OII$/$\Hb.
    In particular, candidate DCBH in a primeval ISM, identified through the previously suggested diagrams, can be distinguished well from more evolved AGN through the optical BPT diagrams. The trends of the UV lines are more complex, but we have identified tight sequences in the \CIV$\lambda 1549$ $/$\CIII$\lambda 1909$ vs.~(\CIII$\lambda 1909$+\CIV$\lambda 1549$)$/$\HeII$\lambda 1640$ diagram that can diagnose the ISM metallicities embedding PopIII and DCBHs.
    
    \item Interestingly, the \OIII$\lambda 5007$ optical transition remains strong even in very low metallicity regimes, by being as strong as H$\beta$ at $\rm Z\sim 10^{-4}$ and only $\sim$10 times fainter than H$\beta$ at $\rm Z\sim 10^{-5}$. Similarly, the \CIV$\lambda1549$ UV transition can be as strong as \HeII$\lambda 1640$ even at metallicities below $\rm Z\sim 10^{-4}$, and, potentially, only $\sim$10 times fainter than \HeII\ at $\rm Z\sim 10^{-5}$. Therefore, galaxies with \OIII$\lambda 5007$/H$\beta \leq$0.1 and \CIV$\lambda 1549$/\HeII$\lambda 1640$ $\leq$0.1 are good candidates for hosting PopIII or DCBH in a pristine or slightly enriched environment.
    
    \item Finally, the \HeI$\lambda 5876$/H$\beta$ ratio, together with the \HeII$\lambda 4686$/H$\beta$ ratio, is a good diagnostic of the gas electron density in the very low metallicity regime (where other density diagnostics associated with metal lines are not detectable).
    
\end{itemize}

While the focus of this work is to provide diagnostics for data that will be obtained in the early universe by forthcoming facilities (e.g. JWST and ELT), some of the currently available observations of local and high redshift galaxies can already be interpreted in the contest of these diagnostics. Such a comparison will be presented in a forthcoming, parallel paper.

\section*{Acknowledgements}

We thank the anonymous referee for helpful comments and discussions that improved our manuscript.
KN acknowledges support from JSPS KAKENHI Grant JP20K22373.
RM acknowledges ERC Advanced Grant 695671 QUENCH, and support from the UK Science and Technology Facilities Council (STFC). RM also acknowledges funding from a research professorship from the Royal Society.

\section*{Data Availability}

The data presented in this article will be shared 
on reasonable request to the corresponding author.

\bibliographystyle{mnras}{}
\bibliography{Refs_paper.bib}{}

\begin{thebibliography}{}
\makeatletter
\relax
\def\mn@urlcharsother{\let\do\@makeother \do\$\do\&\do\#\do\^\do\_\do\%\do\~}
\def\mn@doi{\begingroup\mn@urlcharsother \@ifnextchar [ {\mn@doi@}
  {\mn@doi@[]}}
\def\mn@doi@[#1]#2{\def\@tempa{#1}\ifx\@tempa\@empty \href
  {http://dx.doi.org/#2} {doi:#2}\else \href {http://dx.doi.org/#2} {#1}\fi
  \endgroup}
\def\mn@eprint#1#2{\mn@eprint@#1:#2::\@nil}
\def\mn@eprint@arXiv#1{\href {http://arxiv.org/abs/#1} {{\tt arXiv:#1}}}
\def\mn@eprint@dblp#1{\href {http://dblp.uni-trier.de/rec/bibtex/#1.xml}
  {dblp:#1}}
\def\mn@eprint@#1:#2:#3:#4\@nil{\def\@tempa {#1}\def\@tempb {#2}\def\@tempc
  {#3}\ifx \@tempc \@empty \let \@tempc \@tempb \let \@tempb \@tempa \fi \ifx
  \@tempb \@empty \def\@tempb {arXiv}\fi \@ifundefined
  {mn@eprint@\@tempb}{\@tempb:\@tempc}{\expandafter \expandafter \csname
  mn@eprint@\@tempb\endcsname \expandafter{\@tempc}}}

\bibitem[\protect\citeauthoryear{{Abel}, {Bryan}  \& {Norman}}{{Abel}
  et~al.}{2002}]{abel2002}
{Abel} T.,  {Bryan} G.~L.,   {Norman} M.~L.,  2002, \mn@doi [Science]
  {10.1126/science.295.5552.93}, \href
  {https://ui.adsabs.harvard.edu/abs/2002Sci...295...93A} {295, 93}

\bibitem[\protect\citeauthoryear{{Amor{\'\i}n} et~al.,}{{Amor{\'\i}n}
  et~al.}{2017}]{amorin2017}
{Amor{\'\i}n} R.,  et~al., 2017, \mn@doi [Nature Astronomy]
  {10.1038/s41550-017-0052}, \href
  {https://ui.adsabs.harvard.edu/abs/2017NatAs...1E..52A} {1, 0052}

\bibitem[\protect\citeauthoryear{{Andrews} \& {Martini}}{{Andrews} \&
  {Martini}}{2013}]{AM2013}
{Andrews} B.~H.,  {Martini} P.,  2013, \mn@doi [\apj]
  {10.1088/0004-637X/765/2/140}, \href
  {https://ui.adsabs.harvard.edu/abs/2013ApJ...765..140A} {765, 140}

\bibitem[\protect\citeauthoryear{{Asplund}, {Grevesse}, {Sauval}  \&
  {Scott}}{{Asplund} et~al.}{2009}]{asplund2009}
{Asplund} M.,  {Grevesse} N.,  {Sauval} A.~J.,   {Scott} P.,  2009, \mn@doi
  [\araa] {10.1146/annurev.astro.46.060407.145222}, \href
  {https://ui.adsabs.harvard.edu/abs/2009ARA&A..47..481A} {47, 481}

\bibitem[\protect\citeauthoryear{{Aver}, {Olive}  \& {Skillman}}{{Aver}
  et~al.}{2015}]{aver2015}
{Aver} E.,  {Olive} K.~A.,   {Skillman} E.~D.,  2015, \mn@doi [\jcap]
  {10.1088/1475-7516/2015/07/011}, \href
  {https://ui.adsabs.harvard.edu/abs/2015JCAP...07..011A} {2015, 011}

\bibitem[\protect\citeauthoryear{{Baldwin}, {Phillips}  \&
  {Terlevich}}{{Baldwin} et~al.}{1981}]{BPT1981}
{Baldwin} J.~A.,  {Phillips} M.~M.,   {Terlevich} R.,  1981, \mn@doi [\pasp]
  {10.1086/130766}, \href
  {https://ui.adsabs.harvard.edu/abs/1981PASP...93....5B} {93, 5}

\bibitem[\protect\citeauthoryear{{Barrow}, {Aykutalp}  \& {Wise}}{{Barrow}
  et~al.}{2018}]{barrow2018}
{Barrow} K. S.~S.,  {Aykutalp} A.,   {Wise} J.~H.,  2018, \mn@doi [Nature
  Astronomy] {10.1038/s41550-018-0569-y}, \href
  {https://ui.adsabs.harvard.edu/abs/2018NatAs...2..987B} {2, 987}

\bibitem[\protect\citeauthoryear{{Beckmann} et~al.,}{{Beckmann}
  et~al.}{2019}]{beckmann2019}
{Beckmann} R.~S.,  et~al., 2019, \mn@doi [\aap] {10.1051/0004-6361/201936188},
  \href {https://ui.adsabs.harvard.edu/abs/2019A&A...631A..60B} {631, A60}

\bibitem[\protect\citeauthoryear{{Bowler}, {McLure}, {Dunlop}, {McLeod},
  {Stanway}, {Eldridge}  \& {Jarvis}}{{Bowler} et~al.}{2017}]{bowler2017}
{Bowler} R.~A.~A.,  {McLure} R.~J.,  {Dunlop} J.~S.,  {McLeod} D.~J.,
  {Stanway} E.~R.,  {Eldridge} J.~J.,   {Jarvis} M.~J.,  2017, \mn@doi [\mnras]
  {10.1093/mnras/stx839}, \href
  {https://ui.adsabs.harvard.edu/abs/2017MNRAS.469..448B} {469, 448}

\bibitem[\protect\citeauthoryear{{Bromm}, {Coppi}  \& {Larson}}{{Bromm}
  et~al.}{2002}]{bromm2002}
{Bromm} V.,  {Coppi} P.~S.,   {Larson} R.~B.,  2002, \mn@doi [\apj]
  {10.1086/323947}, \href
  {https://ui.adsabs.harvard.edu/abs/2002ApJ...564...23B} {564, 23}

\bibitem[\protect\citeauthoryear{{Chevallard} et~al.,}{{Chevallard}
  et~al.}{2018}]{chevallard2018}
{Chevallard} J.,  et~al., 2018, \mn@doi [\mnras] {10.1093/mnras/sty1461}, \href
  {https://ui.adsabs.harvard.edu/abs/2018MNRAS.479.3264C} {479, 3264}

\bibitem[\protect\citeauthoryear{{Chon}, {Omukai}  \& {Schneider}}{{Chon}
  et~al.}{2021}]{chon2021}
{Chon} S.,  {Omukai} K.,   {Schneider} R.,  2021, \mn@doi [\mnras]
  {10.1093/mnras/stab2497}, \href
  {https://ui.adsabs.harvard.edu/abs/2021MNRAS.508.4175C} {508, 4175}

\bibitem[\protect\citeauthoryear{{Cullen} et~al.,}{{Cullen}
  et~al.}{2019}]{cullen2019}
{Cullen} F.,  et~al., 2019, \mn@doi [\mnras] {10.1093/mnras/stz1402}, \href
  {https://ui.adsabs.harvard.edu/abs/2019MNRAS.487.2038C} {487, 2038}

\bibitem[\protect\citeauthoryear{{Curti}, {Cresci}, {Mannucci}, {Marconi},
  {Maiolino}  \& {Esposito}}{{Curti} et~al.}{2017}]{curti2017}
{Curti} M.,  {Cresci} G.,  {Mannucci} F.,  {Marconi} A.,  {Maiolino} R.,
  {Esposito} S.,  2017, \mn@doi [\mnras] {10.1093/mnras/stw2766}, \href
  {https://ui.adsabs.harvard.edu/abs/2017MNRAS.465.1384C} {465, 1384}

\bibitem[\protect\citeauthoryear{{Curti} et~al.,}{{Curti}
  et~al.}{2021}]{curti2021}
{Curti} M.,  et~al., 2021, arXiv e-prints, \href
  {https://ui.adsabs.harvard.edu/abs/2021arXiv211011841C} {p. arXiv:2110.11841}

\bibitem[\protect\citeauthoryear{{Dopita} et~al.,}{{Dopita}
  et~al.}{2006}]{dopita2006}
{Dopita} M.~A.,  et~al., 2006, \mn@doi [\apjs] {10.1086/508261}, \href
  {https://ui.adsabs.harvard.edu/abs/2006ApJS..167..177D} {167, 177}

\bibitem[\protect\citeauthoryear{{Dors}, {Cardaci}, {H{\"a}gele}  \&
  {Krabbe}}{{Dors} et~al.}{2014}]{dors2014}
{Dors} O.~L.,  {Cardaci} M.~V.,  {H{\"a}gele} G.~F.,   {Krabbe} {\^A}.~C.,
  2014, \mn@doi [\mnras] {10.1093/mnras/stu1218}, \href
  {https://ui.adsabs.harvard.edu/abs/2014MNRAS.443.1291D} {443, 1291}

\bibitem[\protect\citeauthoryear{{Eldridge}, {Stanway}, {Xiao}, {McClelland },
  {Taylor}, {Ng}, {Greis}  \& {Bray}}{{Eldridge} et~al.}{2017}]{eldridge2017}
{Eldridge} J.~J.,  {Stanway} E.~R.,  {Xiao} L.,  {McClelland } L.~A.~S.,
  {Taylor} G.,  {Ng} M.,  {Greis} S.~M.~L.,   {Bray} J.~C.,  2017, \mn@doi
  [\pasa] {10.1017/pasa.2017.51}, \href
  {https://ui.adsabs.harvard.edu/abs/2017PASA...34...58E} {34, e058}

\bibitem[\protect\citeauthoryear{{Elvis}, {Risaliti}  \& {Zamorani}}{{Elvis}
  et~al.}{2002}]{elvis2002}
{Elvis} M.,  {Risaliti} G.,   {Zamorani} G.,  2002, \mn@doi [\apjl]
  {10.1086/339197}, \href
  {https://ui.adsabs.harvard.edu/abs/2002ApJ...565L..75E} {565, L75}

\bibitem[\protect\citeauthoryear{{Feltre}, {Charlot}  \& {Gutkin}}{{Feltre}
  et~al.}{2016}]{feltre2016}
{Feltre} A.,  {Charlot} S.,   {Gutkin} J.,  2016, \mn@doi [\mnras]
  {10.1093/mnras/stv2794}, \href
  {https://ui.adsabs.harvard.edu/abs/2016MNRAS.456.3354F} {456, 3354}

\bibitem[\protect\citeauthoryear{{Ferland}, {Korista}, {Verner}, {Ferguson},
  {Kingdon}  \& {Verner}}{{Ferland} et~al.}{1998}]{ferland1998}
{Ferland} G.~J.,  {Korista} K.~T.,  {Verner} D.~A.,  {Ferguson} J.~W.,
  {Kingdon} J.~B.,   {Verner} E.~M.,  1998, \mn@doi [\pasp] {10.1086/316190},
  \href {https://ui.adsabs.harvard.edu/abs/1998PASP..110..761F} {110, 761}

\bibitem[\protect\citeauthoryear{{Ferland} et~al.,}{{Ferland}
  et~al.}{2013}]{ferland2013}
{Ferland} G.~J.,  et~al., 2013, \rmxaa, \href
  {https://ui.adsabs.harvard.edu/abs/2013RMxAA..49..137F} {49, 137}

\bibitem[\protect\citeauthoryear{{Ferrara}, {Salvadori}, {Yue}  \&
  {Schleicher}}{{Ferrara} et~al.}{2014}]{ferrara2014}
{Ferrara} A.,  {Salvadori} S.,  {Yue} B.,   {Schleicher} D.,  2014, \mn@doi
  [\mnras] {10.1093/mnras/stu1280}, \href
  {https://ui.adsabs.harvard.edu/abs/2014MNRAS.443.2410F} {443, 2410}

\bibitem[\protect\citeauthoryear{{Groves}, {Heckman}  \& {Kauffmann}}{{Groves}
  et~al.}{2006}]{groves2006}
{Groves} B.~A.,  {Heckman} T.~M.,   {Kauffmann} G.,  2006, \mn@doi [\mnras]
  {10.1111/j.1365-2966.2006.10812.x}, \href
  {https://ui.adsabs.harvard.edu/abs/2006MNRAS.371.1559G} {371, 1559}

\bibitem[\protect\citeauthoryear{{Gutkin}, {Charlot}  \& {Bruzual}}{{Gutkin}
  et~al.}{2016}]{gutkin2016}
{Gutkin} J.,  {Charlot} S.,   {Bruzual} G.,  2016, \mn@doi [\mnras]
  {10.1093/mnras/stw1716}, \href
  {https://ui.adsabs.harvard.edu/abs/2016MNRAS.462.1757G} {462, 1757}

\bibitem[\protect\citeauthoryear{{Habouzit}, {Volonteri}, {Somerville},
  {Dubois}, {Peirani}, {Pichon}  \& {Devriendt}}{{Habouzit}
  et~al.}{2019}]{habouzit2019}
{Habouzit} M.,  {Volonteri} M.,  {Somerville} R.~S.,  {Dubois} Y.,  {Peirani}
  S.,  {Pichon} C.,   {Devriendt} J.,  2019, \mn@doi [\mnras]
  {10.1093/mnras/stz2105}, \href
  {https://ui.adsabs.harvard.edu/abs/2019MNRAS.489.1206H} {489, 1206}

\bibitem[\protect\citeauthoryear{{Harikane}, {Laporte}, {Ellis}  \&
  {Matsuoka}}{{Harikane} et~al.}{2020}]{harikane2020_abs}
{Harikane} Y.,  {Laporte} N.,  {Ellis} R.~S.,   {Matsuoka} Y.,  2020, \mn@doi
  [\apj] {10.3847/1538-4357/abb597}, \href
  {https://ui.adsabs.harvard.edu/abs/2020ApJ...902..117H} {902, 117}

\bibitem[\protect\citeauthoryear{{Hartwig} et~al.,}{{Hartwig}
  et~al.}{2016}]{hartwig2016}
{Hartwig} T.,  et~al., 2016, \mn@doi [\mnras] {10.1093/mnras/stw1775}, \href
  {https://ui.adsabs.harvard.edu/abs/2016MNRAS.462.2184H} {462, 2184}

\bibitem[\protect\citeauthoryear{{Hayden-Pawson} et~al.,}{{Hayden-Pawson}
  et~al.}{2021}]{hayden-pawson2022}
{Hayden-Pawson} C.,  et~al., 2021, arXiv e-prints, \href
  {https://ui.adsabs.harvard.edu/abs/2021arXiv211000033H} {p. arXiv:2110.00033}

\bibitem[\protect\citeauthoryear{{Hirano}, {Hosokawa}, {Yoshida}, {Omukai}  \&
  {Yorke}}{{Hirano} et~al.}{2015}]{hirano2015}
{Hirano} S.,  {Hosokawa} T.,  {Yoshida} N.,  {Omukai} K.,   {Yorke} H.~W.,
  2015, \mn@doi [\mnras] {10.1093/mnras/stv044}, \href
  {https://ui.adsabs.harvard.edu/abs/2015MNRAS.448..568H} {448, 568}

\bibitem[\protect\citeauthoryear{{Hirschmann}, {Charlot}, {Feltre}, {Naab},
  {Somerville}  \& {Choi}}{{Hirschmann} et~al.}{2019}]{hirschmann2019}
{Hirschmann} M.,  {Charlot} S.,  {Feltre} A.,  {Naab} T.,  {Somerville} R.~S.,
   {Choi} E.,  2019, \mn@doi [\mnras] {10.1093/mnras/stz1256}, \href
  {https://ui.adsabs.harvard.edu/abs/2019MNRAS.487..333H} {487, 333}

\bibitem[\protect\citeauthoryear{{Hsyu}, {Cooke}, {Prochaska}  \&
  {Bolte}}{{Hsyu} et~al.}{2020}]{hsyu2020}
{Hsyu} T.,  {Cooke} R.~J.,  {Prochaska} J.~X.,   {Bolte} M.,  2020, \mn@doi
  [\apj] {10.3847/1538-4357/ab91af}, \href
  {https://ui.adsabs.harvard.edu/abs/2020ApJ...896...77H} {896, 77}

\bibitem[\protect\citeauthoryear{{Inayoshi}, {Li}  \& {Haiman}}{{Inayoshi}
  et~al.}{2018}]{inayoshi2018}
{Inayoshi} K.,  {Li} M.,   {Haiman} Z.,  2018, \mn@doi [\mnras]
  {10.1093/mnras/sty1720}, \href
  {https://ui.adsabs.harvard.edu/abs/2018MNRAS.479.4017I} {479, 4017}

\bibitem[\protect\citeauthoryear{{Inayoshi}, {Visbal}  \& {Haiman}}{{Inayoshi}
  et~al.}{2020}]{inayoshi2020}
{Inayoshi} K.,  {Visbal} E.,   {Haiman} Z.,  2020, \mn@doi [\araa]
  {10.1146/annurev-astro-120419-014455}, \href
  {https://ui.adsabs.harvard.edu/abs/2020ARA&A..58...27I} {58, 27}

\bibitem[\protect\citeauthoryear{{Inoue}}{{Inoue}}{2011}]{inoue2011_metal_poor}
{Inoue} A.~K.,  2011, \mn@doi [\mnras] {10.1111/j.1365-2966.2011.18906.x},
  \href {https://ui.adsabs.harvard.edu/abs/2011MNRAS.415.2920I} {415, 2920}

\bibitem[\protect\citeauthoryear{{Izotov}, {Thuan}  \& {Guseva}}{{Izotov}
  et~al.}{2014}]{izotov2014}
{Izotov} Y.~I.,  {Thuan} T.~X.,   {Guseva} N.~G.,  2014, \mn@doi [\mnras]
  {10.1093/mnras/stu1771}, \href
  {https://ui.adsabs.harvard.edu/abs/2014MNRAS.445..778I} {445, 778}

\bibitem[\protect\citeauthoryear{{Jaacks}, {Thompson}, {Finkelstein}  \&
  {Bromm}}{{Jaacks} et~al.}{2018}]{jaacks2018}
{Jaacks} J.,  {Thompson} R.,  {Finkelstein} S.~L.,   {Bromm} V.,  2018, \mn@doi
  [\mnras] {10.1093/mnras/sty062}, \href
  {https://ui.adsabs.harvard.edu/abs/2018MNRAS.475.4396J} {475, 4396}

\bibitem[\protect\citeauthoryear{{Jaskot} \& {Ravindranath}}{{Jaskot} \&
  {Ravindranath}}{2016}]{JR2016}
{Jaskot} A.~E.,  {Ravindranath} S.,  2016, \mn@doi [\apj]
  {10.3847/1538-4357/833/2/136}, \href
  {https://ui.adsabs.harvard.edu/abs/2016ApJ...833..136J} {833, 136}

\bibitem[\protect\citeauthoryear{{Jeon}, {Bromm}, {Pawlik}  \&
  {Milosavljevi{\'c}}}{{Jeon} et~al.}{2015}]{jeon2015}
{Jeon} M.,  {Bromm} V.,  {Pawlik} A.~H.,   {Milosavljevi{\'c}} M.,  2015,
  \mn@doi [\mnras] {10.1093/mnras/stv1353}, \href
  {https://ui.adsabs.harvard.edu/abs/2015MNRAS.452.1152J} {452, 1152}

\bibitem[\protect\citeauthoryear{{Jiang} et~al.,}{{Jiang}
  et~al.}{2021}]{jiang2021}
{Jiang} L.,  et~al., 2021, \mn@doi [Nature Astronomy]
  {10.1038/s41550-020-01275-y}, \href
  {https://ui.adsabs.harvard.edu/abs/2021NatAs...5..256J} {5, 256}

\bibitem[\protect\citeauthoryear{{Kauffmann} et~al.,}{{Kauffmann}
  et~al.}{2003}]{kauffmann2003}
{Kauffmann} G.,  et~al., 2003, \mn@doi [\mnras]
  {10.1111/j.1365-2966.2003.07154.x}, \href
  {https://ui.adsabs.harvard.edu/abs/2003MNRAS.346.1055K} {346, 1055}

\bibitem[\protect\citeauthoryear{{Kewley} \& {Dopita}}{{Kewley} \&
  {Dopita}}{2002}]{KD2002}
{Kewley} L.~J.,  {Dopita} M.~A.,  2002, \mn@doi [\apjs] {10.1086/341326}, \href
  {https://ui.adsabs.harvard.edu/abs/2002ApJS..142...35K} {142, 35}

\bibitem[\protect\citeauthoryear{{Kewley}, {Dopita}, {Sutherland}, {Heisler}
  \& {Trevena}}{{Kewley} et~al.}{2001}]{kewley2001}
{Kewley} L.~J.,  {Dopita} M.~A.,  {Sutherland} R.~S.,  {Heisler} C.~A.,
  {Trevena} J.,  2001, \mn@doi [\apj] {10.1086/321545}, \href
  {https://ui.adsabs.harvard.edu/abs/2001ApJ...556..121K} {556, 121}

\bibitem[\protect\citeauthoryear{{Kewley}, {Dopita}, {Leitherer}, {Dav{\'e}},
  {Yuan}, {Allen}, {Groves}  \& {Sutherland}}{{Kewley}
  et~al.}{2013}]{kewley2013_theory}
{Kewley} L.~J.,  {Dopita} M.~A.,  {Leitherer} C.,  {Dav{\'e}} R.,  {Yuan} T.,
  {Allen} M.,  {Groves} B.,   {Sutherland} R.,  2013, \mn@doi [\apj]
  {10.1088/0004-637X/774/2/100}, \href
  {https://ui.adsabs.harvard.edu/abs/2013ApJ...774..100K} {774, 100}

\bibitem[\protect\citeauthoryear{{Kroupa}}{{Kroupa}}{2001}]{kroupa2001}
{Kroupa} P.,  2001, \mn@doi [\mnras] {10.1046/j.1365-8711.2001.04022.x}, \href
  {https://ui.adsabs.harvard.edu/abs/2001MNRAS.322..231K} {322, 231}

\bibitem[\protect\citeauthoryear{{Lamareille}}{{Lamareille}}{2010}]{lamareille2010}
{Lamareille} F.,  2010, \mn@doi [\aap] {10.1051/0004-6361/200913168}, \href
  {https://ui.adsabs.harvard.edu/abs/2010A&A...509A..53L} {509, A53}

\bibitem[\protect\citeauthoryear{{Laporte}, {Nakajima}, {Ellis}, {Zitrin},
  {Stark}, {Mainali}  \& {Roberts-Borsani}}{{Laporte}
  et~al.}{2017}]{laporte2017_agn}
{Laporte} N.,  {Nakajima} K.,  {Ellis} R.~S.,  {Zitrin} A.,  {Stark} D.~P.,
  {Mainali} R.,   {Roberts-Borsani} G.~W.,  2017, \mn@doi [\apj]
  {10.3847/1538-4357/aa96a8}, \href
  {https://ui.adsabs.harvard.edu/abs/2017ApJ...851...40L} {851, 40}

\bibitem[\protect\citeauthoryear{{Liu} \& {Bromm}}{{Liu} \&
  {Bromm}}{2020}]{liu2020}
{Liu} B.,  {Bromm} V.,  2020, \mn@doi [\mnras] {10.1093/mnras/staa2143}, \href
  {https://ui.adsabs.harvard.edu/abs/2020MNRAS.497.2839L} {497, 2839}

\bibitem[\protect\citeauthoryear{{L{\'o}pez-S{\'a}nchez}, {Dopita}, {Kewley},
  {Zahid}, {Nicholls}  \& {Scharw{\"a}chter}}{{L{\'o}pez-S{\'a}nchez}
  et~al.}{2012}]{lopez-sanchez2012}
{L{\'o}pez-S{\'a}nchez} {\'A}.~R.,  {Dopita} M.~A.,  {Kewley} L.~J.,  {Zahid}
  H.~J.,  {Nicholls} D.~C.,   {Scharw{\"a}chter} J.,  2012, \mn@doi [\mnras]
  {10.1111/j.1365-2966.2012.21145.x}, \href
  {https://ui.adsabs.harvard.edu/abs/2012MNRAS.426.2630L} {426, 2630}

\bibitem[\protect\citeauthoryear{{Maiolino} \& {Mannucci}}{{Maiolino} \&
  {Mannucci}}{2019}]{MM2019}
{Maiolino} R.,  {Mannucci} F.,  2019, \mn@doi [\aapr]
  {10.1007/s00159-018-0112-2}, \href
  {https://ui.adsabs.harvard.edu/abs/2019A&ARv..27....3M} {27, 3}

\bibitem[\protect\citeauthoryear{{Maiolino} et~al.,}{{Maiolino}
  et~al.}{2008}]{maiolino2008}
{Maiolino} R.,  et~al., 2008, \mn@doi [\aap] {10.1051/0004-6361:200809678},
  \href {https://ui.adsabs.harvard.edu/abs/2008A&A...488..463M} {488, 463}

\bibitem[\protect\citeauthoryear{{Matsumoto} et~al.,}{{Matsumoto}
  et~al.}{2022}]{matsumoto2022}
{Matsumoto} A.,  et~al., 2022, arXiv e-prints, \href
  {https://ui.adsabs.harvard.edu/abs/2022arXiv220309617M} {p. arXiv:2203.09617}

\bibitem[\protect\citeauthoryear{{Mezcua} \& {Dom{\'\i}nguez
  S{\'a}nchez}}{{Mezcua} \& {Dom{\'\i}nguez S{\'a}nchez}}{2020}]{mezcua2020}
{Mezcua} M.,  {Dom{\'\i}nguez S{\'a}nchez} H.,  2020, \mn@doi [\apjl]
  {10.3847/2041-8213/aba199}, \href
  {https://ui.adsabs.harvard.edu/abs/2020ApJ...898L..30M} {898, L30}

\bibitem[\protect\citeauthoryear{{Mezcua}, {Civano}, {Fabbiano}, {Miyaji}  \&
  {Marchesi}}{{Mezcua} et~al.}{2016}]{mezcua2016}
{Mezcua} M.,  {Civano} F.,  {Fabbiano} G.,  {Miyaji} T.,   {Marchesi} S.,
  2016, \mn@doi [\apj] {10.3847/0004-637X/817/1/20}, \href
  {https://ui.adsabs.harvard.edu/abs/2016ApJ...817...20M} {817, 20}

\bibitem[\protect\citeauthoryear{{Nagao}, {Marconi}  \& {Maiolino}}{{Nagao}
  et~al.}{2006}]{nagao2006_agn}
{Nagao} T.,  {Marconi} A.,   {Maiolino} R.,  2006, \mn@doi [\aap]
  {10.1051/0004-6361:20054024}, \href
  {https://ui.adsabs.harvard.edu/abs/2006A&A...447..157N} {447, 157}

\bibitem[\protect\citeauthoryear{{Nakajima} et~al.,}{{Nakajima}
  et~al.}{2018}]{nakajima2018_vuds}
{Nakajima} K.,  et~al., 2018, \mn@doi [\aap] {10.1051/0004-6361/201731935},
  \href {https://ui.adsabs.harvard.edu/abs/2018A&A...612A..94N} {612, A94}

\bibitem[\protect\citeauthoryear{{Natarajan}, {Pacucci}, {Ferrara}, {Agarwal},
  {Ricarte}, {Zackrisson}  \& {Cappelluti}}{{Natarajan}
  et~al.}{2017}]{natarajan2017}
{Natarajan} P.,  {Pacucci} F.,  {Ferrara} A.,  {Agarwal} B.,  {Ricarte} A.,
  {Zackrisson} E.,   {Cappelluti} N.,  2017, \mn@doi [\apj]
  {10.3847/1538-4357/aa6330}, \href
  {https://ui.adsabs.harvard.edu/abs/2017ApJ...838..117N} {838, 117}

\bibitem[\protect\citeauthoryear{{Oesch} et~al.,}{{Oesch}
  et~al.}{2016}]{oesch2016}
{Oesch} P.~A.,  et~al., 2016, \mn@doi [\apj] {10.3847/0004-637X/819/2/129},
  \href {https://ui.adsabs.harvard.edu/abs/2016ApJ...819..129O} {819, 129}

\bibitem[\protect\citeauthoryear{{Pacucci}, {Ferrara}, {Grazian}, {Fiore},
  {Giallongo}  \& {Puccetti}}{{Pacucci} et~al.}{2016}]{pacucci2016}
{Pacucci} F.,  {Ferrara} A.,  {Grazian} A.,  {Fiore} F.,  {Giallongo} E.,
  {Puccetti} S.,  2016, \mn@doi [\mnras] {10.1093/mnras/stw725}, \href
  {https://ui.adsabs.harvard.edu/abs/2016MNRAS.459.1432P} {459, 1432}

\bibitem[\protect\citeauthoryear{{Pacucci}, {Pallottini}, {Ferrara}  \&
  {Gallerani}}{{Pacucci} et~al.}{2017}]{pacucci2017}
{Pacucci} F.,  {Pallottini} A.,  {Ferrara} A.,   {Gallerani} S.,  2017, \mn@doi
  [\mnras] {10.1093/mnrasl/slx029}, \href
  {https://ui.adsabs.harvard.edu/abs/2017MNRAS.468L..77P} {468, L77}

\bibitem[\protect\citeauthoryear{{Pacucci}, {Mezcua}  \& {Regan}}{{Pacucci}
  et~al.}{2021}]{pacucci2021}
{Pacucci} F.,  {Mezcua} M.,   {Regan} J.~A.,  2021, \mn@doi [\apj]
  {10.3847/1538-4357/ac1595}, \href
  {https://ui.adsabs.harvard.edu/abs/2021ApJ...920..134P} {920, 134}

\bibitem[\protect\citeauthoryear{{Pawlik}, {Milosavljevi{\'c}}  \&
  {Bromm}}{{Pawlik} et~al.}{2011}]{pawlik2011}
{Pawlik} A.~H.,  {Milosavljevi{\'c}} M.,   {Bromm} V.,  2011, \mn@doi [\apj]
  {10.1088/0004-637X/731/1/54}, \href
  {https://ui.adsabs.harvard.edu/abs/2011ApJ...731...54P} {731, 54}

\bibitem[\protect\citeauthoryear{{P{\'e}rez-Montero} \&
  {Amor{\'\i}n}}{{P{\'e}rez-Montero} \&
  {Amor{\'\i}n}}{2017}]{perez-montero2017}
{P{\'e}rez-Montero} E.,  {Amor{\'\i}n} R.,  2017, \mn@doi [\mnras]
  {10.1093/mnras/stx186}, \href
  {https://ui.adsabs.harvard.edu/abs/2017MNRAS.467.1287P} {467, 1287}

\bibitem[\protect\citeauthoryear{{Raiter}, {Schaerer}  \& {Fosbury}}{{Raiter}
  et~al.}{2010}]{raiter2010}
{Raiter} A.,  {Schaerer} D.,   {Fosbury} R.~A.~E.,  2010, \mn@doi [\aap]
  {10.1051/0004-6361/201015236}, \href
  {https://ui.adsabs.harvard.edu/abs/2010A&A...523A..64R} {523, A64}

\bibitem[\protect\citeauthoryear{{Salpeter}}{{Salpeter}}{1955}]{salpeter1955}
{Salpeter} E.~E.,  1955, \mn@doi [\apj] {10.1086/145971}, \href
  {https://ui.adsabs.harvard.edu/abs/1955ApJ...121..161S} {121, 161}

\bibitem[\protect\citeauthoryear{{Sanders} et~al.,}{{Sanders}
  et~al.}{2016}]{sanders2016}
{Sanders} R.~L.,  et~al., 2016, \mn@doi [\apj] {10.3847/0004-637X/816/1/23},
  \href {https://ui.adsabs.harvard.edu/abs/2016ApJ...816...23S} {816, 23}

\bibitem[\protect\citeauthoryear{{Sanders} et~al.,}{{Sanders}
  et~al.}{2020}]{sanders2020}
{Sanders} R.~L.,  et~al., 2020, \mn@doi [\mnras] {10.1093/mnras/stz3032}, \href
  {https://ui.adsabs.harvard.edu/abs/2020MNRAS.491.1427S} {491, 1427}

\bibitem[\protect\citeauthoryear{{Saxena} et~al.,}{{Saxena}
  et~al.}{2020}]{saxena2020}
{Saxena} A.,  et~al., 2020, \mn@doi [\aap] {10.1051/0004-6361/201937170}, \href
  {https://ui.adsabs.harvard.edu/abs/2020A&A...636A..47S} {636, A47}

\bibitem[\protect\citeauthoryear{{Schaerer}}{{Schaerer}}{2003}]{schaerer2003}
{Schaerer} D.,  2003, \mn@doi [\aap] {10.1051/0004-6361:20021525}, \href
  {https://ui.adsabs.harvard.edu/abs/2003A&A...397..527S} {397, 527}

\bibitem[\protect\citeauthoryear{{Schaerer}, {Fragos}  \& {Izotov}}{{Schaerer}
  et~al.}{2019}]{schaerer2019}
{Schaerer} D.,  {Fragos} T.,   {Izotov} Y.~I.,  2019, \mn@doi [\aap]
  {10.1051/0004-6361/201935005}, \href
  {https://ui.adsabs.harvard.edu/abs/2019A&A...622L..10S} {622, L10}

\bibitem[\protect\citeauthoryear{{Schneider}, {Salvaterra}, {Ferrara}  \&
  {Ciardi}}{{Schneider} et~al.}{2006}]{schneider2006}
{Schneider} R.,  {Salvaterra} R.,  {Ferrara} A.,   {Ciardi} B.,  2006, \mn@doi
  [\mnras] {10.1111/j.1365-2966.2006.10331.x}, \href
  {https://ui.adsabs.harvard.edu/abs/2006MNRAS.369..825S} {369, 825}

\bibitem[\protect\citeauthoryear{{Schutte} \& {Reines}}{{Schutte} \&
  {Reines}}{2022}]{schutte2022}
{Schutte} Z.,  {Reines} A.,  2022, arXiv e-prints, \href
  {https://ui.adsabs.harvard.edu/abs/2022arXiv220108396S} {p. arXiv:2201.08396}

\bibitem[\protect\citeauthoryear{{Senchyna} et~al.,}{{Senchyna}
  et~al.}{2017}]{senchyna2017}
{Senchyna} P.,  et~al., 2017, \mn@doi [\mnras] {10.1093/mnras/stx2059}, \href
  {https://ui.adsabs.harvard.edu/abs/2017MNRAS.472.2608S} {472, 2608}

\bibitem[\protect\citeauthoryear{{Senchyna}, {Stark}, {Mirocha}, {Reines},
  {Charlot}, {Jones}  \& {Mulchaey}}{{Senchyna} et~al.}{2020}]{senchyna2020}
{Senchyna} P.,  {Stark} D.~P.,  {Mirocha} J.,  {Reines} A.~E.,  {Charlot} S.,
  {Jones} T.,   {Mulchaey} J.~S.,  2020, \mn@doi [\mnras]
  {10.1093/mnras/staa586}, \href
  {https://ui.adsabs.harvard.edu/abs/2020MNRAS.494..941S} {494, 941}

\bibitem[\protect\citeauthoryear{{Shapley} et~al.,}{{Shapley}
  et~al.}{2015}]{shapley2015}
{Shapley} A.~E.,  et~al., 2015, \mn@doi [\apj] {10.1088/0004-637X/801/2/88},
  \href {https://ui.adsabs.harvard.edu/abs/2015ApJ...801...88S} {801, 88}

\bibitem[\protect\citeauthoryear{{Shibuya} et~al.,}{{Shibuya}
  et~al.}{2018}]{shibuya2018_spec}
{Shibuya} T.,  et~al., 2018, \mn@doi [\pasj] {10.1093/pasj/psx107}, \href
  {https://ui.adsabs.harvard.edu/abs/2018PASJ...70S..15S} {70, S15}

\bibitem[\protect\citeauthoryear{{Shirazi}, {Brinchmann}  \&
  {Rahmati}}{{Shirazi} et~al.}{2014}]{shirazi2014}
{Shirazi} M.,  {Brinchmann} J.,   {Rahmati} A.,  2014, \mn@doi [\apj]
  {10.1088/0004-637X/787/2/120}, \href
  {https://ui.adsabs.harvard.edu/abs/2014ApJ...787..120S} {787, 120}

\bibitem[\protect\citeauthoryear{{Sobral}, {Matthee}, {Darvish}, {Schaerer},
  {Mobasher}, {R{\"o}ttgering}, {Santos}  \& {Hemmati}}{{Sobral}
  et~al.}{2015}]{sobral2015}
{Sobral} D.,  {Matthee} J.,  {Darvish} B.,  {Schaerer} D.,  {Mobasher} B.,
  {R{\"o}ttgering} H. J.~A.,  {Santos} S.,   {Hemmati} S.,  2015, \mn@doi
  [\apj] {10.1088/0004-637X/808/2/139}, \href
  {https://ui.adsabs.harvard.edu/abs/2015ApJ...808..139S} {808, 139}

\bibitem[\protect\citeauthoryear{{Sobral} et~al.,}{{Sobral}
  et~al.}{2019}]{sobral2019}
{Sobral} D.,  et~al., 2019, \mn@doi [\mnras] {10.1093/mnras/sty2779}, \href
  {https://ui.adsabs.harvard.edu/abs/2019MNRAS.482.2422S} {482, 2422}

\bibitem[\protect\citeauthoryear{{Stanway} \& {Eldridge}}{{Stanway} \&
  {Eldridge}}{2018}]{stanway2018}
{Stanway} E.~R.,  {Eldridge} J.~J.,  2018, \mn@doi [\mnras]
  {10.1093/mnras/sty1353}, \href
  {https://ui.adsabs.harvard.edu/abs/2018MNRAS.479...75S} {479, 75}

\bibitem[\protect\citeauthoryear{{Steidel} et~al.,}{{Steidel}
  et~al.}{2014}]{steidel2014}
{Steidel} C.~C.,  et~al., 2014, \mn@doi [\apj] {10.1088/0004-637X/795/2/165},
  \href {https://ui.adsabs.harvard.edu/abs/2014ApJ...795..165S} {795, 165}

\bibitem[\protect\citeauthoryear{{Steidel}, {Strom}, {Pettini}, {Rudie},
  {Reddy}  \& {Trainor}}{{Steidel} et~al.}{2016}]{steidel2016}
{Steidel} C.~C.,  {Strom} A.~L.,  {Pettini} M.,  {Rudie} G.~C.,  {Reddy} N.~A.,
    {Trainor} R.~F.,  2016, \mn@doi [\apj] {10.3847/0004-637X/826/2/159}, \href
  {https://ui.adsabs.harvard.edu/abs/2016ApJ...826..159S} {826, 159}

\bibitem[\protect\citeauthoryear{{Stiavelli} \& {Trenti}}{{Stiavelli} \&
  {Trenti}}{2010}]{stiavelli2010}
{Stiavelli} M.,  {Trenti} M.,  2010, \mn@doi [\apjl]
  {10.1088/2041-8205/716/2/L190}, \href
  {https://ui.adsabs.harvard.edu/abs/2010ApJ...716L.190S} {716, L190}

\bibitem[\protect\citeauthoryear{{Strom}, {Steidel}, {Rudie}, {Trainor},
  {Pettini}  \& {Reddy}}{{Strom} et~al.}{2017}]{strom2017}
{Strom} A.~L.,  {Steidel} C.~C.,  {Rudie} G.~C.,  {Trainor} R.~F.,  {Pettini}
  M.,   {Reddy} N.~A.,  2017, \mn@doi [\apj] {10.3847/1538-4357/836/2/164},
  \href {https://ui.adsabs.harvard.edu/abs/2017ApJ...836..164S} {836, 164}

\bibitem[\protect\citeauthoryear{{Tornatore}, {Ferrara}  \&
  {Schneider}}{{Tornatore} et~al.}{2007}]{tornatore2007}
{Tornatore} L.,  {Ferrara} A.,   {Schneider} R.,  2007, \mn@doi [\mnras]
  {10.1111/j.1365-2966.2007.12215.x}, \href
  {https://ui.adsabs.harvard.edu/abs/2007MNRAS.382..945T} {382, 945}

\bibitem[\protect\citeauthoryear{{Tumlinson}}{{Tumlinson}}{2006}]{tumlinson2006}
{Tumlinson} J.,  2006, \mn@doi [\apj] {10.1086/500383}, \href
  {https://ui.adsabs.harvard.edu/abs/2006ApJ...641....1T} {641, 1}

\bibitem[\protect\citeauthoryear{{Umeda}, {Ouchi}, {Nakajima}, {Isobe},
  {Aoyama}, {Harikane}, {Ono}  \& {Matsumoto}}{{Umeda}
  et~al.}{2022}]{umeda2022}
{Umeda} H.,  {Ouchi} M.,  {Nakajima} K.,  {Isobe} Y.,  {Aoyama} S.,  {Harikane}
  Y.,  {Ono} Y.,   {Matsumoto} A.,  2022, arXiv e-prints, \href
  {https://ui.adsabs.harvard.edu/abs/2022arXiv220106593U} {p. arXiv:2201.06593}

\bibitem[\protect\citeauthoryear{{Valiante}, {Schneider}, {Volonteri}  \&
  {Omukai}}{{Valiante} et~al.}{2016}]{valiante2016}
{Valiante} R.,  {Schneider} R.,  {Volonteri} M.,   {Omukai} K.,  2016, \mn@doi
  [\mnras] {10.1093/mnras/stw225}, \href
  {https://ui.adsabs.harvard.edu/abs/2016MNRAS.457.3356V} {457, 3356}

\bibitem[\protect\citeauthoryear{{Valiante}, {Schneider}, {Zappacosta},
  {Graziani}, {Pezzulli}  \& {Volonteri}}{{Valiante}
  et~al.}{2018}]{valiante2018}
{Valiante} R.,  {Schneider} R.,  {Zappacosta} L.,  {Graziani} L.,  {Pezzulli}
  E.,   {Volonteri} M.,  2018, \mn@doi [\mnras] {10.1093/mnras/sty213}, \href
  {https://ui.adsabs.harvard.edu/abs/2018MNRAS.476..407V} {476, 407}

\bibitem[\protect\citeauthoryear{{Vikaeus}, {Zackrisson}, {Schaerer}, {Visbal},
  {Fransson}, {Malhotra}, {Rhoads}  \& {Sahl{\'e}n}}{{Vikaeus}
  et~al.}{2021}]{vikaeus2021}
{Vikaeus} A.,  {Zackrisson} E.,  {Schaerer} D.,  {Visbal} E.,  {Fransson} E.,
  {Malhotra} S.,  {Rhoads} J.,   {Sahl{\'e}n} M.,  2021, arXiv e-prints, \href
  {https://ui.adsabs.harvard.edu/abs/2021arXiv210701230V} {p. arXiv:2107.01230}

\bibitem[\protect\citeauthoryear{{Visbal}, {Bryan}  \& {Haiman}}{{Visbal}
  et~al.}{2020}]{visbal2020}
{Visbal} E.,  {Bryan} G.~L.,   {Haiman} Z.,  2020, \mn@doi [\apj]
  {10.3847/1538-4357/ab994e}, \href
  {https://ui.adsabs.harvard.edu/abs/2020ApJ...897...95V} {897, 95}

\bibitem[\protect\citeauthoryear{{Volonteri}}{{Volonteri}}{2012}]{volonteri2012}
{Volonteri} M.,  2012, \mn@doi [Science] {10.1126/science.1220843}, \href
  {https://ui.adsabs.harvard.edu/abs/2012Sci...337..544V} {337, 544}

\bibitem[\protect\citeauthoryear{{Volonteri}, {Reines}, {Atek}, {Stark}  \&
  {Trebitsch}}{{Volonteri} et~al.}{2017}]{volonteri2017}
{Volonteri} M.,  {Reines} A.~E.,  {Atek} H.,  {Stark} D.~P.,   {Trebitsch} M.,
  2017, \mn@doi [\apj] {10.3847/1538-4357/aa93f1}, \href
  {https://ui.adsabs.harvard.edu/abs/2017ApJ...849..155V} {849, 155}

\bibitem[\protect\citeauthoryear{{Whalen}, {Surace}, {Bernhardt}, {Zackrisson},
  {Pacucci}, {Ziegler}  \& {Hirschmann}}{{Whalen} et~al.}{2020}]{whalen2020}
{Whalen} D.~J.,  {Surace} M.,  {Bernhardt} C.,  {Zackrisson} E.,  {Pacucci} F.,
   {Ziegler} B.,   {Hirschmann} M.,  2020, \mn@doi [\apjl]
  {10.3847/2041-8213/ab9d29}, \href
  {https://ui.adsabs.harvard.edu/abs/2020ApJ...897L..16W} {897, L16}

\bibitem[\protect\citeauthoryear{{Yue}, {Ferrara}, {Salvaterra}, {Xu}  \&
  {Chen}}{{Yue} et~al.}{2013}]{yue2013}
{Yue} B.,  {Ferrara} A.,  {Salvaterra} R.,  {Xu} Y.,   {Chen} X.,  2013,
  \mn@doi [\mnras] {10.1093/mnras/stt826}, \href
  {https://ui.adsabs.harvard.edu/abs/2013MNRAS.433.1556Y} {433, 1556}

\bibitem[\protect\citeauthoryear{{Zamorani} et~al.,}{{Zamorani}
  et~al.}{1981}]{zamorani1981}
{Zamorani} G.,  et~al., 1981, \mn@doi [\apj] {10.1086/158815}, \href
  {https://ui.adsabs.harvard.edu/abs/1981ApJ...245..357Z} {245, 357}

\makeatother
\end{thebibliography}


\appendix

\section{Photoionisation model results assuming different gas densities} 
\label{sec_app:results_different_densities}

In the main text we present results of photoionisation models by mostly adopting a gas density
fixed to the fiducial value of $\rm 10^3 cm^{-3}$. In this appendix, we present the results with different gas densities
and examine how the diagnostics are (not) impacted by the choice of gas density.

\begin{figure*}
  \centering
    \begin{tabular}{c}
      \begin{minipage}{0.24\hsize}
        \begin{center}
         \includegraphics[bb=18 143 555 680, width=0.95\textwidth]{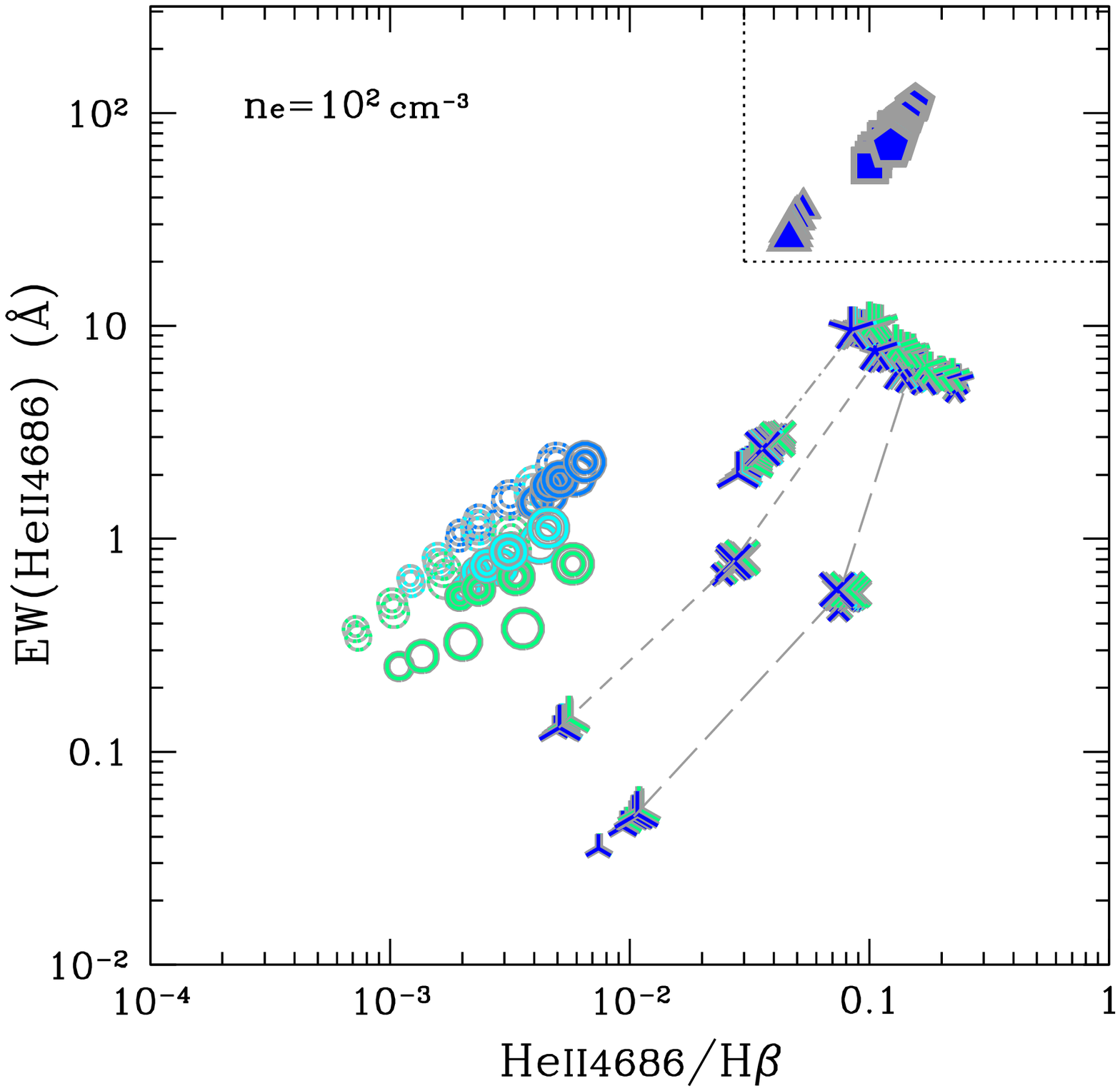}
        \end{center}
      \end{minipage}
      \begin{minipage}{0.24\hsize}
        \begin{center}
         \includegraphics[bb=18 143 555 680, width=0.95\textwidth]{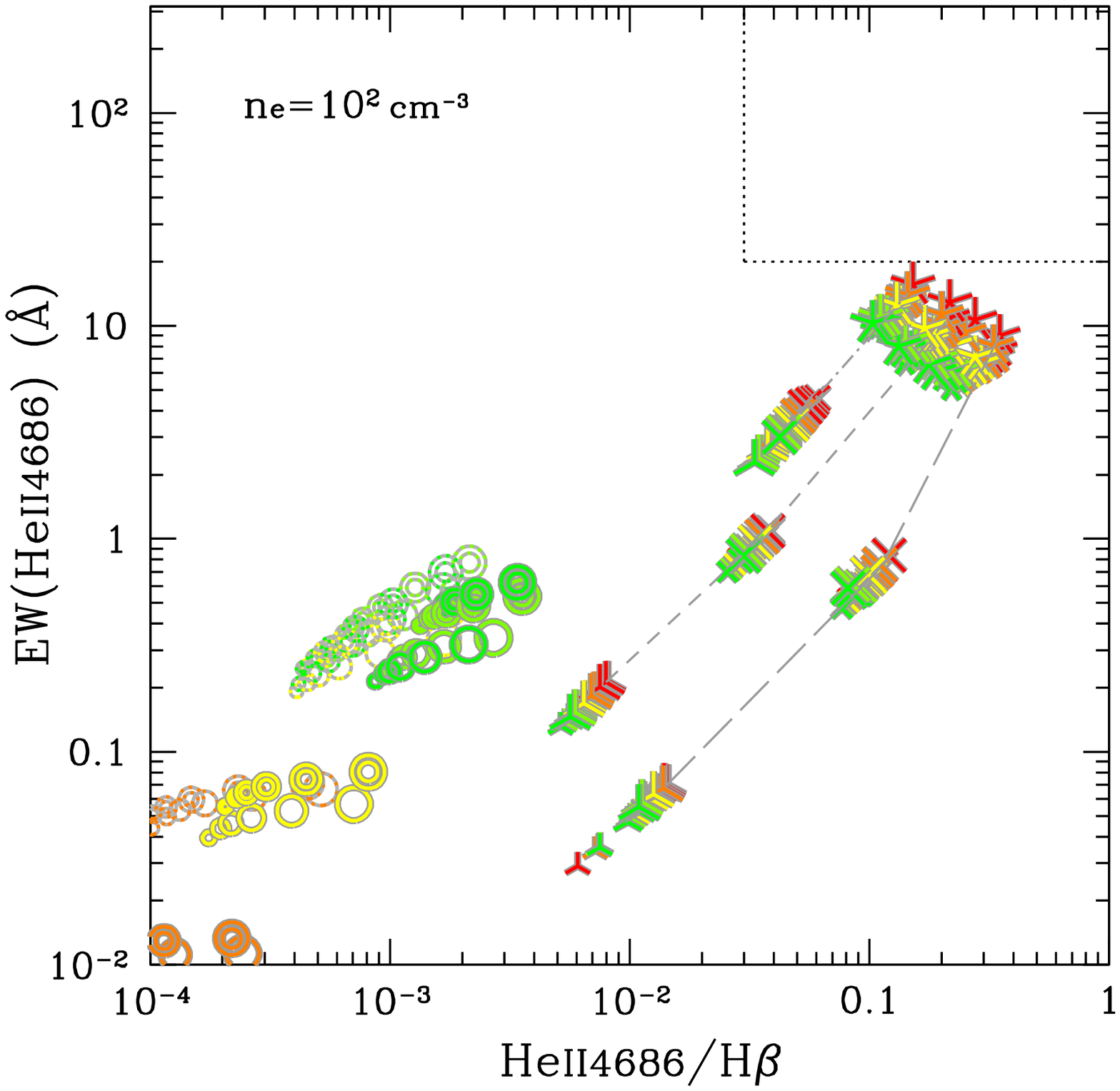}
        \end{center}
      \end{minipage}
      \hspace{0.02\hsize}
      \begin{minipage}{0.24\hsize}
        \begin{center}
         \includegraphics[bb=18 143 555 680, width=0.95\textwidth]{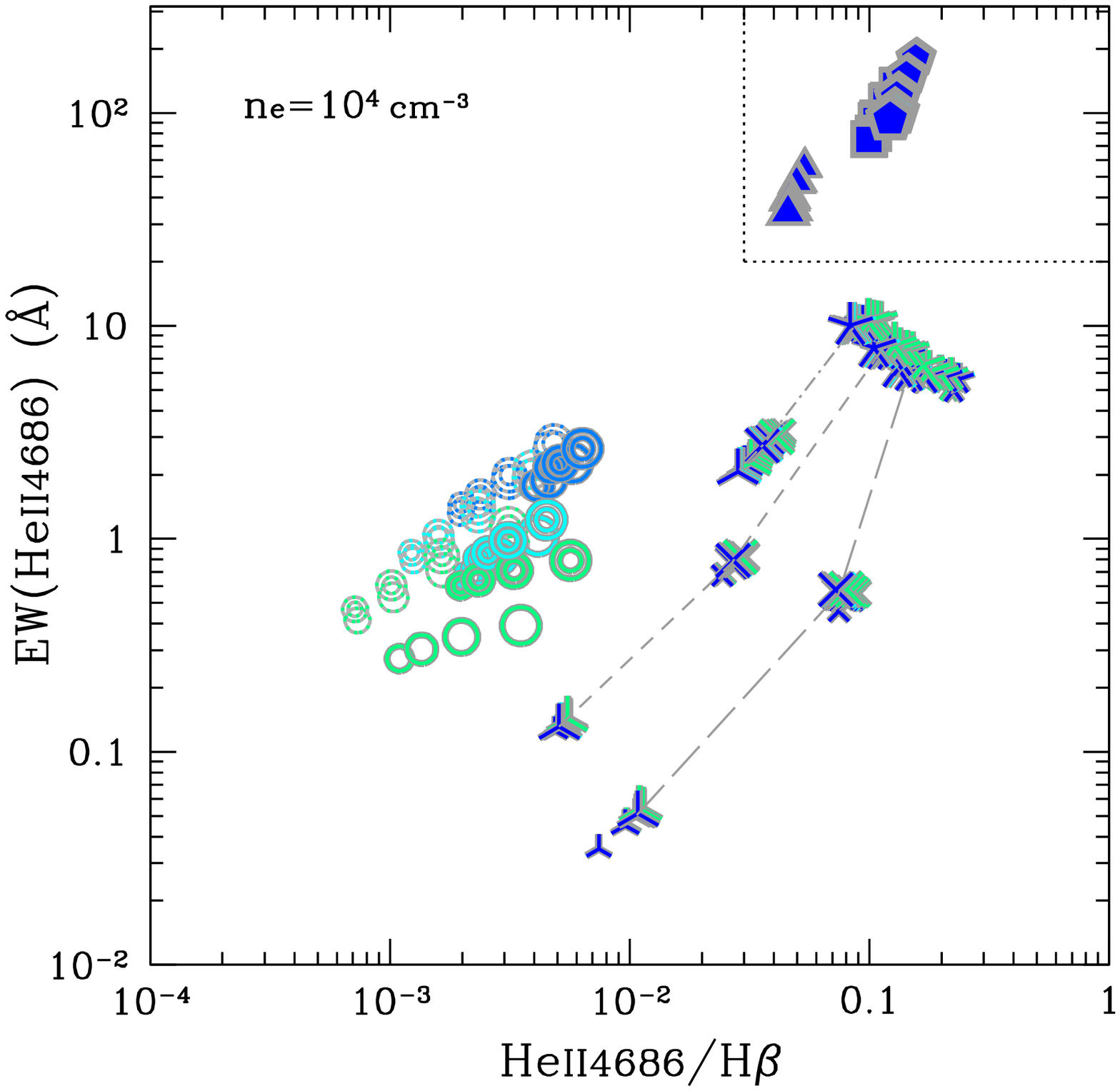}
        \end{center}
      \end{minipage}
      \begin{minipage}{0.24\hsize}
        \begin{center}
         \includegraphics[bb=18 143 555 680, width=0.95\textwidth]{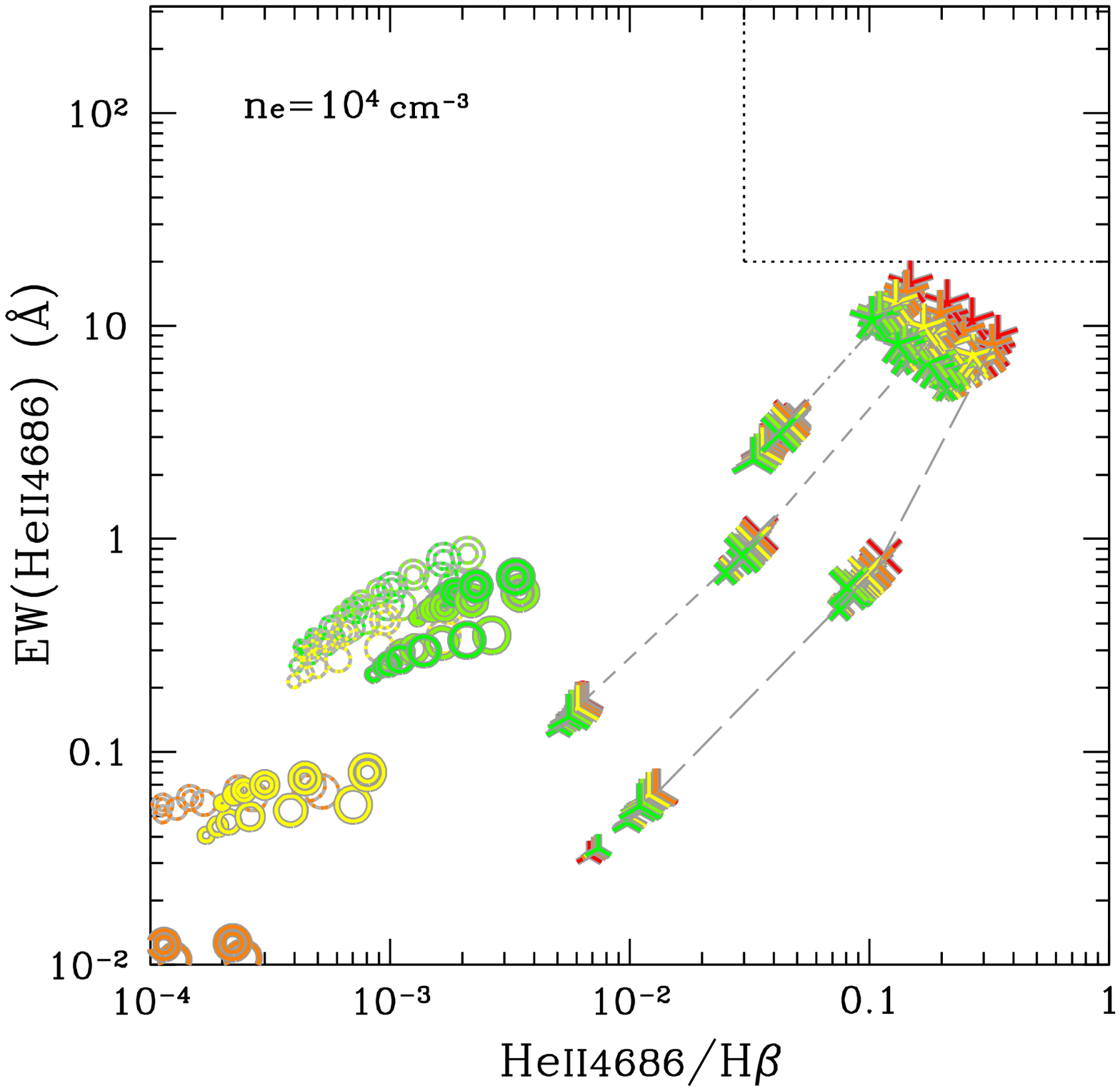}
        \end{center}
      \end{minipage}
    \end{tabular}
    \caption{%
    	    	Same as Fig.~\ref{fig:ewhe2_he2hb}, but with a different gas density of 
	    	$10^2$ (left two panels) and $10^4$\,cm$^{-3}$ (right two panels).
    }
    \label{fig:ewhe2_he2hb_appendix}
\end{figure*}

\begin{figure*}
  \centering
    \begin{tabular}{c}
      \begin{minipage}{0.24\hsize}
        \begin{center}
         \includegraphics[bb=18 143 555 680, width=0.95\textwidth]{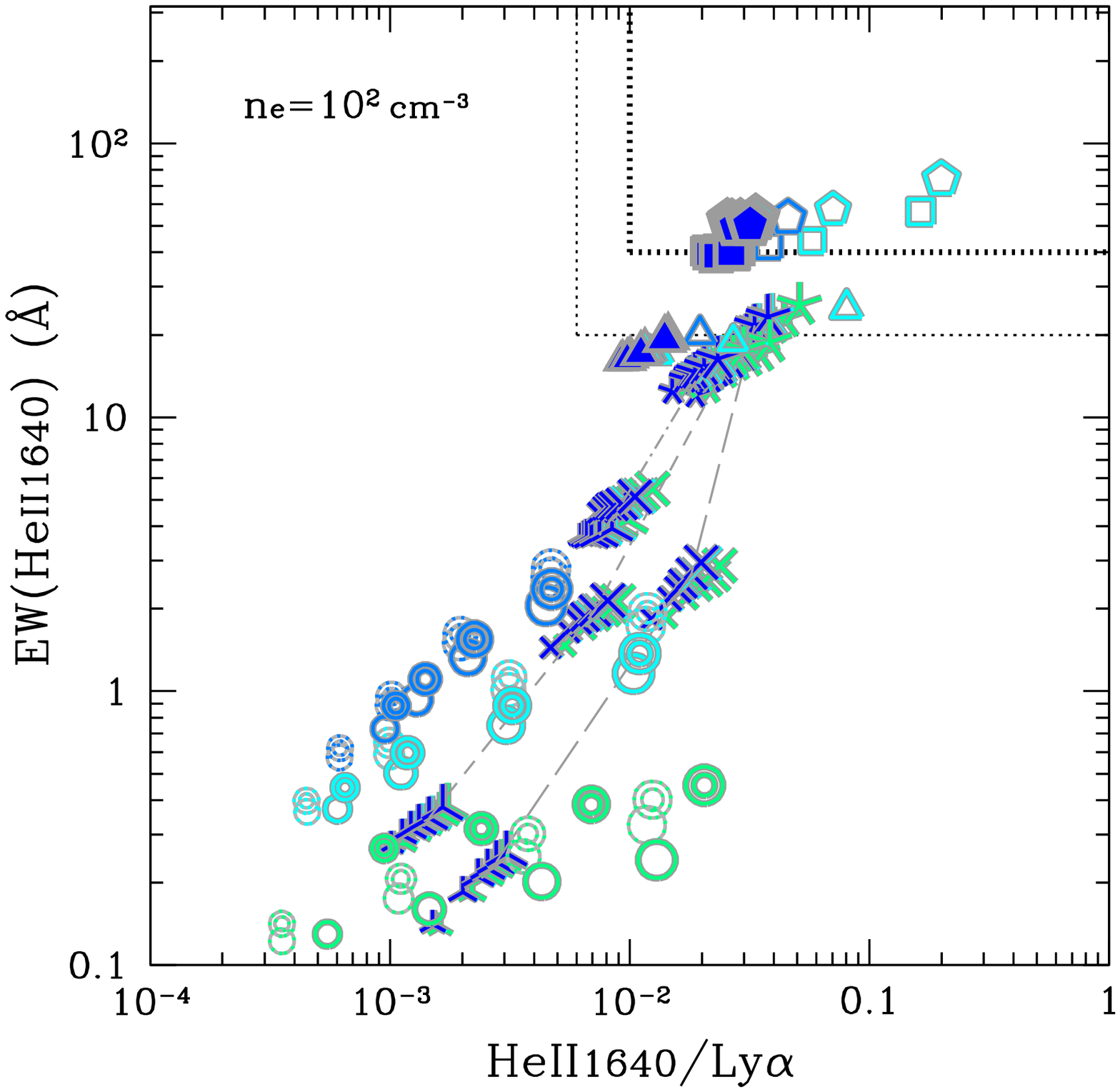}
        \end{center}
      \end{minipage}
      \begin{minipage}{0.24\hsize}
        \begin{center}
         \includegraphics[bb=18 143 555 680, width=0.95\textwidth]{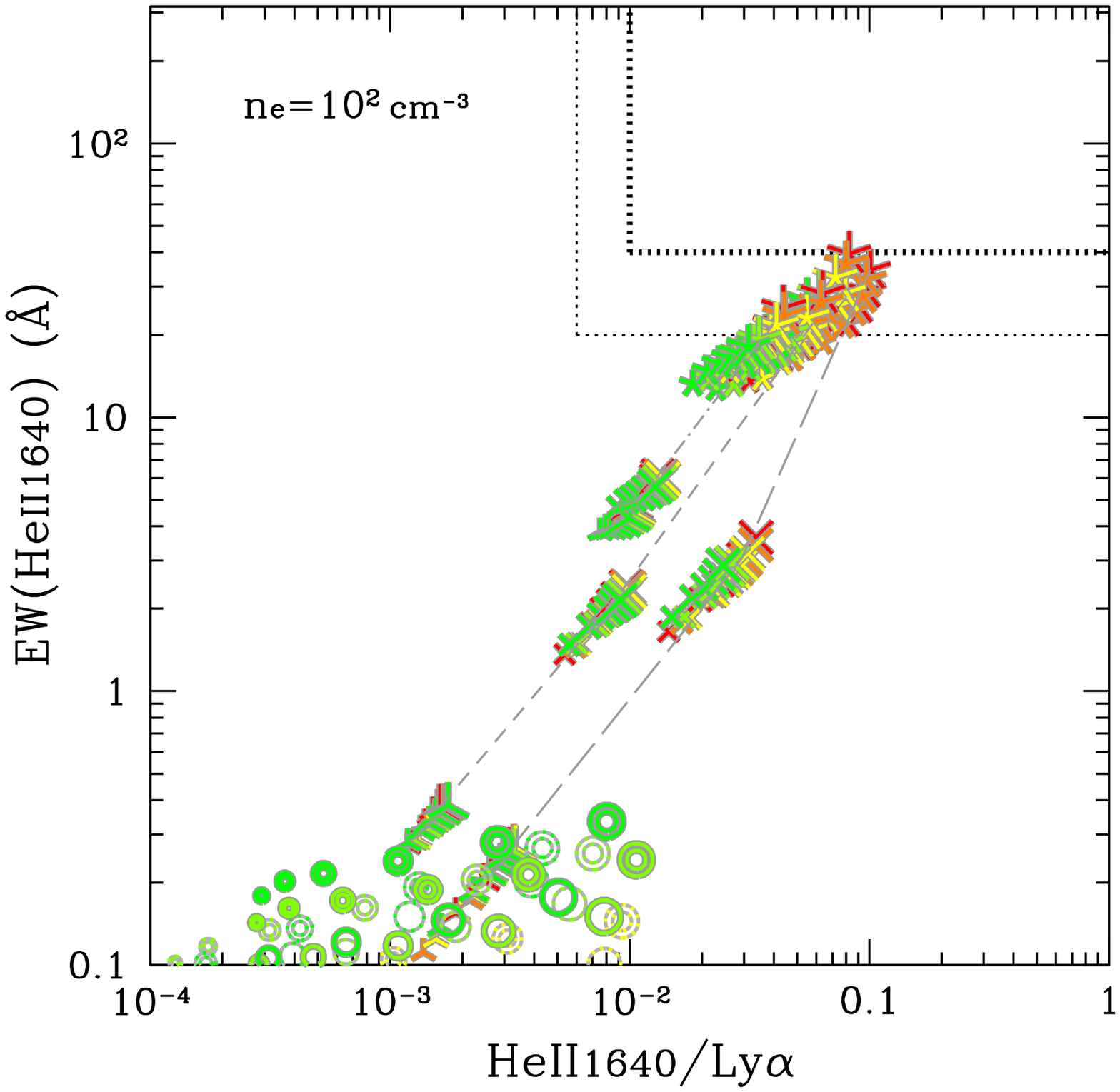}
        \end{center}
      \end{minipage}
      \hspace{0.02\hsize}
      \begin{minipage}{0.24\hsize}
        \begin{center}
         \includegraphics[bb=18 143 555 680, width=0.95\textwidth]{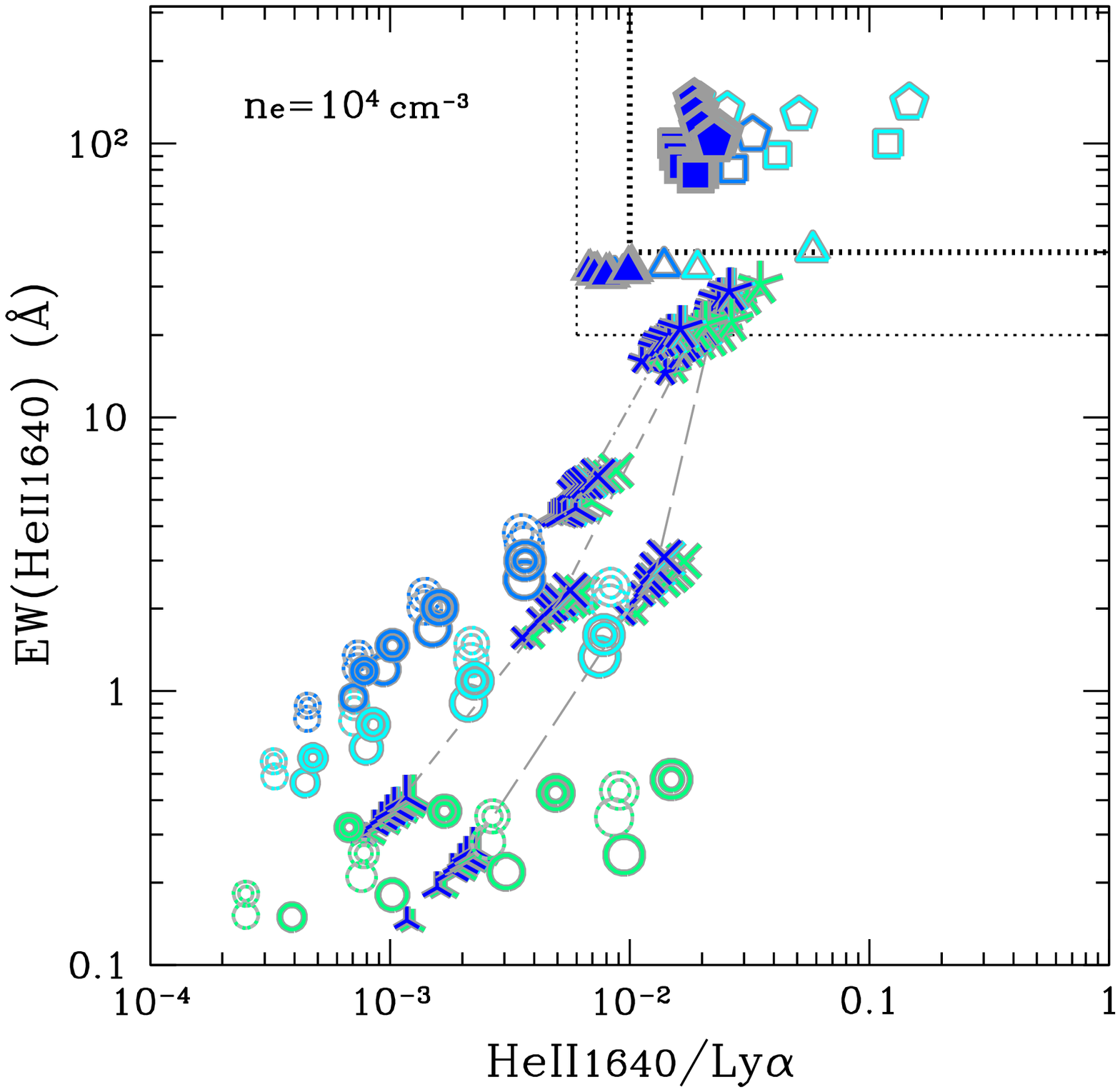}
        \end{center}
      \end{minipage}
      \begin{minipage}{0.24\hsize}
        \begin{center}
         \includegraphics[bb=18 143 555 680, width=0.95\textwidth]{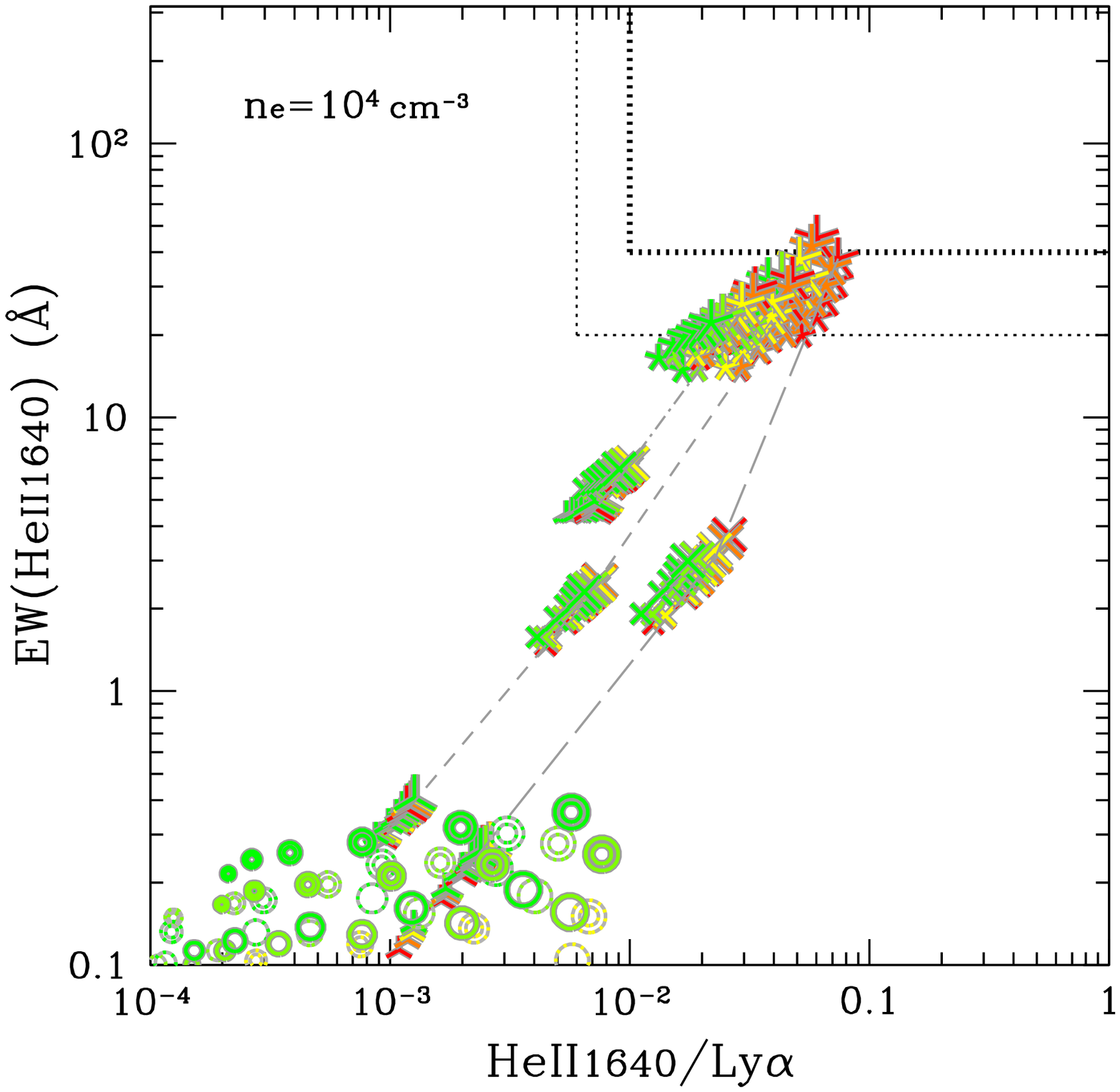}
        \end{center}
      \end{minipage}
    \end{tabular}
    \caption{%
    	    	Same as Fig.~\ref{fig:ewhe2_he2lya}, but with a different gas density of 
	    	$10^2$ (left two panels) and $10^4$\,cm$^{-3}$ (right two panels).
    }
    \label{fig:ewhe2_he2lya_appendix}
\end{figure*}

\begin{figure*}
  \centering
    \begin{tabular}{c}
      \begin{minipage}{0.24\hsize}
        \begin{center}
         \includegraphics[bb=18 143 555 680, width=0.95\textwidth]{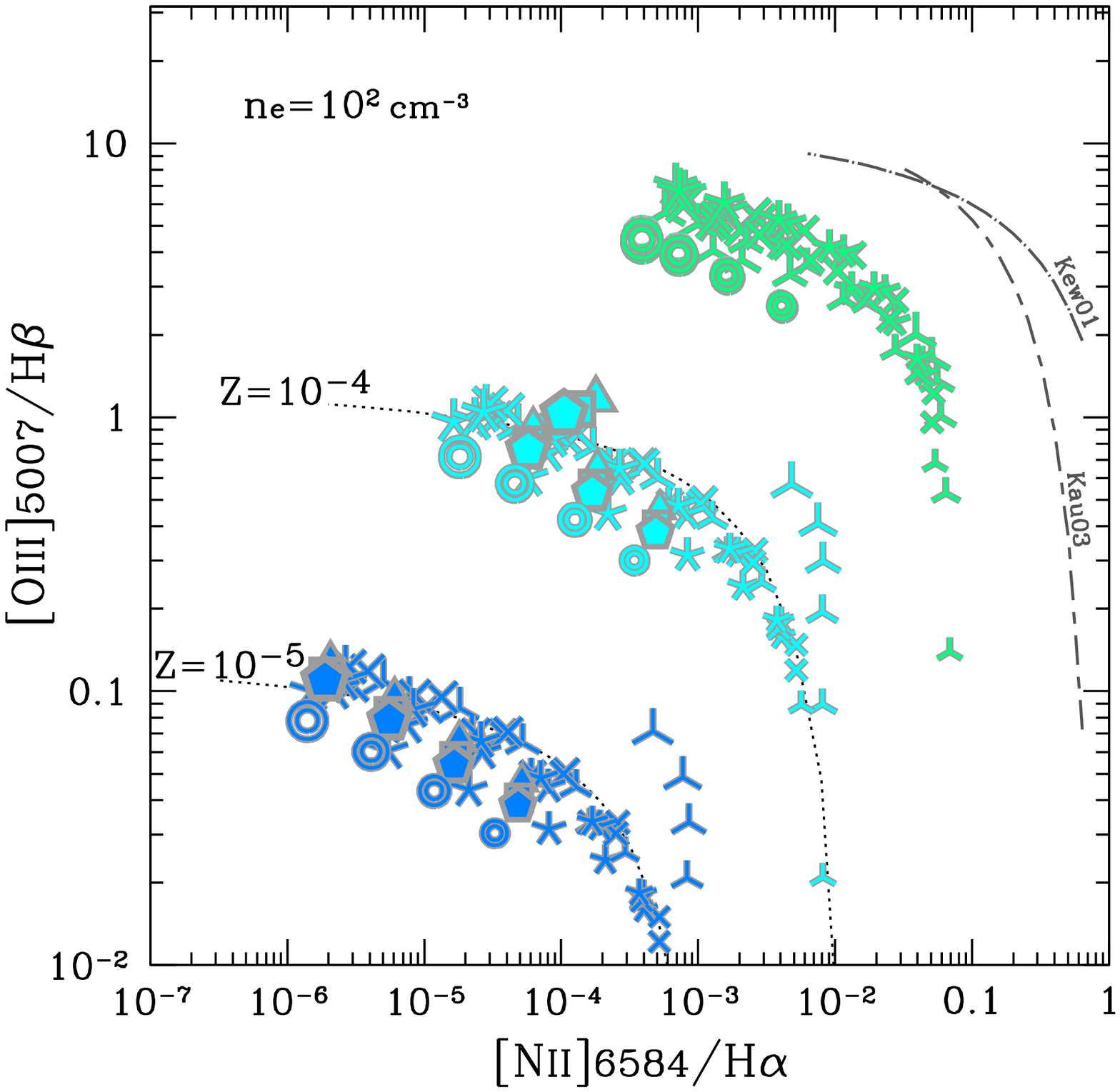}
        \end{center}
      \end{minipage}
      \begin{minipage}{0.24\hsize}
        \begin{center}
         \includegraphics[bb=18 143 555 680, width=0.95\textwidth]{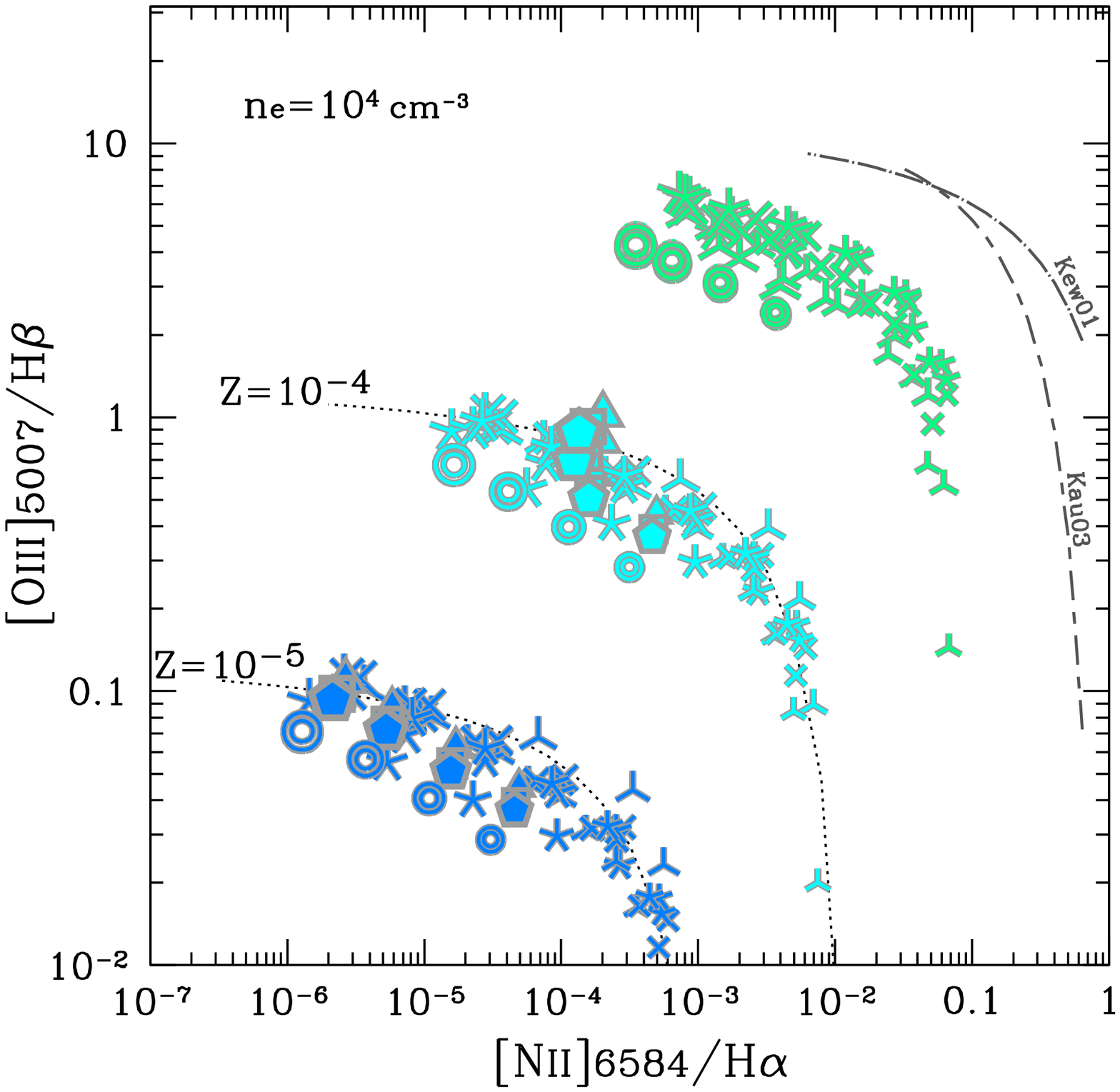}
        \end{center}
      \end{minipage}
      \hspace{0.02\hsize}
      \begin{minipage}{0.24\hsize}
        \begin{center}
         \includegraphics[bb=18 143 555 680, width=0.95\textwidth]{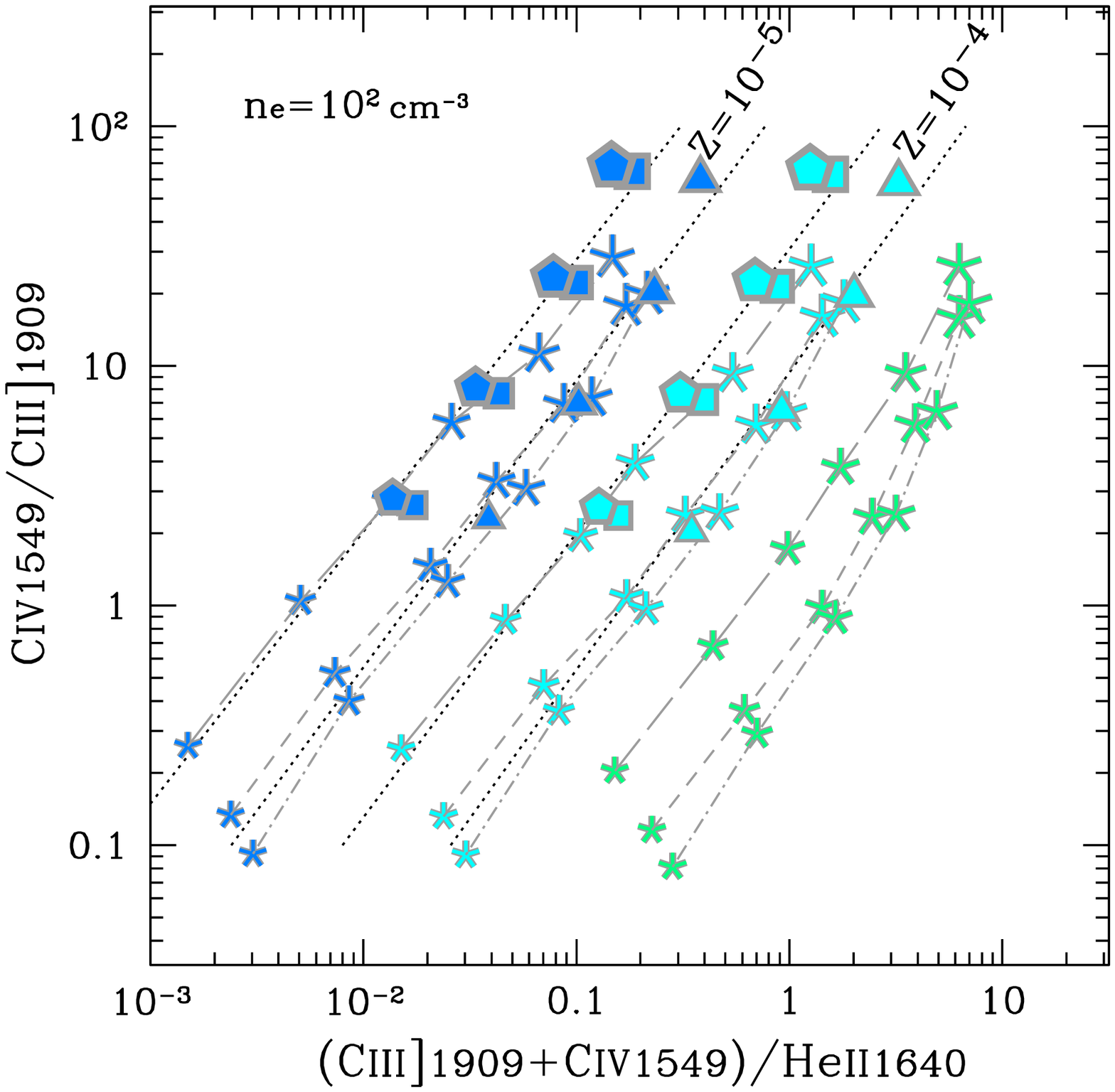}
        \end{center}
      \end{minipage}
      \begin{minipage}{0.24\hsize}
        \begin{center}
         \includegraphics[bb=18 143 555 680, width=0.95\textwidth]{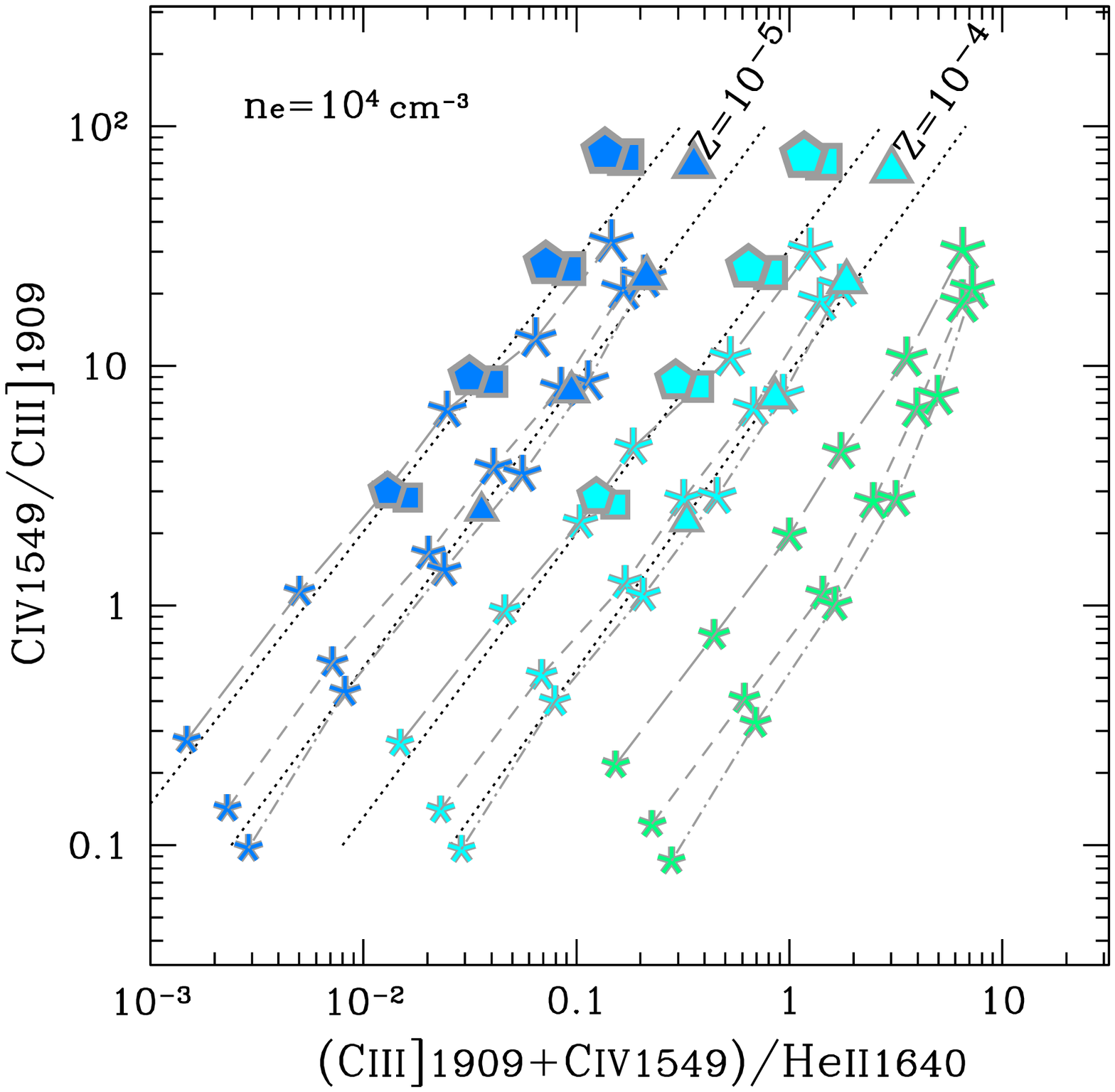}
        \end{center}
      \end{minipage}
    \end{tabular}
    \caption{%
    		Metallicity diagnostics. 
    		(Left 2 panels:) The \NII\ BPT diagram (same as in top-left in Fig.~\ref{fig:metals_optical})
		with a different gas density of $10^2$ (left) and $10^4$\,cm$^{-3}$ (right).
		(Right 2 panels:) The C4C3-C34 diagram (same as in left in Fig.~\ref{fig:c4c3_c34})
		with a different gas density of $10^2$ (left) and $10^4$\,cm$^{-3}$ (right).
    }
    \label{fig:metals_appendix}
\end{figure*}

Fig.~\ref{fig:ewhe2_he2hb_appendix} presents the optical \HeII\ diagnostic
for a gas density of $10^2$ (left panels) and $10^4$\,cm$^{-3}$ (right panels). 
This confirms that the recombination lines ratio \HeII$/$\Hb\ is almost unchanged 
for any of the populations over the reasonable range of gas density from $10^2$ to $10^4$\,cm$^{-3}$.
EW(\HeII$\lambda 4686$) shows a weak dependence on density, 
getting stronger by a factor of $\sim 1.2$ if a gas density higher by a factor of 10 is assumed. 
This dependence does not change the selection criteria for PopIII galaxies on this diagram (Eq.~\ref{eq:ewhe2_he2hb}).
We therefore reaffirm that the optical \HeII\ diagram is the best discriminator of PopIII galaxies.

Fig.~\ref{fig:ewhe2_he2lya_appendix} shows the UV \HeII\ diagnostics for different gas densities.
A higher gas density leads to a stronger EW(\HeII$\lambda 1640$), by a factor of $\sim 1.3$ 
if assuming a gas density ten times higher, in a similar way as seen for the optical \HeII\ emission.
The 	strict selection criteria (Eq.~\ref{eq:ewhe2_he2lya_strict}) of EW(\HeII$\lambda 1640$) $>40$\,\AA\
is confirmed valid even for the low gas density case. 
Although some metal-rich AGNs with a hard ionising spectrum ($\alpha=-1.2$) and 
a high density ($\sim 10^4$\,cm$^{-3}$) can present an EW(\HeII$\lambda 1640)$
as strong as $\sim 40$\,\AA, such objects can be well-discriminated based on the diagnostics
using the UV metal lines (\S\ref{sssec:results_uv_metals}).
On the other hand, the relaxed selection criteria (Eq.~\ref{eq:ewhe2_he2lya_relaxed})
of EW(\HeII$\lambda 1640$) $>20$\,\AA\ may miss some PopIII galaxies with a modest IMF
if the gas density is as low as $\sim 10^2$\,cm$^{-3}$. 
This brings us another caveat when using the UV \HeII\ diagnostic.
The conservative lower-limits on \HeII$/$\Lya\ ratio remain valid irrespective of the gas density.

In Fig.~\ref{fig:metals_appendix}, we show some key diagnostic emission line ratios using metal lines to assess their dependences on gas density. We only present the low-metallicity models for the \NII\ BPT diagram and C4C3-C34 diagram as representatives. The dotted curves tracing the $\rm Z = 10^{-5}$ and $10^{-4}$ are as determined based on the fiducial gas density models (Eqs.~\ref{eq:o3hb_n2ha_zem5}--\ref{eq:o3hb_n2ha_zem4}, \ref{eq:c4c3_c34_A_zem5}--\ref{eq:c4c3_c34_B_zem4}). Fig.~\ref{fig:metals_appendix} demonstrates that these dotted curves work well for the different gas density models, confirming that the gas density has little effect on the diagnostic emission line ratios, and hence on the metallicity estimations based on the diagrams. Although these metal lines are collisional excitation lines, the effect is minimal at the low metallicity regime, especially when we use line ratios.

We also emphasise that our primary purpose here is to explore variations of metallicity spanning orders of magnitude; compared with the effects of such large metallicity variations, the effects of gas density variations on the diagnostics is essentially unimportant (with the exception of the \HeI\ lines, as shown in Fig.~\ref{fig:he2hb_he1hb_5876_density}). We can therefore conclude that the diagnostics developed in the main text work for different gas densities, and hence for a system with a possible internal variation of gas density, within the reasonable range of $10^2-10^4$\,cm$^{-3}$, are solid for constraining (when possible) the nature of different classes of sources and their metallicities.


\label{lastpage}
\end{document}